\documentclass[usenatbib,usegraphicx]{mn2e}

\usepackage{times,mathptm}
\newlength{\greywidth}
\newlength{\contwidth}
\newlength{\profwidth}
\title[\textit B band luminosities of QSO host galaxies]
{The \textbfit{B\/} band luminosities of QSO host galaxies
\thanks{Based on observations made at the European Southern
Observatory, La Silla, Chile} }
\author[K. Jahnke \& L. Wisotzki] 
       {Knud Jahnke$^{1,2}$ and Lutz Wisotzki$^{1,3}$ \\
	$^1$Astrophysikalisches Institut Potsdam, An der Sternwarte
16, 14482 Potsdam, Germany\\
        $^2$Hamburger Sternwarte, Gojenbergsweg 112, 21029 Hamburg, Germany\\
	$^3$Universit\"at Potsdam, Am Neuen Palais 10, 14469 Potsdam, Germany}

\begin{document}

\date{}

\pagerange{\pageref{firstpage}--\pageref{lastpage}} \pubyear{2003}

\maketitle

\label{firstpage}

\begin{abstract}
We report on the analysis of $B$ band imaging data of 57 low-redshift
QSOs and Seyfert~1 galaxies selected from the Hamburg/ESO-Survey, for
which host galaxy dependent selection biases are greatly reduced
compared to other optical surveys. Only one object in the sample is
known to be radio-loud.

We adopted a procedure to remove the AGN contribution by subtracting a
scaled point spread functions from each QSO image. To reclaim the
integrated host galaxy flux we correct for oversubtraction based on
simulations. This method shows to be quite insensitive to the host
galaxy morphological type, which we can unambiguously established for
15 of the 57 objects.

The quasar host galaxies are detected in all cases. The hosts are very
luminous, ranging in absolute magnitude $M_B$ from $-19.0$ to $-23.8$,
with an average of $M_{B,\rmn{gal}} = -21.5$, considerably above $L^*$
for field galaxies. For the luminous QSO subsample with $M_B < -23$
the average host absolute magnitude is $M_{B,\rmn{gal}} = -23.0$,
while for the complementary low-luminosity AGN we get $M_{B,\rmn{gal}}
= -21.2$, roughly equal to $L^*$.

The luminous host galaxies in the sample are typically $\sim$1~mag
brighter than expected when inferring $B$ band luminosities from
studies of similar objects at longer wavebands. We argue that this
mismatch is not likely to be explained by selection effects, but favor
host galaxy colours significanlty bluer than those of inactive
galaxies. Although published $B$ band data are scant, this result and
the findings of other authors are in good agreement.

\end{abstract}


\section{Introduction}\label{sec:intro}
Imaging studies of QSO host galaxies show a wide variety of galaxy
types and luminosities: some are large and very luminous ellipticals,
others are perfect spirals, yet others show strong evidence for tidal
distortions and merging. A few QSOs reside in rich clusters while the
majority prefers loose groups. So far, there are very few noticeable
correlations between nuclear and host properties. One of the most
frequently quoted cases is the dichotomy of radio-loud and radio-quiet
QSOs, the former claimed to be harboured by ellipticals, the latter by
spiral galaxies. Recent investigations have shown this to be a strong
oversimplification. Also many radio-quiet QSOs are hosted in
elliptical or bulge dominated galaxies \citep{tayl96,bahc97,dunl03},
the host galaxy type is more a function of nuclear luminosity than
radio properties.

The most fundamental relation found in in recent years, between the
mass of central black holes and the bulge mass of the surrounding
galaxies \citep{mago98,gebh00,ferr00}, seems to be valid for both
inactive and active galaxies \citep{mclu02}. This relation links the
nuclear and host galaxy luminosities and explains the correlation of
these parameters found in several studies to date
\citep[e.g.][]{mcle95b,scha00}.

There is ample evidence that QSOs are generally found in galaxies of
luminosities -- and thus masses -- above the average field population
\citep[e.g.,][]{dunl93,mcle94a,mcle94b,roen96}, although the effect
seems to be less pronounced for low-luminosity Seyfert galaxies
\citep{koti94,mcle95b}. Other properties are much less
constrained. Obtaining reliable morphological information beyond a
binary Hubble-type classification, such as bulge/disc ratios or scale
lengths, is compromised by the presence of the bright AGN in the
centre, often outshining the entire galaxy. HST has brought major
advancement in this respect \citep{bahc97,mclu99,scha00}, allowing to
better resolve substructures in the hosts.

A severe limitation in interpreting the relation of active to inactive
galaxies is the fact that spectral information is sparse. Especially
among the more luminous QSO hosts, colours are available for only very
few objects. Yet, if galaxy interactions and mergers are important in
triggering nuclear activity as repeatedly suggested
\citep[e.g.,][]{stoc82,hern89,kauf00}, this should be reflected in the
host galaxy colours, possibly tinted blue by enhanced star-forming
activity. In fact, there have been some claims that QSO hosts have
bluer colours than average field galaxies
\citep{hutc89,mcle95a,roen96,cana00a}, but conclusive evidence is
certainly weak as studies with contradictory results exist
\citep[e.g.][]{koti94,scha00,dunl03}.

A general shortcoming of many previous investigations is the
definition of observed samples. QSO surveys always impose certain
selection criteria, some of which may affect the average host
properties; an obvious example is the rejection of `non-stellar
objects' frequently applied in optical QSO surveys, immediately
imposing redshift-dependent morphological biases in the
samples. Constructing mixed samples from large catalogues of
inhomogeneous composition does by no means ensure that such effects
are eliminated, and it is conceivable that some of the past
disagreements on host galaxy properties were due to artefacts of
improper -- i.e., non-representative -- samples.

In this paper we adress the question of potentially blue colours of
quasar host galaxies. We present an observational study of
low-redshift QSOs and Seyfert galaxies selected in the course of the
Hamburg/ESO Survey. The study is one of the few conducted in the
optical $B$ band, being particularly sensitive to star forming
components in low-redshift galaxies. With 57 objects it is also one of
the largest host galaxy samples investigated altogether.

In the following we describe the sample and observations
(Sect.~\ref{sec:data}) and our adopted method to extract the host
galaxy from the initial images (Sect.~\ref{sec:analysis}). The derived
host galaxy properties are reported in Section~\ref{sec:results} and
compared to existing studies in
Section~\ref{sec:others}. Section~\ref{sec:discussion} provides a
discussion of the results and our conclusions are presented in
Section~\ref{sec:conclusions}. Throughout the paper we adopt a world
model with $H_0 = 50$~km~s$^{-1}$~Mpc$^{-1}$, $\Omega = 1$, and
$\Lambda = 0$.

\section{Data} \label{sec:data}

\subsection{Sample selection}\label{sec:selection}

The objects discussed in this paper were originally selected on
photographic Schmidt plates obtained in the course of the Hamburg/ESO
survey for bright QSOs \citep[HES,][]{wiso96,reim96,wiso00}. The HES
is a wide-angle survey covering the entire southern extragalactic
hemisphere with an average limiting magnitude of $B\la 17.5$, based on
automated quasar candidate selection with digitised objective prism
plates. One important feature of the HES selection procedure is the
treatment of low-redshift QSOs and AGN. In contrast to most other
optical surveys, there is no discrimination against objects with
extended morphological structure. At the same time, specific selection
criteria ensure maximum completeness also for lower-luminosity
Seyfert~1 galaxies and objects on the classical QSO/Seyfert
borderline. Among other results, the survey yielded a direct estimate
of the local luminosity function of QSOs and Seyfert~1 nuclei
\citep{koeh97,wiso00c}, showing that the space density of luminous
low-$z$ QSOs is actually much higher than previously assumed.

45 of the objects form a random subsample with $z<0.2$ of the full set
of now more than 350 QSOs with $z<0.2$ selected by the HES, only a
small fraction of which were previously listed in QSO catalogues. A
synopsis of redshifts and magnitudes is given in Table
\ref{tab:objekte1}. Only one of the objects is known to be a powerful
radio emitter (PKS\,1020--103). While it is too early to make strong
statements about the radio properties of HES-selected quasars, a
systematic radio follow-up programme of the HES being in progress, the
sample is certainly dominated by radio-quiet QSOs (RQQs). The in- or
exclusion of a few radio-loud objects (RLQs) has no effect on any of
the statistical conclusions drawn below.

An additional set of 12 mostly low-luminosity AGN was selected in the
course of an identification program for ROSAT X-ray sources, using the
HES database of digital objective prism spectra to provide basic
spectral information about possible optical counterparts. Further
details are given by \citet{bade95}. Again, since the identification
was performed irrespective of optical morphology, no corresponding
bias in the subsample is expected. The objects of this subsample are
marked as `RX' sources in Table~\ref{tab:objekte1}. In the remainder
of this paper no distinction is made between these and the optically
selected objects, referring to the combined dataset as `the HES
sample'.

\begin{figure}
\begin{center}
\includegraphics[bb = 70 584 323 770,clip,angle=0,width=\columnwidth]{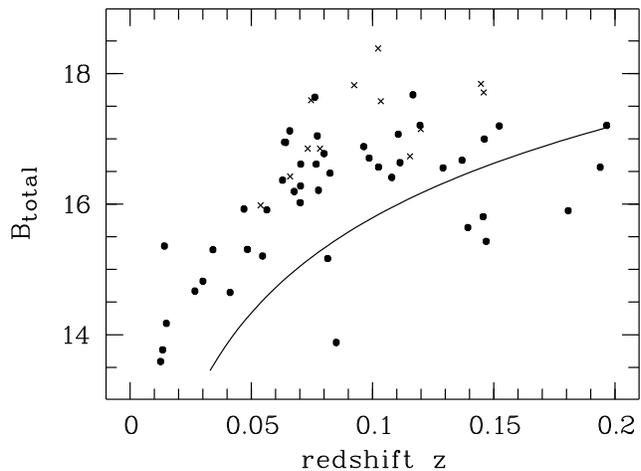}
\end{center}
\caption{
Hubble diagram for the sample. Circles mark the objects drawn from the
HES, crosses the RX objects selected as ROSAT sources on HES
plates. The solid line denotes the apparent magnitude of a QSO with
$M_B = -23$ at the given redshift.
}
\label{fig:B_z_distribution1}
\end{figure}

\begin{table*}
\begin{small}
\caption{Summary of observing campaigns.}
\begin{tabular}{cccccccc}
\hline
Campaign& ID& Telescope& Instrument& CCD&Scale& \# Obj.& Exposure\\ 
&&ESO&&ESO&[\arcsec/pix]&&[seconds]\\
\hline
\hline
12/90 & a& 3.6\,m& EFOSC1 & \#8 &0.34&8& 30\\
02/92 & b& 2.2\,m& EFOSC2 & \#19 &0.34&4& 60\\
02/92 & c& 3.6\,m& EFOSC1 & \#26 &0.61&20& 30\\
03/93 & d& 3.6\,m& EFOSC1 & \#26 &0.61&25& 30\\
\hline
\end{tabular}
\label{tab:campaigns}
\end{small}
\end{table*}

\subsection{Observations}\label{sec:obs}

The data were obtained during several observing runs de\-di\-ca\-ted
to low resolution slit spectroscopy with the principal aim to confirm
the AGN nature of candidates \citep{reim96}. During the early phase of
the survey (between 1990 and 1993) the main instrument for this
purpose was the ESO 3.6\,m telescope with its focal reducer
spectrograph EFOSC1. As a part of the acquisition procedure, taking at
least one direct CCD image was mandatory; these images form the input
dataset for the present analysis. To enable photometric comparison
with our blue-sensitive photographic survey material, we always used a
$B$ band filter. Typical exposure time was 30\,s per image, under
seeing conditions varying between $1\farcs 1$ and $1\farcs 7$. Some
properties of the observational campaigns are listed in
Table~\ref{tab:campaigns}.

\begin{figure*}
\setlength{\unitlength}{1cm}
\begin{picture}(17.7,5.85)
\put(0.0,3.35){%
\includegraphics[bb = 98 98 299 299,clip,angle=0,width=\greywidth]{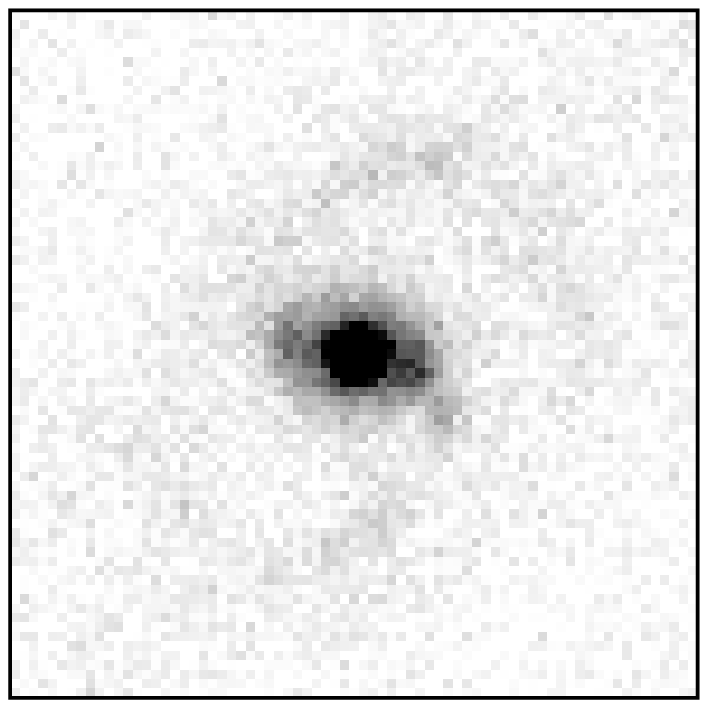}}
\put(0.0,0.15){%
\includegraphics[bb = 98 98 299 299,clip,angle=0,width=\greywidth]{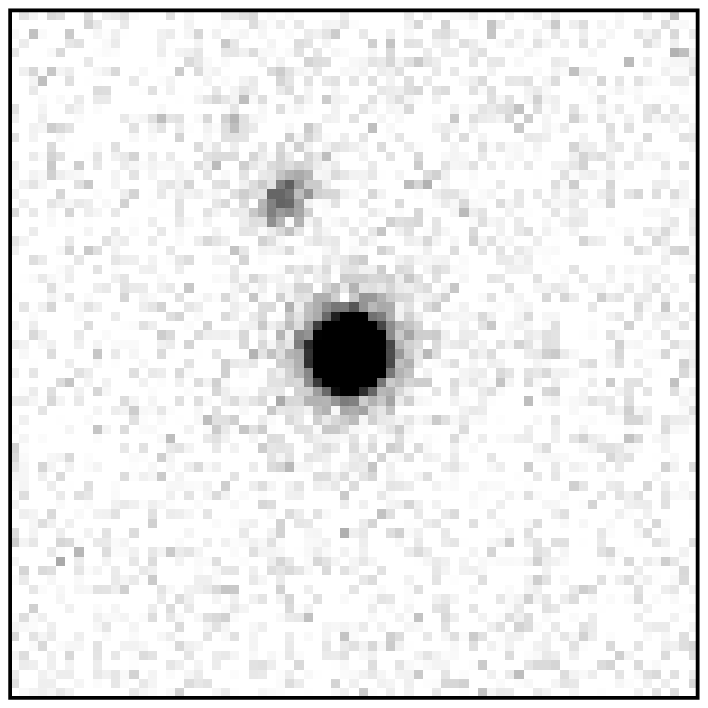}}
\put(2.8,3.2){%
\includegraphics[bb = 90 558 308 770,clip,angle=0,width=\contwidth]{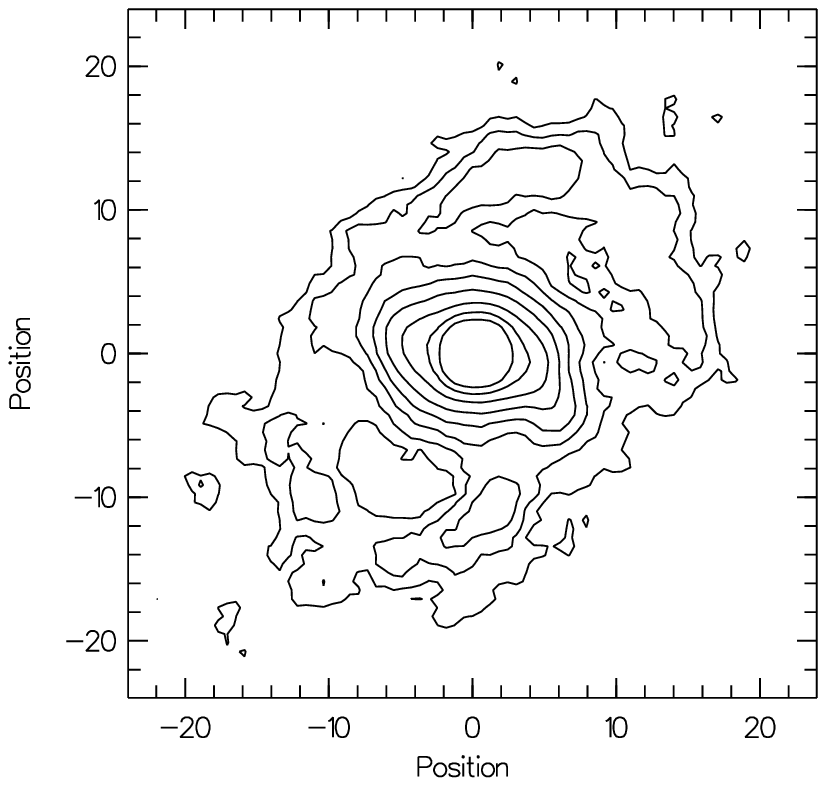}}
\put(2.8,0.0){%
\includegraphics[bb = 90 558 308 770,clip,angle=0,width=\contwidth]{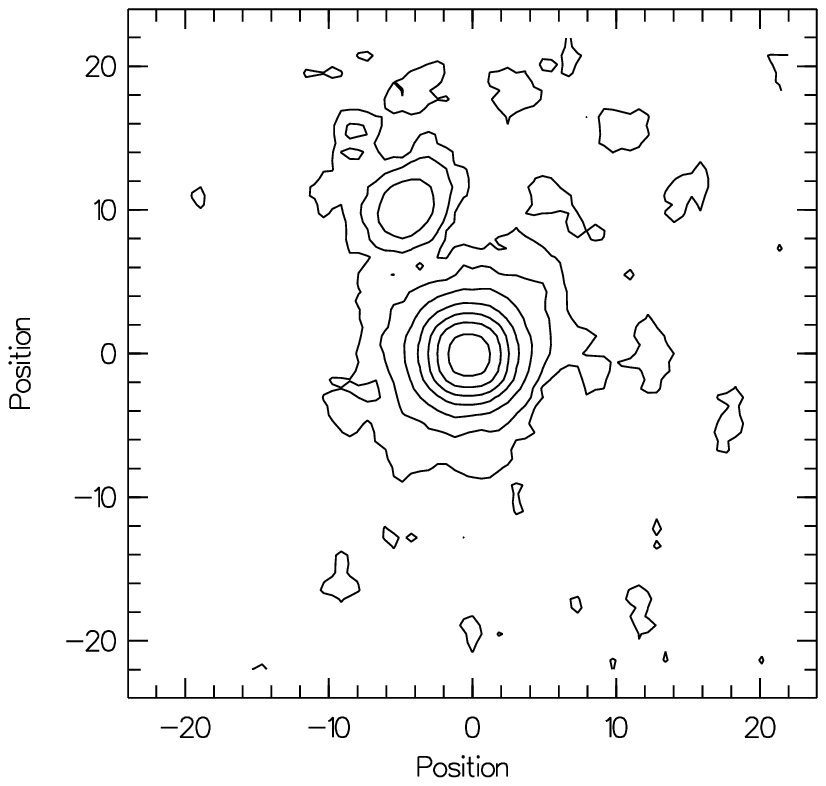}}
\put(6.1,3.35){%
\includegraphics[bb = 98 98 299 299,clip,angle=0,width=\greywidth]{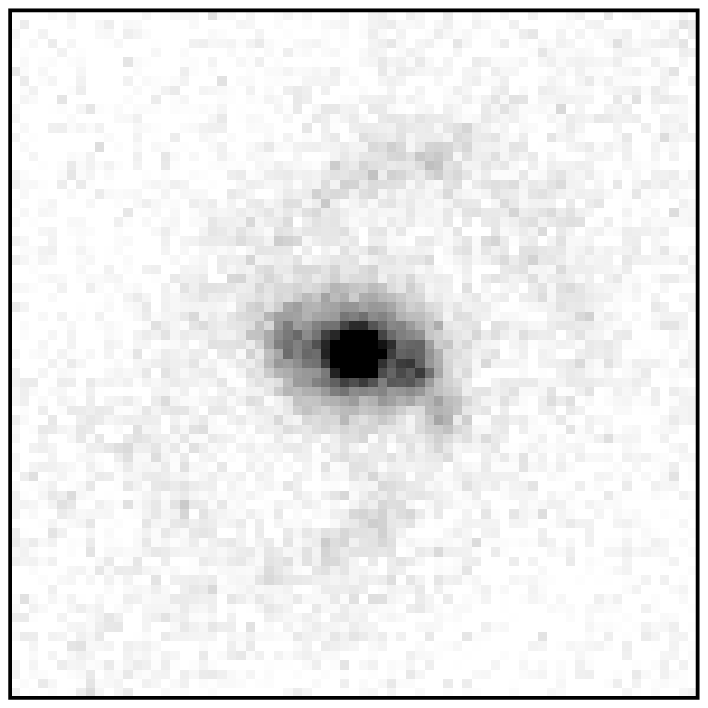}}
\put(6.1,0.15){%
\includegraphics[bb = 98 98 299 299,clip,angle=0,width=\greywidth]{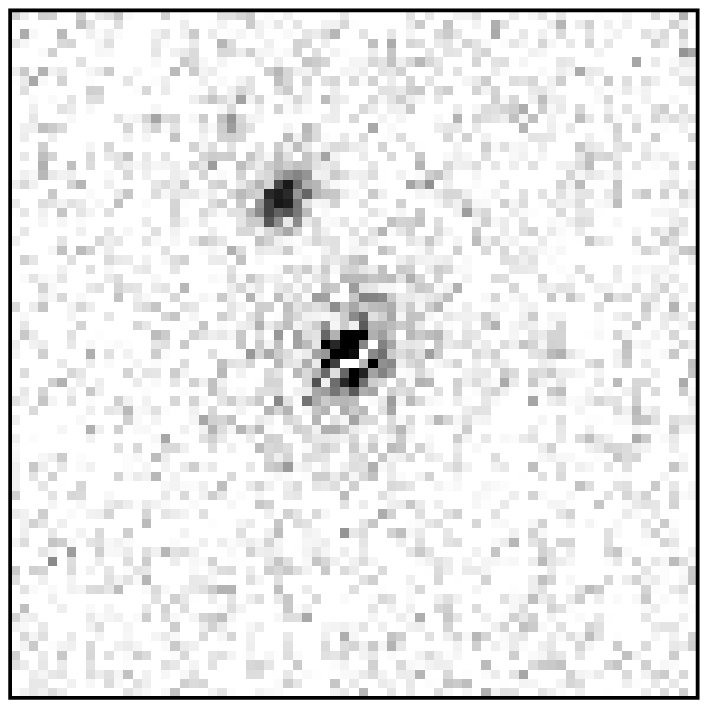}}
\put(8.9,3.2){%
\includegraphics[bb = 97 101 314 314,clip,angle=0,width=\contwidth]{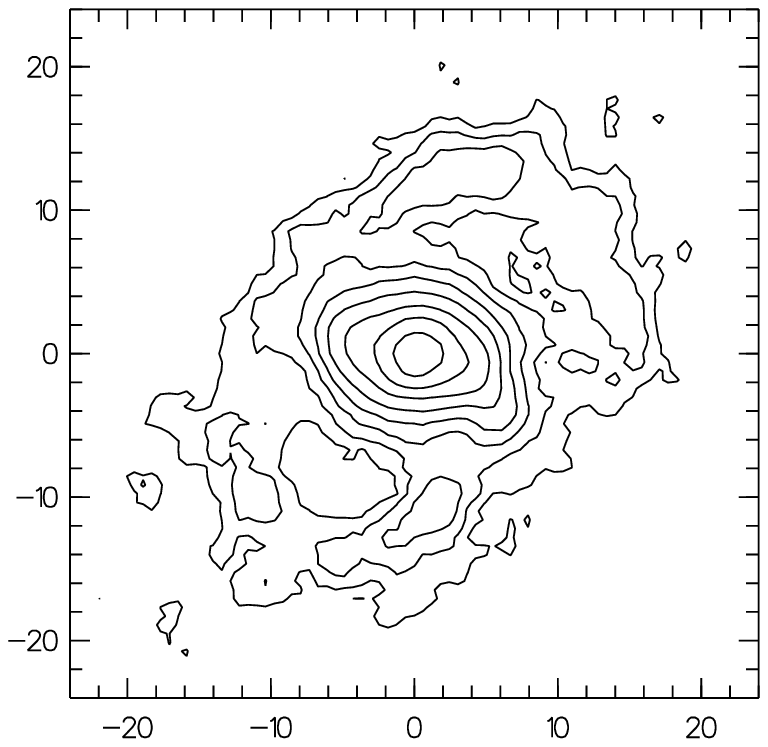}}
\put(8.9,0.0){%
\includegraphics[bb = 97 101 314 314,clip,angle=0,width=\contwidth]{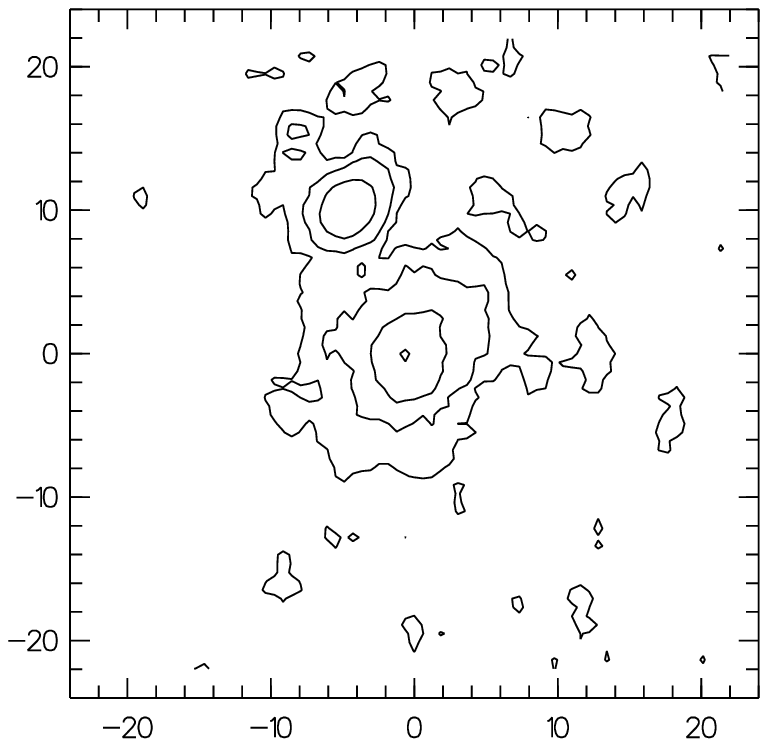}}
\put(11.95,3.2){%
\includegraphics[bb = 96 558 308 770,clip,angle=0,width=\profwidth]{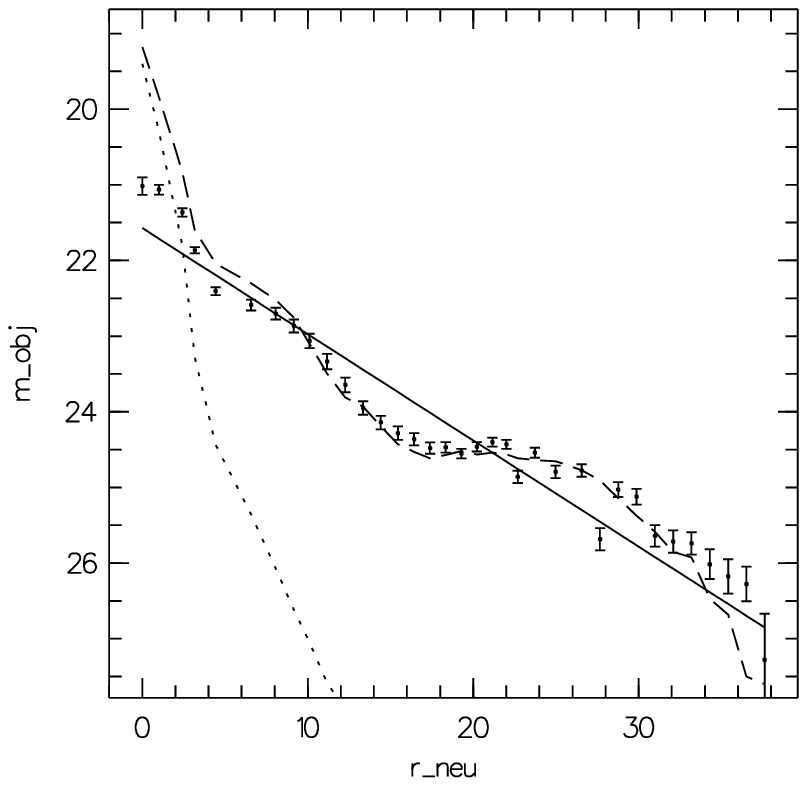}}
\put(11.95,0.0){%
\includegraphics[bb = 96 558 308 770,clip,angle=0,width=\profwidth]{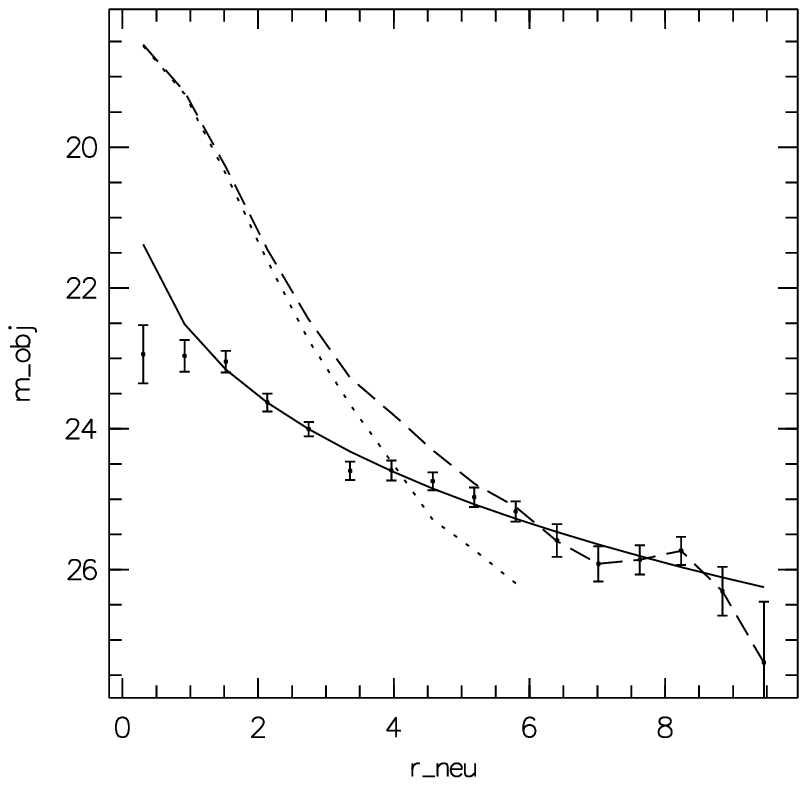}}
\put(15.0,3.2){%
\includegraphics[bb = 97 618 251 770,clip,angle=0,width=\profwidth]{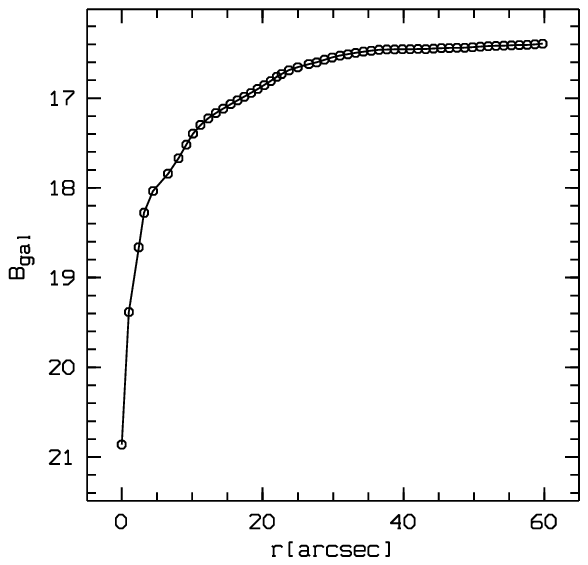}}
\put(15.0,0.0){%
\includegraphics[bb = 97 618 251 770,clip,angle=0,width=\profwidth]{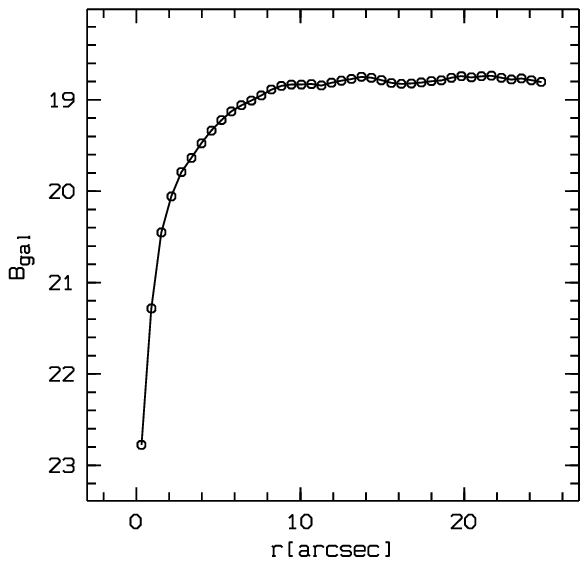}}
\put(2.1,2.9){Input image}
\put(7.8,2.9){Host galaxy image}
\put(12.6,2.9){Radial profile}
\put(15.6,2.9){Growth curve}

\end{picture}
\caption{
Typical image quality for two redshifts. Top: HE\,1043--1346
($z=0.068$), bottom: HE\,1110--1910 ($z=0.111$), all observed with
EFOSC1 at the ESO~3.6m telescope. From left: image and contour
including the nucleus, the same with the nucleus subtracted, radial
profile (dashed line: QSO including nucleus, dotted line: PSF, points:
residual galaxy, solid line: fits for disc and/or spheroid galaxy),
curve of growth in apparent magnitudes. The lowest contour in each
object is 25 and 26~mag arcsec$^{-2}$, respectively, with contour
spacings of 0.5, and 1~mag arcsec$^{-2}$. Coordinates are given in
arcsec.
}\label{fig:images}
\end{figure*}

Although the imaging data were not initially meant to be used for host
galaxy studies and the data are clearly low signal-to-noise, we found
that extended `fuzz' around pointlike nuclei could easily be detected
already by eye in a good fraction of the low-redshift objects. Example
images illustrating the typical data quality are displayed in
Fig.~\ref{fig:images}, the full set of host galaxies is shown in
Fig.~\ref{fig:allobj}. While detailed structural information generally
cannot be extracted from these short exposure images, it turned out
that estimation of integrated luminosities was indeed feasible.

\begin{figure*}
\setlength{\unitlength}{1cm}
\begin{picture}(16.95,22.65)

\put(0.0,19.95){%
\includegraphics[bb = 98 98 299 299,clip,angle=0,width=2.7cm]{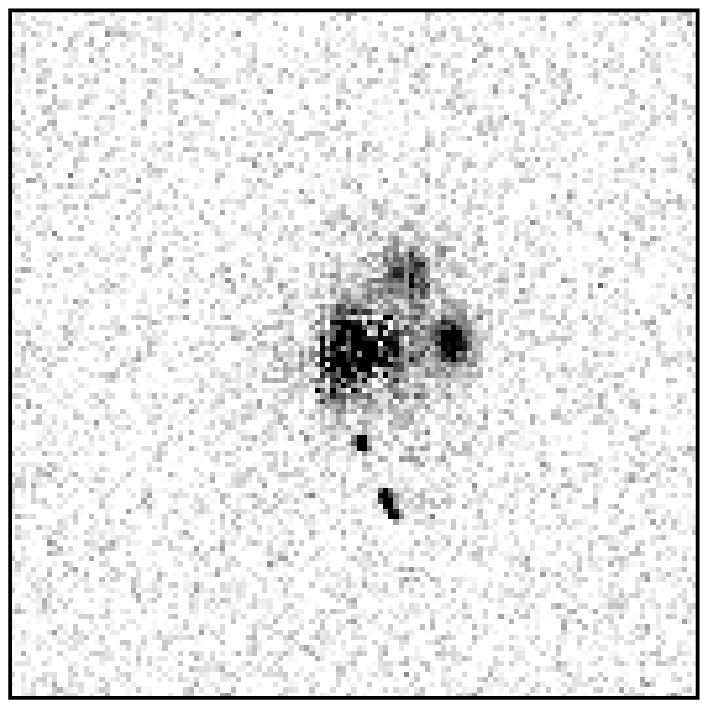}}
\put(0.15,20.1){HE\,0317--2638}
\put(2.85,19.95){%
\includegraphics[bb = 98 98 299 299,clip,angle=0,width=2.7cm]{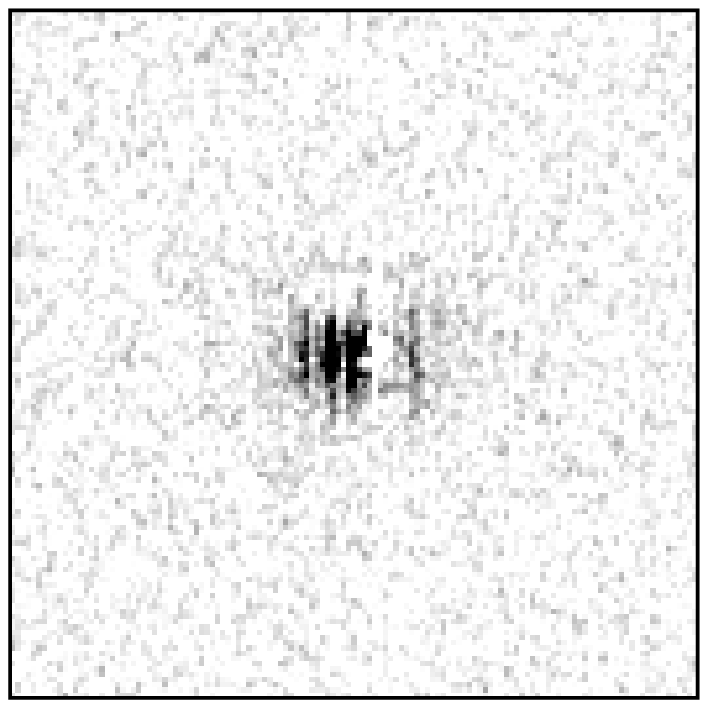}}
\put(3.0,20.1){IR\,03450+0055}
\put(5.7,19.95){%
\includegraphics[bb = 98 98 299 299,clip,angle=0,width=2.7cm]{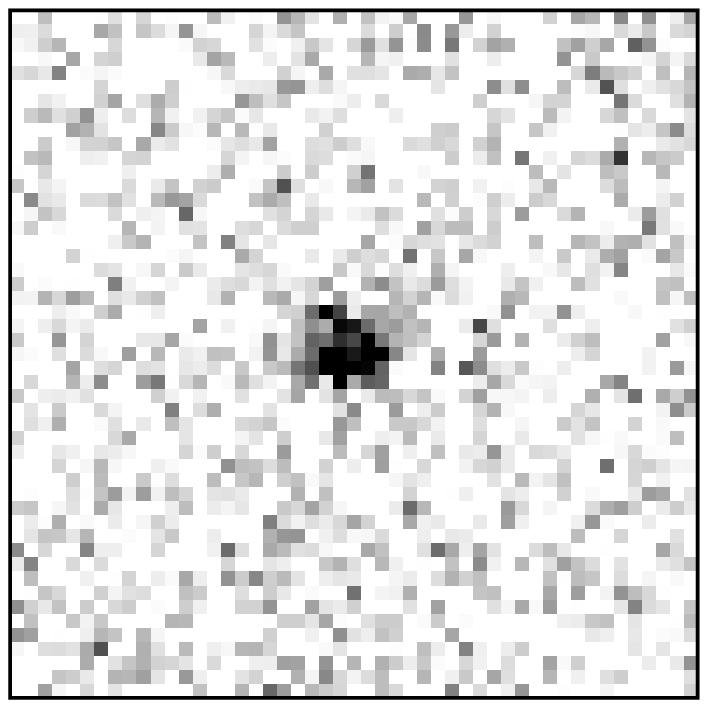}}
\put(5.85,20.1){HE\,0348--2226}
\put(8.55,19.95){%
\includegraphics[bb = 98 98 299 299,clip,angle=0,width=2.7cm]{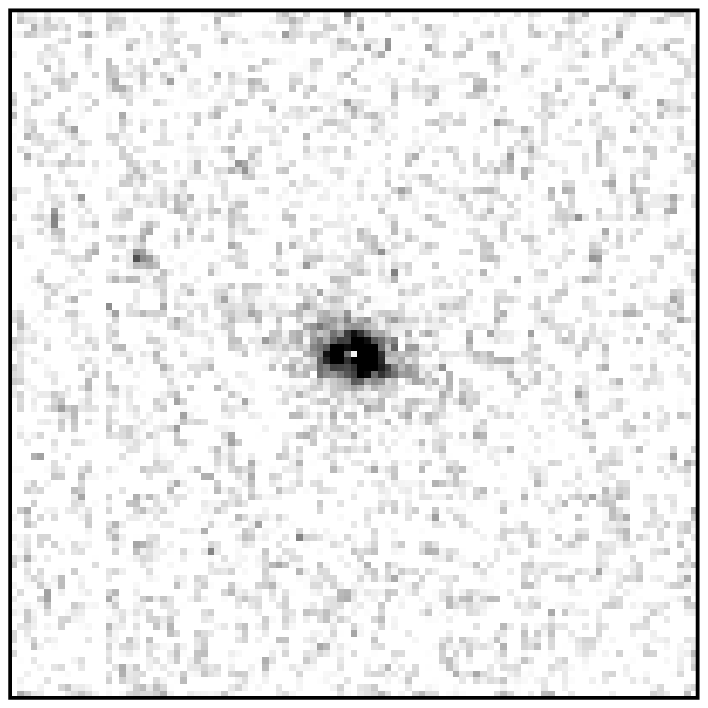}}
\put(8.7,20.1){HE\,0403--3719}
\put(11.4,19.95){%
\includegraphics[bb = 98 98 299 299,clip,angle=0,width=2.7cm]{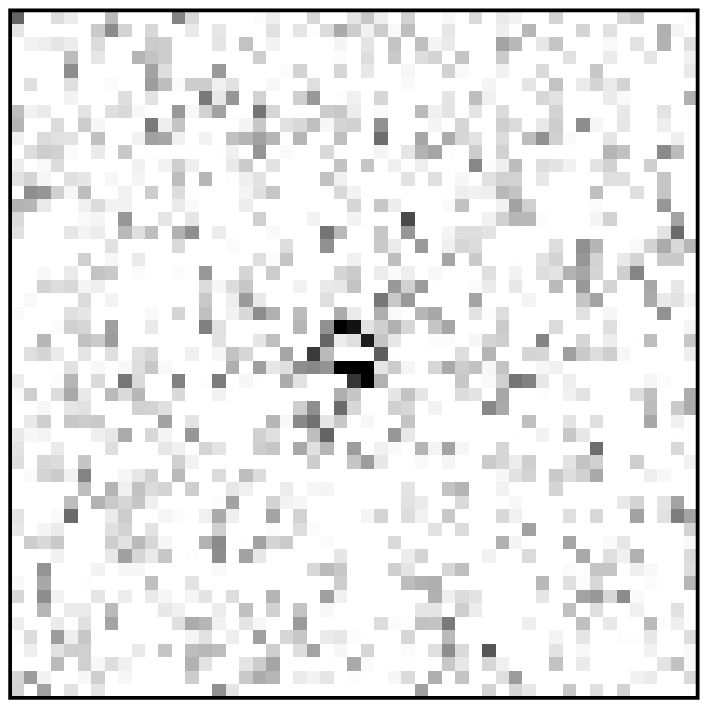}}
\put(11.55,20.1){HE\,0414--2552}
\put(14.25,19.95){%
\includegraphics[bb = 98 98 299 299,clip,angle=0,width=2.7cm]{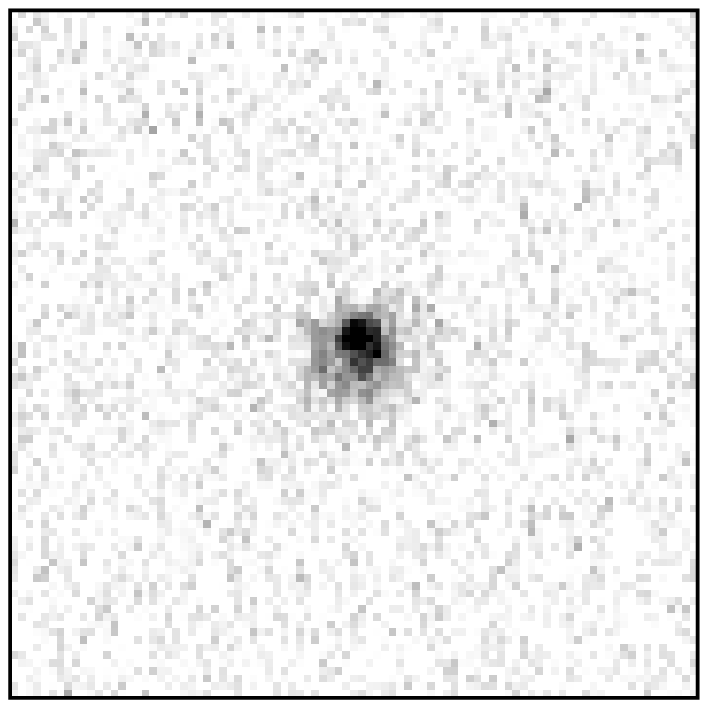}}
\put(14.4,20.1){HE\,0427--2701}
\put(0.0,17.1){%
\includegraphics[bb = 98 98 299 299,clip,angle=0,width=2.7cm]{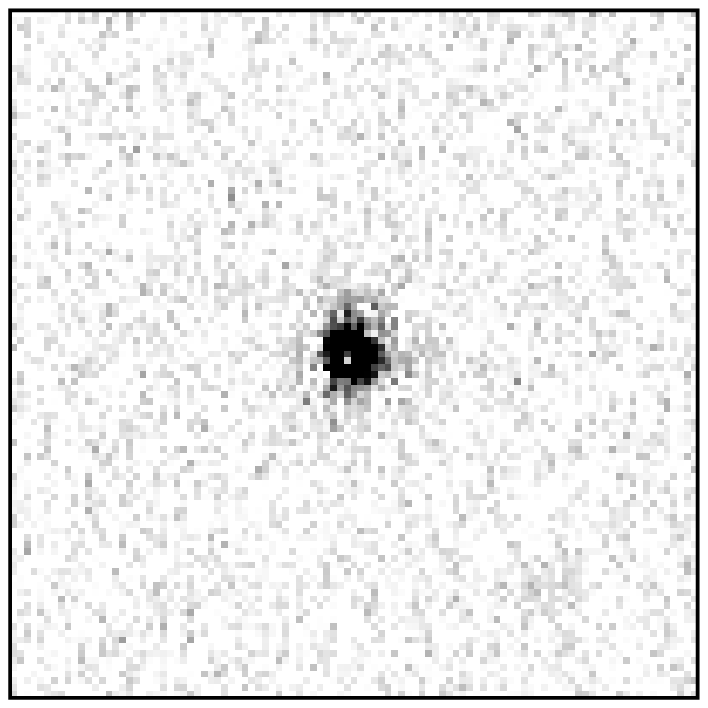}}
\put(0.15,17.25){HE\,0444--3449}
\put(2.85,17.1){%
\includegraphics[bb = 98 98 299 299,clip,angle=0,width=2.7cm]{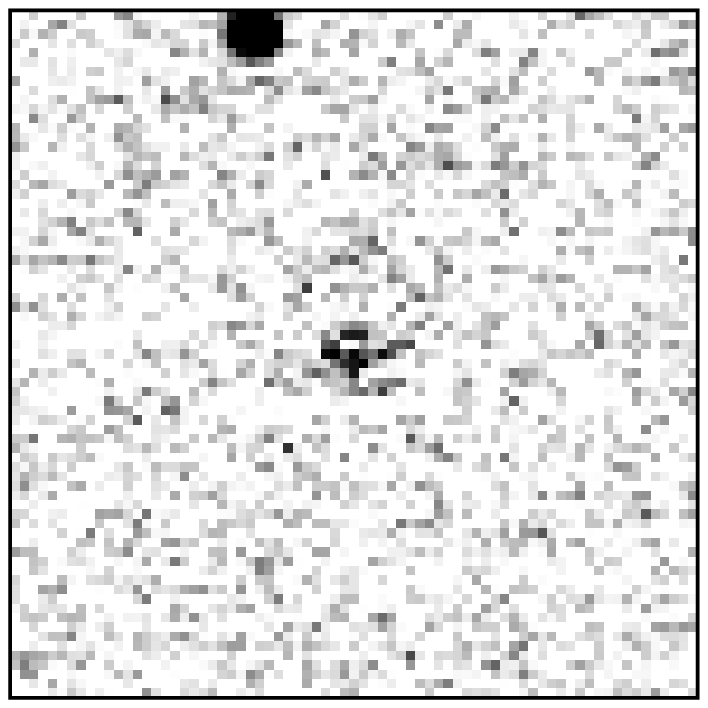}}
\put(3.0,17.25){HE\,0444--3900}
\put(5.7,17.1){%
\includegraphics[bb = 98 98 299 299,clip,angle=0,width=2.7cm]{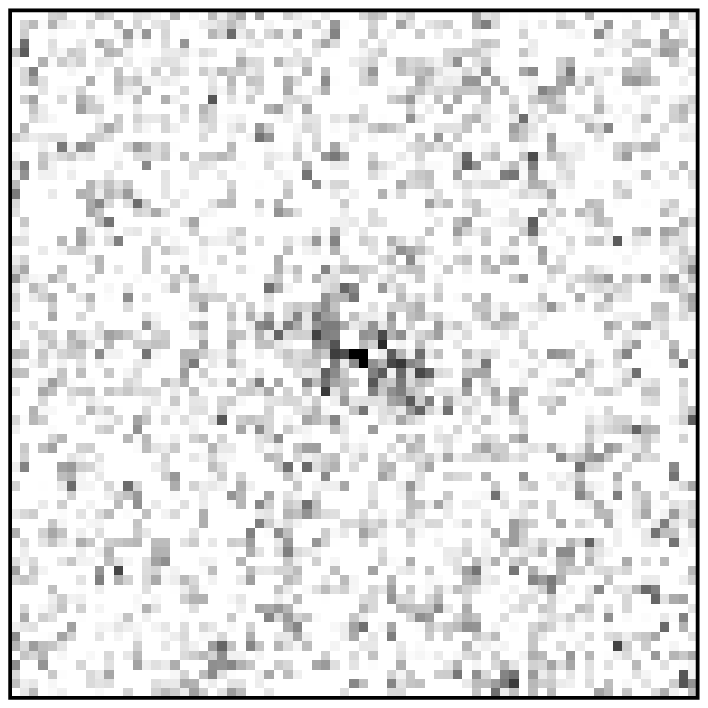}}
\put(5.85,17.25){HE\,0507--2710}
\put(8.55,17.1){%
\includegraphics[bb = 98 98 299 299,clip,angle=0,width=2.7cm]{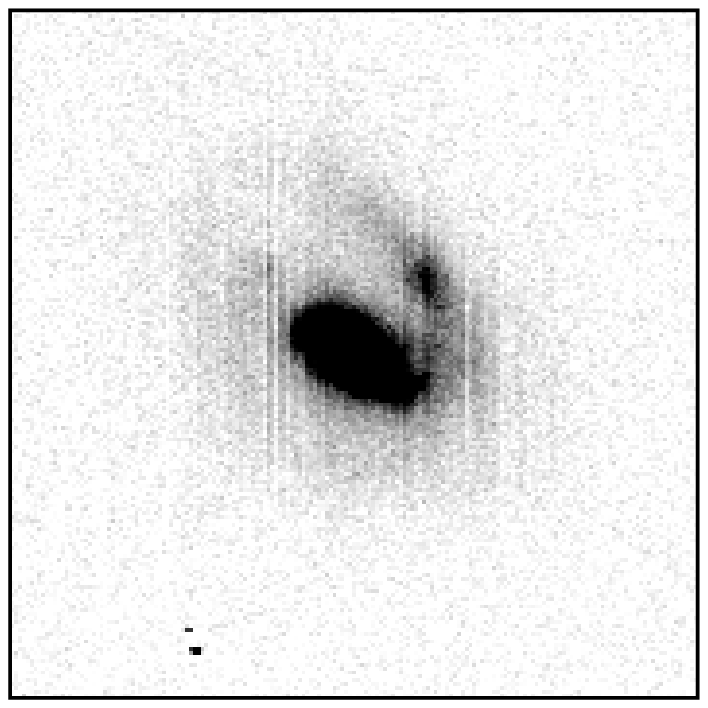}}
\put(8.7,17.25){HE\,0517--3243}
\put(11.4,17.1){%
\includegraphics[bb = 98 98 299 299,clip,angle=0,width=2.7cm]{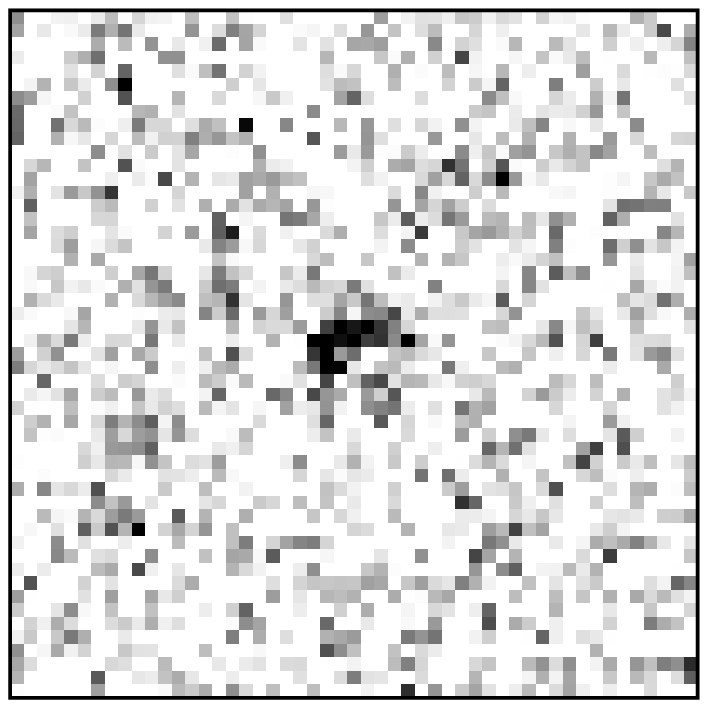}}
\put(11.55,17.25){HE\,0526--4148}
\put(14.25,17.1){%
\includegraphics[bb = 98 98 299 299,clip,angle=0,width=2.7cm]{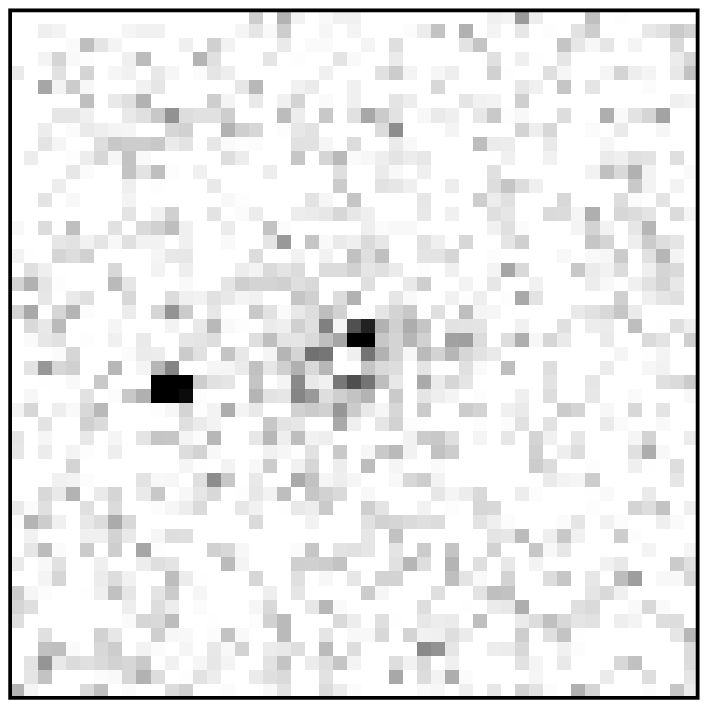}}
\put(14.4,17.25){HE\,0529--3918}
\put(0.0,14.25){%
\includegraphics[bb = 98 98 299 299,clip,angle=0,width=2.7cm]{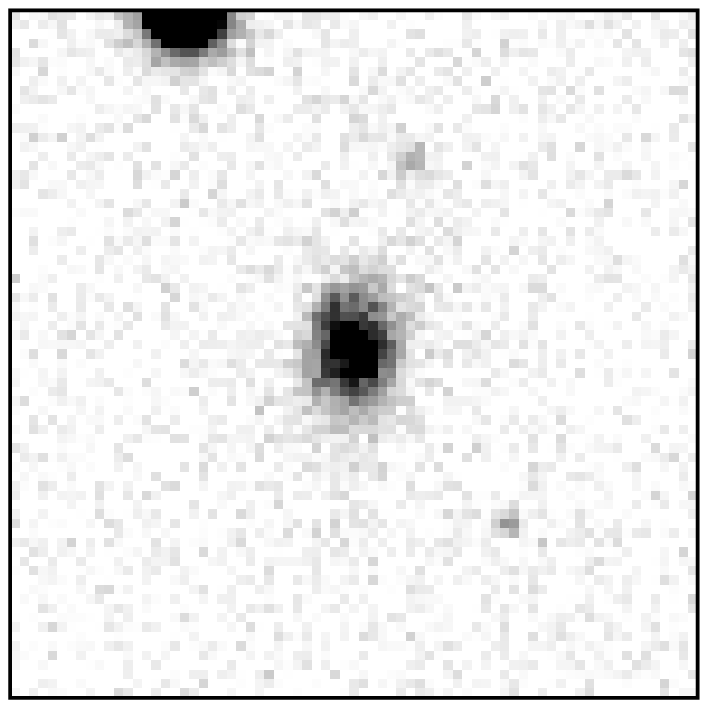}}
\put(0.15,14.4){HE\,0952--1552}
\put(2.85,14.25){%
\includegraphics[bb = 98 98 299 299,clip,angle=0,width=2.7cm]{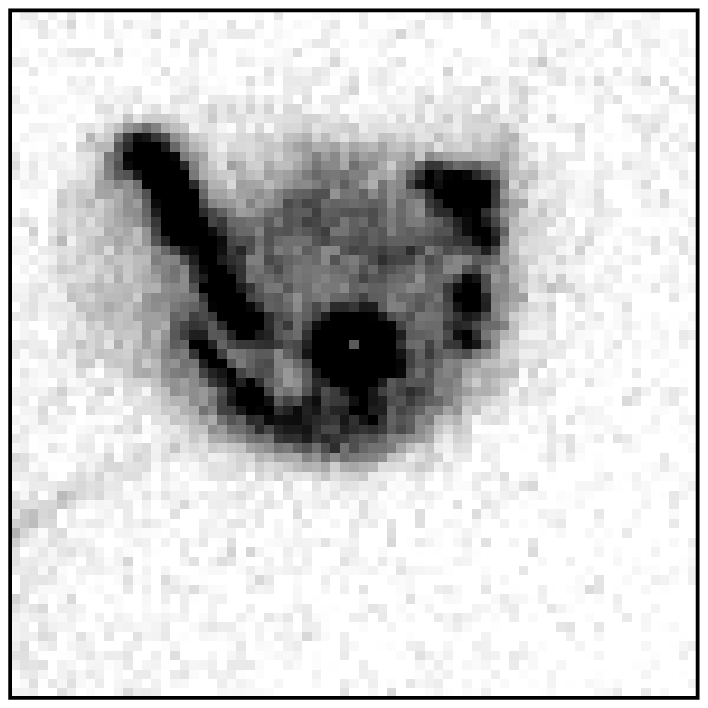}}
\put(3.0,14.4){IR\,09595--0755}
\put(5.7,14.25){%
\includegraphics[bb = 98 98 299 299,clip,angle=0,width=2.7cm]{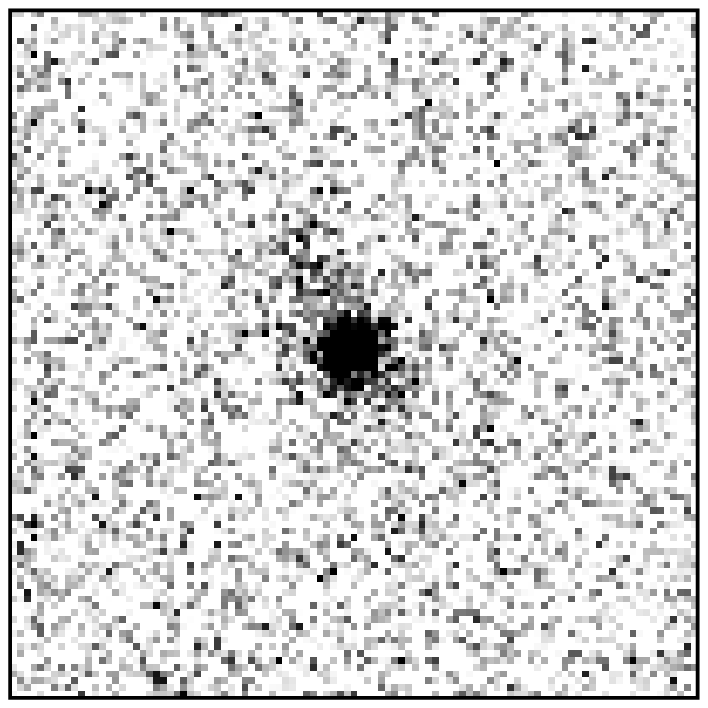}}
\put(5.85,14.4){HE\,1019--1414}
\put(8.55,14.25){%
\includegraphics[bb = 98 98 299 299,clip,angle=0,width=2.7cm]{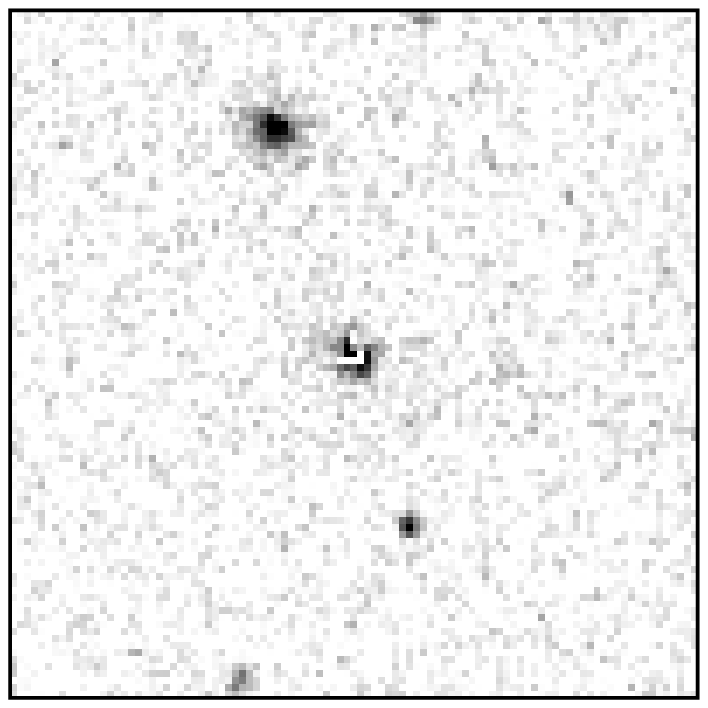}}
\put(8.7,14.4){PKS\,1020--103}
\put(11.4,14.25){%
\includegraphics[bb = 98 98 299 299,clip,angle=0,width=2.7cm]{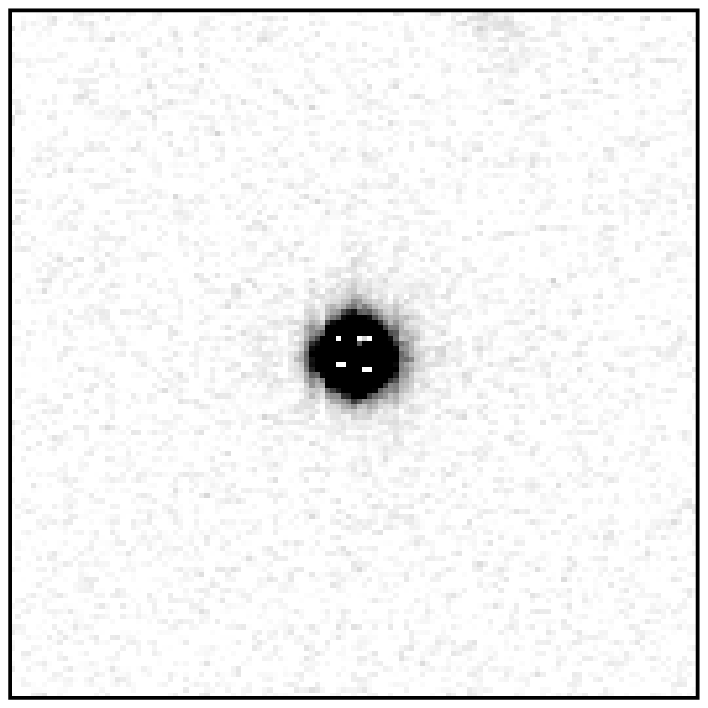}}
\put(11.55,14.4){HE\,1029--1401}
\put(14.25,14.25){%
\includegraphics[bb = 98 98 299 299,clip,angle=0,width=2.7cm]{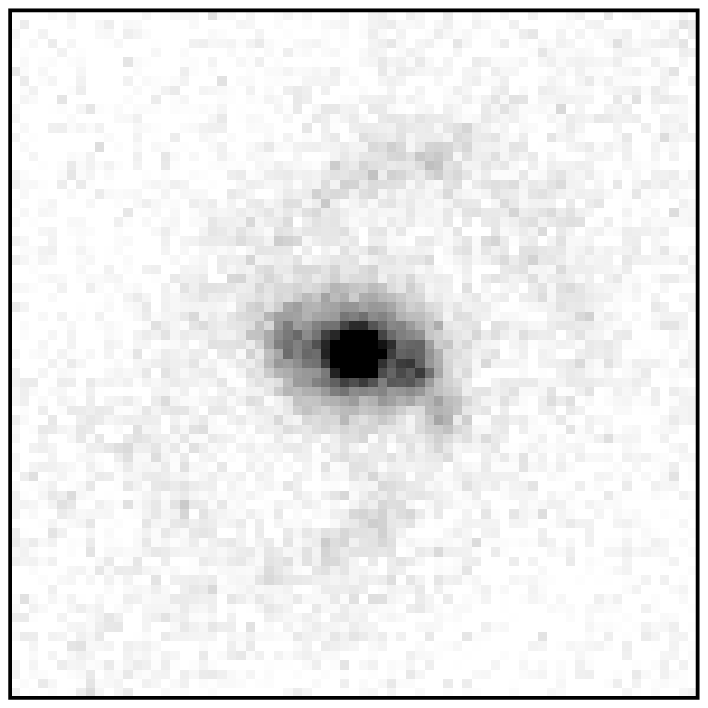}}
\put(14.4,14.4){HE\,1043--1346}
\put(0.0,11.4){%
\includegraphics[bb = 98 98 299 299,clip,angle=0,width=2.7cm]{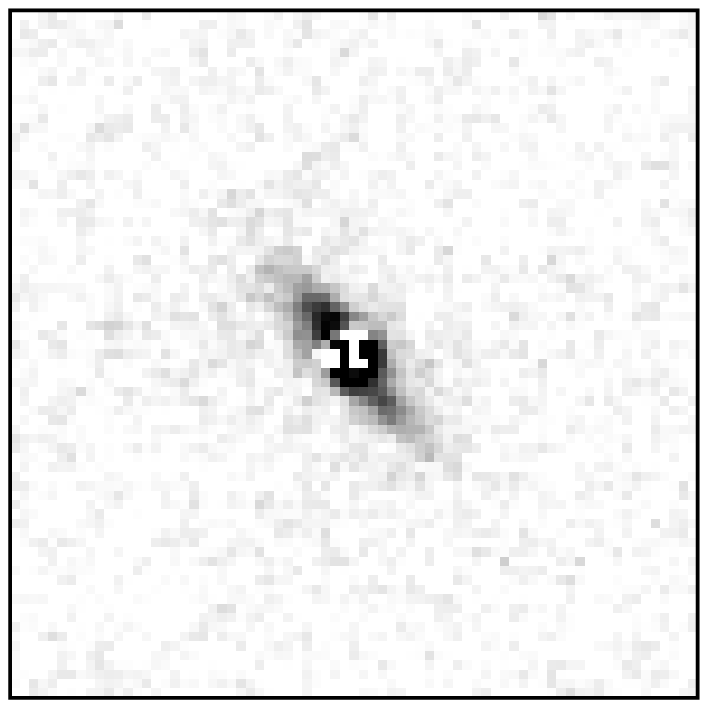}}
\put(0.15,11.55){HE\,1106--2321}
\put(2.85,11.4){%
\includegraphics[bb = 98 98 299 299,clip,angle=0,width=2.7cm]{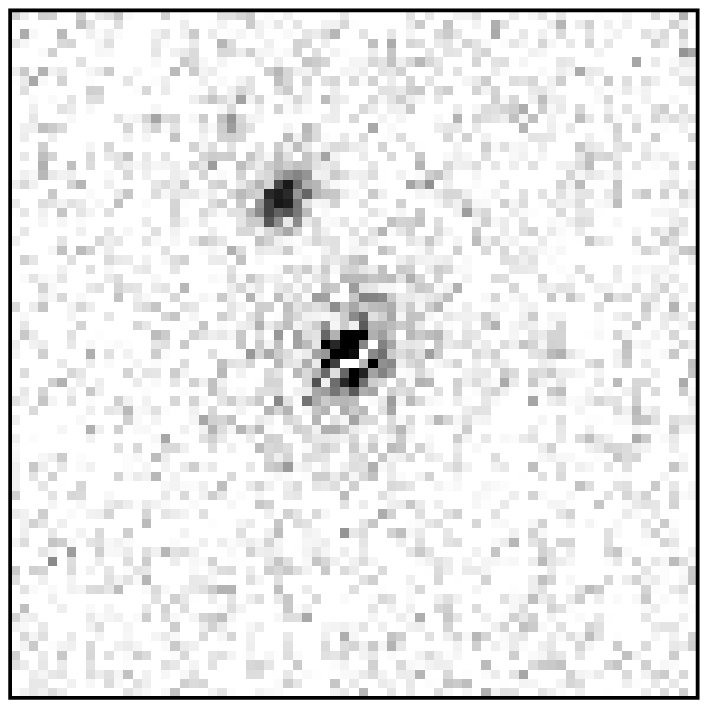}}
\put(3.0,11.55){HE\,1110--1910}
\put(5.7,11.4){%
\includegraphics[bb = 98 98 299 299,clip,angle=0,width=2.7cm]{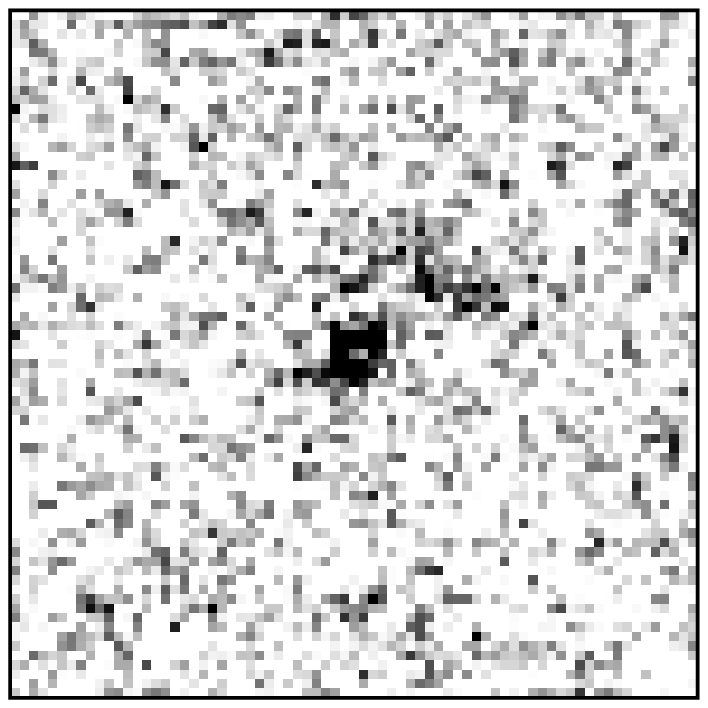}}
\put(5.85,11.55){Q\,1114--2846}
\put(8.55,11.4){%
\includegraphics[bb = 98 98 299 299,clip,angle=0,width=2.7cm]{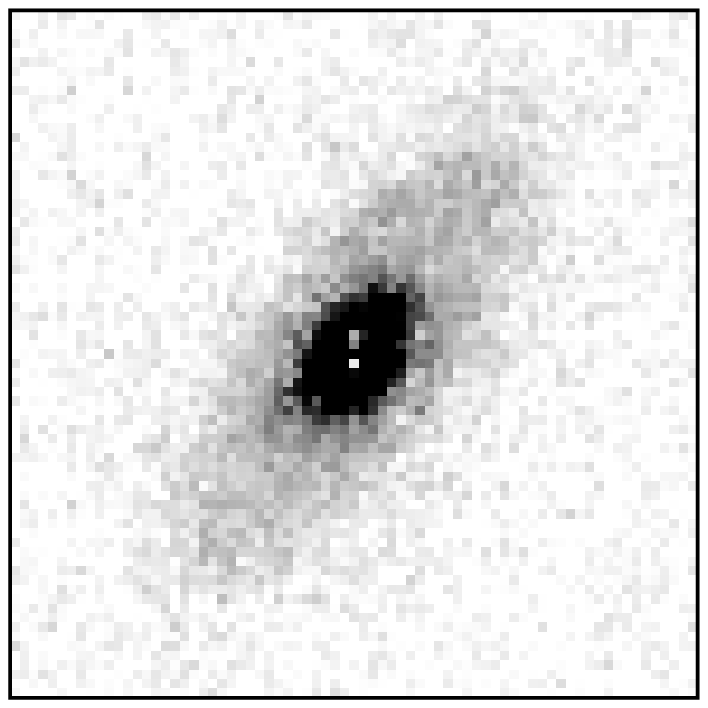}}
\put(8.7,11.55){PG\,1149--1105}
\put(11.4,11.4){%
\includegraphics[bb = 98 98 299 299,clip,angle=0,width=2.7cm]{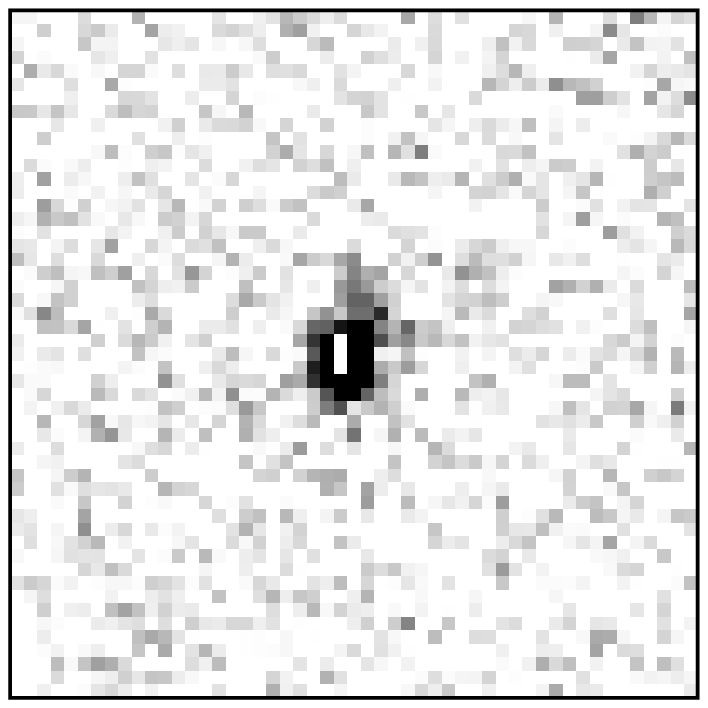}}
\put(11.55,11.55){HE\,1201--2409}
\put(14.25,11.4){%
\includegraphics[bb = 98 98 299 299,clip,angle=0,width=2.7cm]{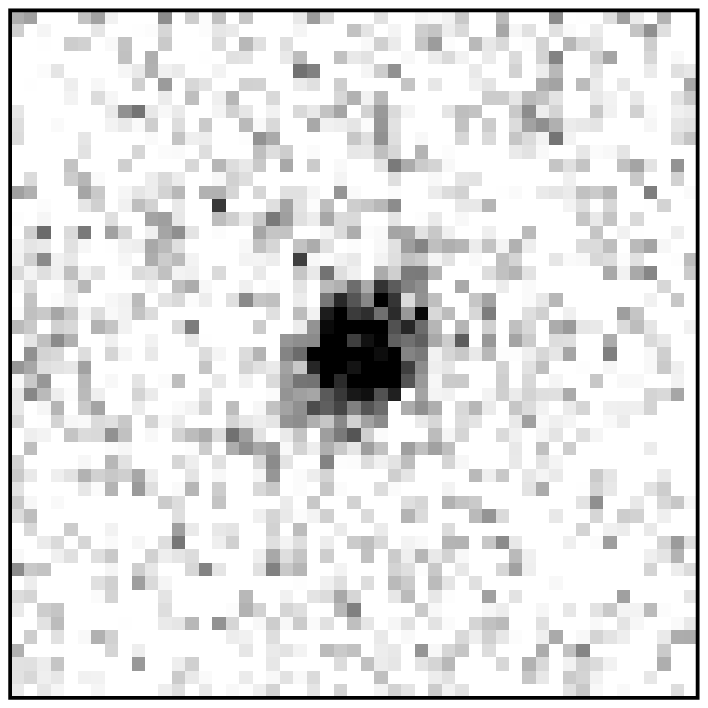}}
\put(14.4,11.55){HE\,1213--1633}
\put(0.0,8.55){%
\includegraphics[bb = 98 98 299 299,clip,angle=0,width=2.7cm]{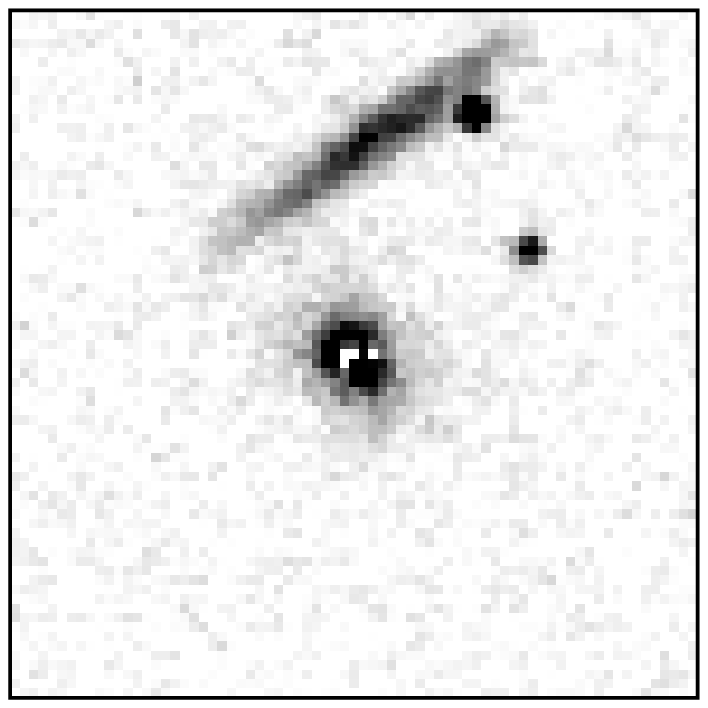}}
\put(0.15,8.7){HE\,1217--1340}
\put(2.85,8.55){%
\includegraphics[bb = 98 98 299 299,clip,angle=0,width=2.7cm]{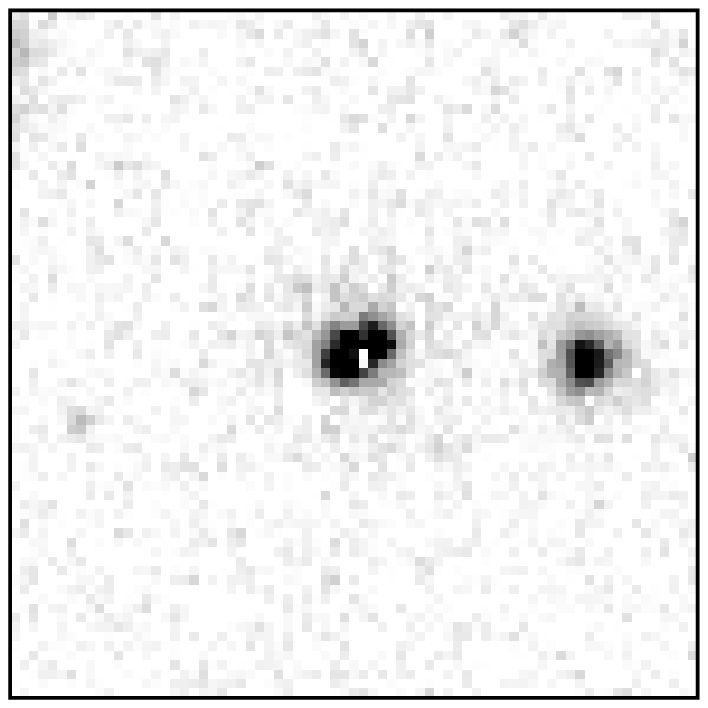}}
\put(3.0,8.7){HE\,1228--1637}
\put(5.7,8.55){%
\includegraphics[bb = 98 98 299 299,clip,angle=0,width=2.7cm]{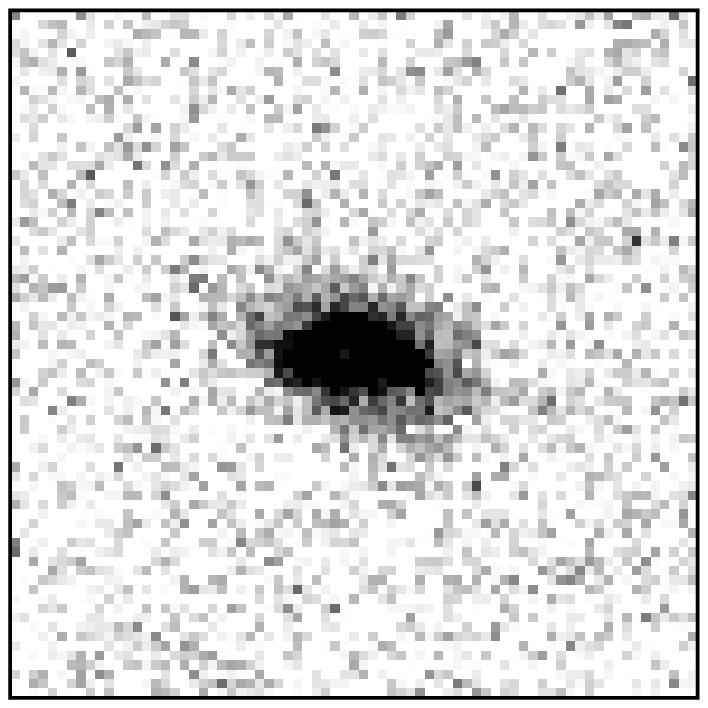}}
\put(5.85,8.7){HE\,1235--0857}
\put(8.55,8.55){%
\includegraphics[bb = 98 98 299 299,clip,angle=0,width=2.7cm]{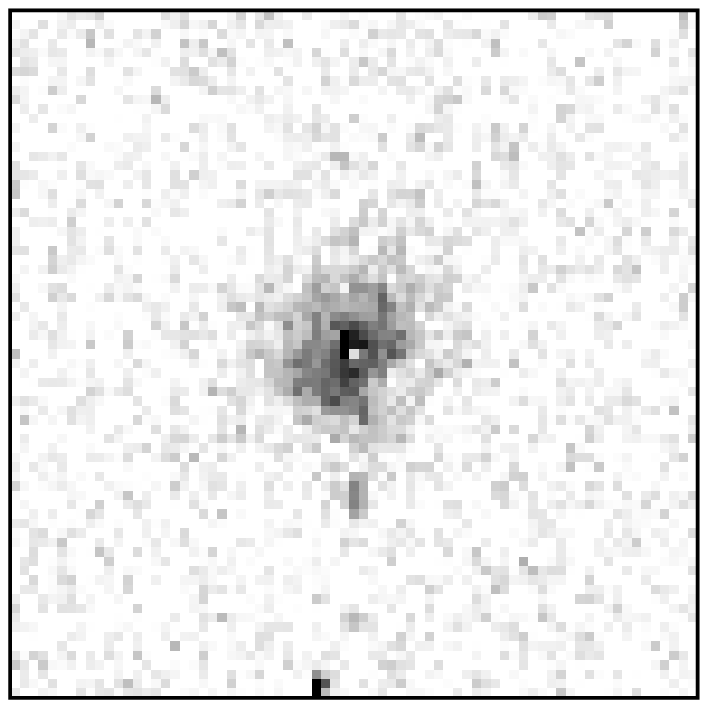}}
\put(8.7,8.7){HE\,1237--2252}
\put(11.4,8.55){%
\includegraphics[bb = 98 98 299 299,clip,angle=0,width=2.7cm]{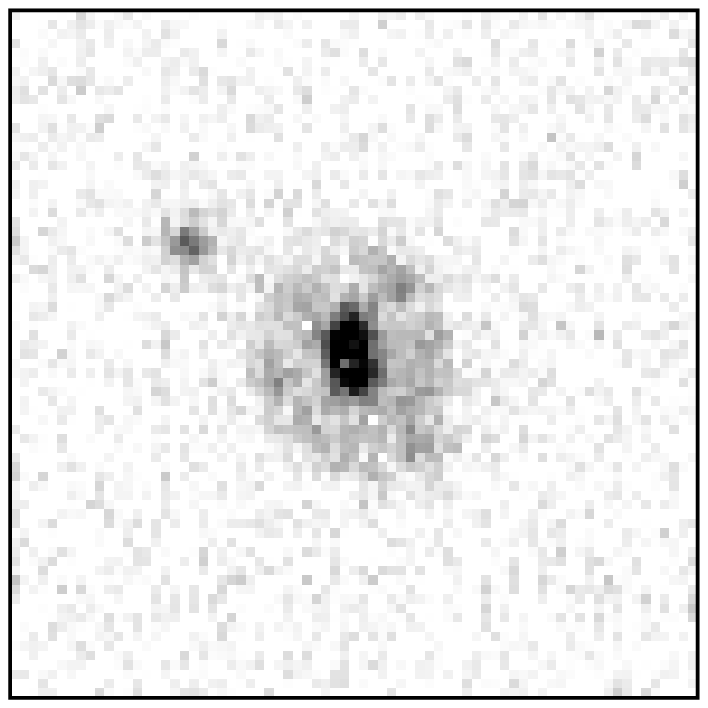}}
\put(11.55,8.7){HE\,1239--2426}
\put(14.25,8.55){%
\includegraphics[bb = 98 98 299 299,clip,angle=0,width=2.7cm]{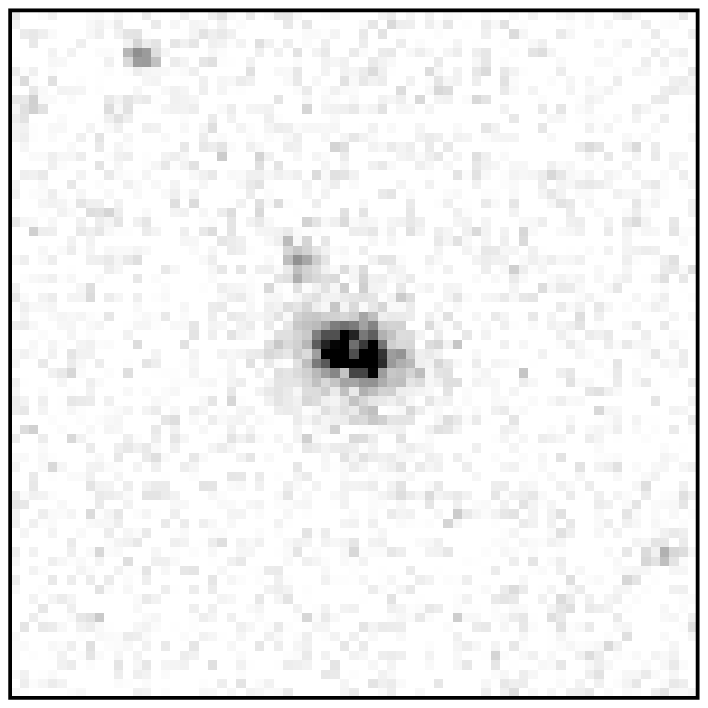}}
\put(14.4,8.7){HE\,1245--2709}
\put(0.0,5.7){%
\includegraphics[bb = 98 98 299 299,clip,angle=0,width=2.7cm]{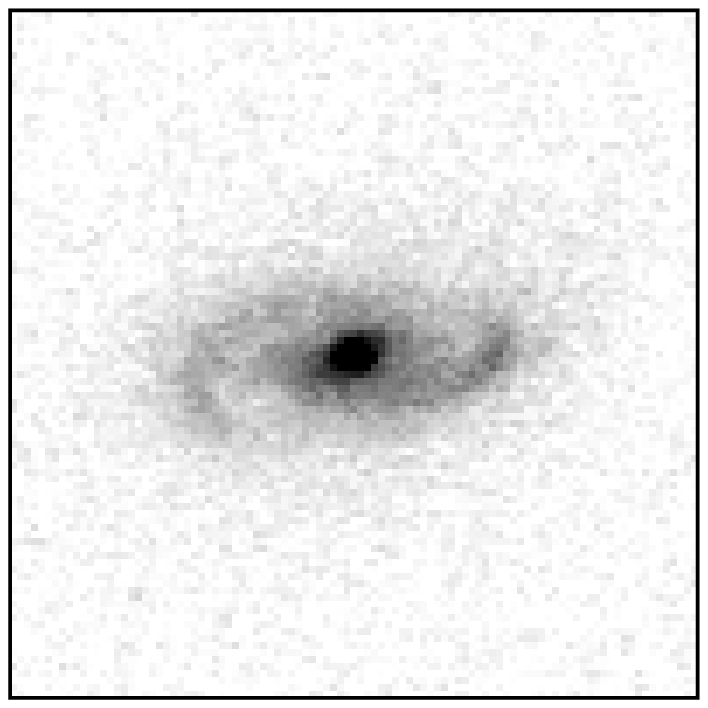}}
\put(0.15,5.85){HE\,1248--1357}
\put(2.85,5.7){%
\includegraphics[bb = 98 98 299 299,clip,angle=0,width=2.7cm]{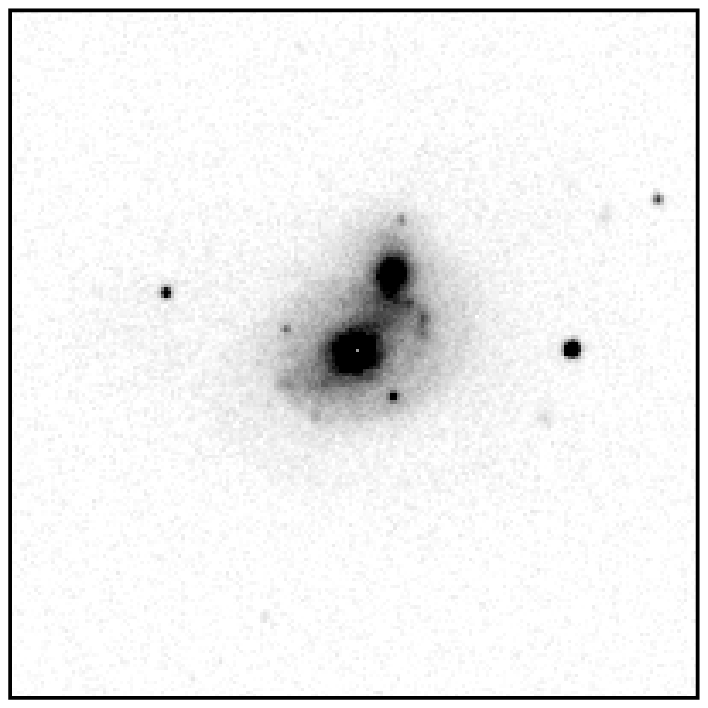}}
\put(3.0,5.85){IR\,12495--1308}
\put(5.7,5.7){%
\includegraphics[bb = 98 98 299 299,clip,angle=0,width=2.7cm]{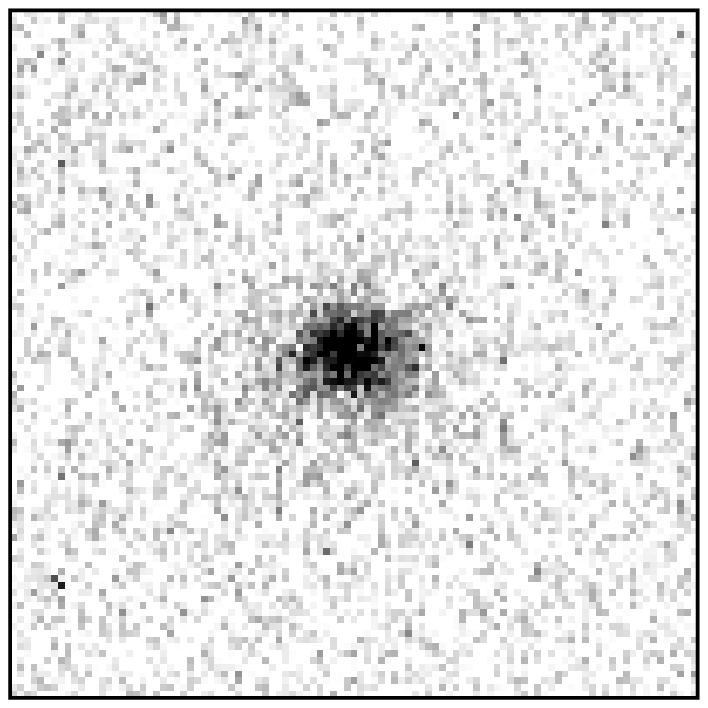}}
\put(5.85,5.85){HE\,1300--1325}
\put(8.55,5.7){%
\includegraphics[bb = 98 98 299 299,clip,angle=0,width=2.7cm]{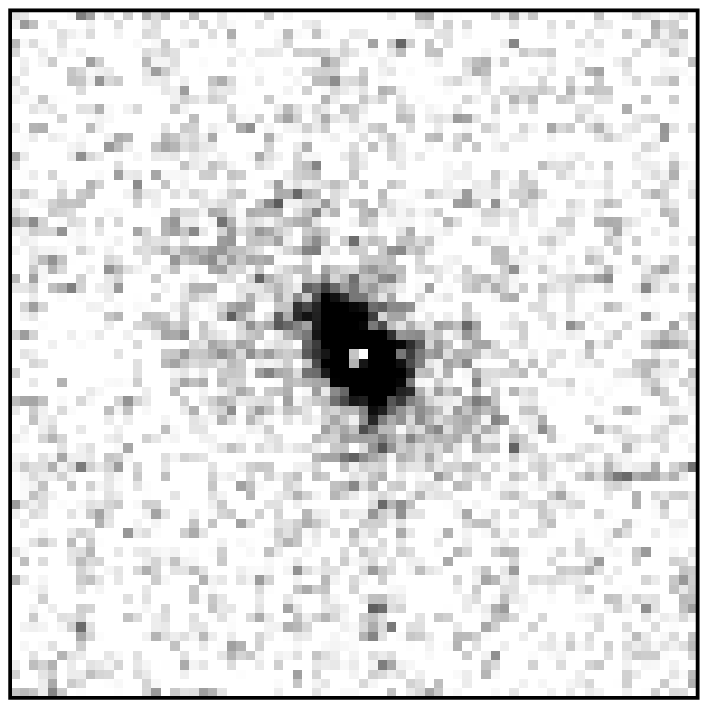}}
\put(8.7,5.85){HE\,1309--2501}
\put(11.4,5.7){%
\includegraphics[bb = 98 98 299 299,clip,angle=0,width=2.7cm]{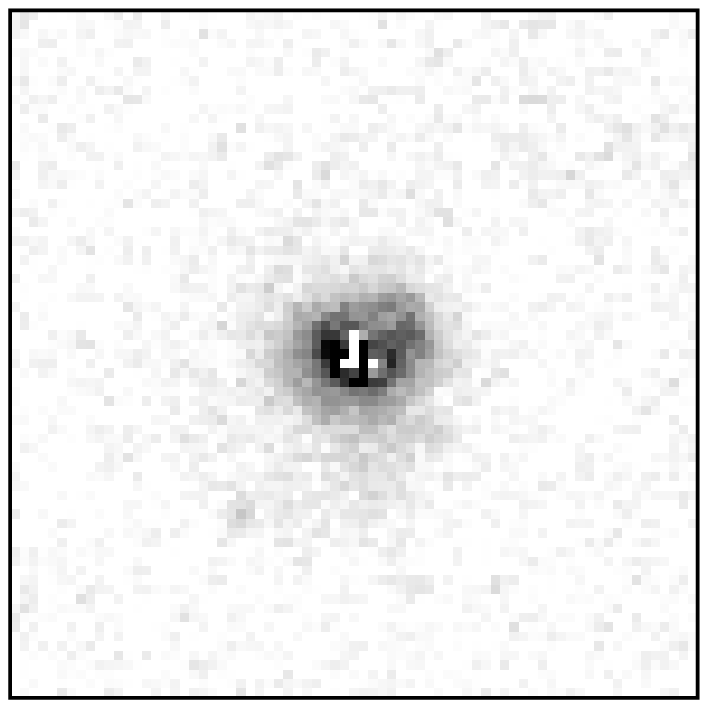}}
\put(11.55,5.85){PG\,1310--1051}
\put(14.25,5.7){%
\includegraphics[bb = 98 98 299 299,clip,angle=0,width=2.7cm]{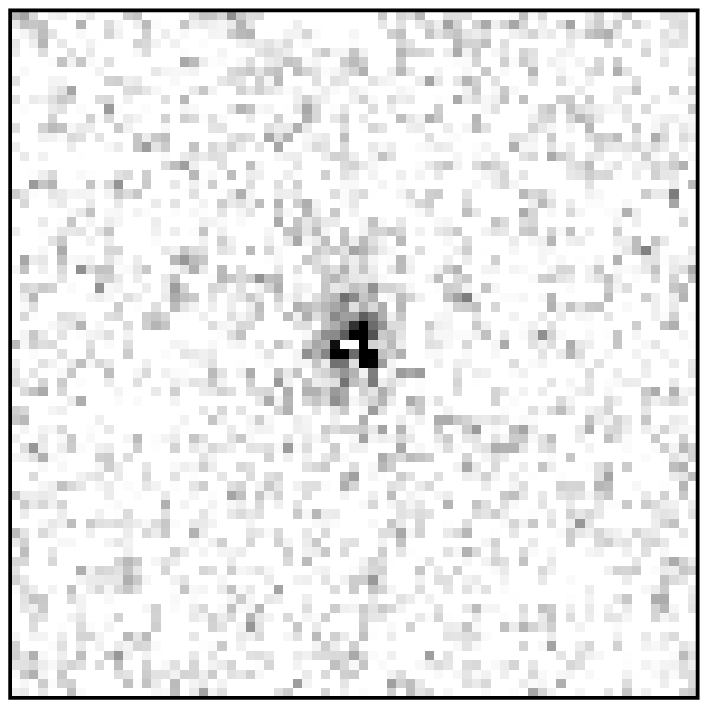}}
\put(14.4,5.85){HE\,1315--1028}
\put(0.0,2.85){%
\includegraphics[bb = 98 98 299 299,clip,angle=0,width=2.7cm]{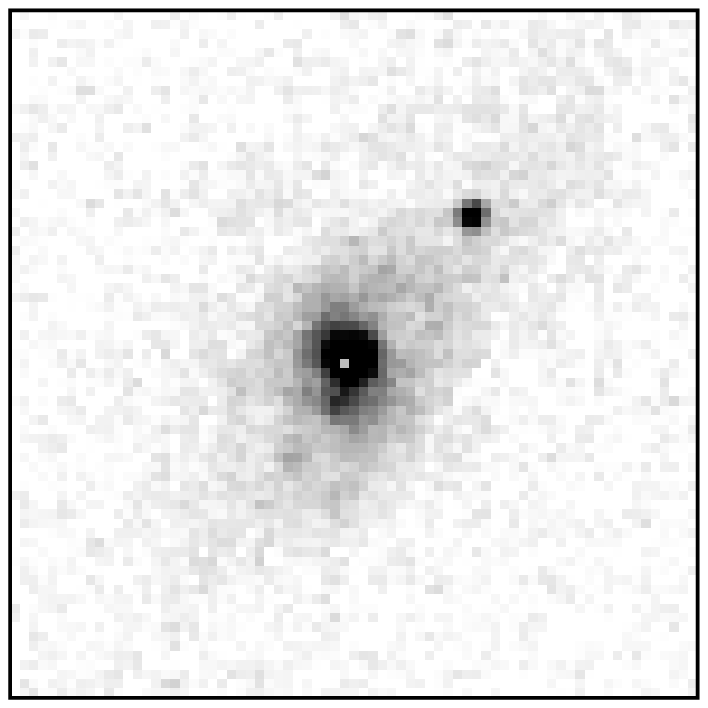}}
\put(0.15,3.0){HE\,1328--2509}
\put(2.85,2.85){%
\includegraphics[bb = 98 98 299 299,clip,angle=0,width=2.7cm]{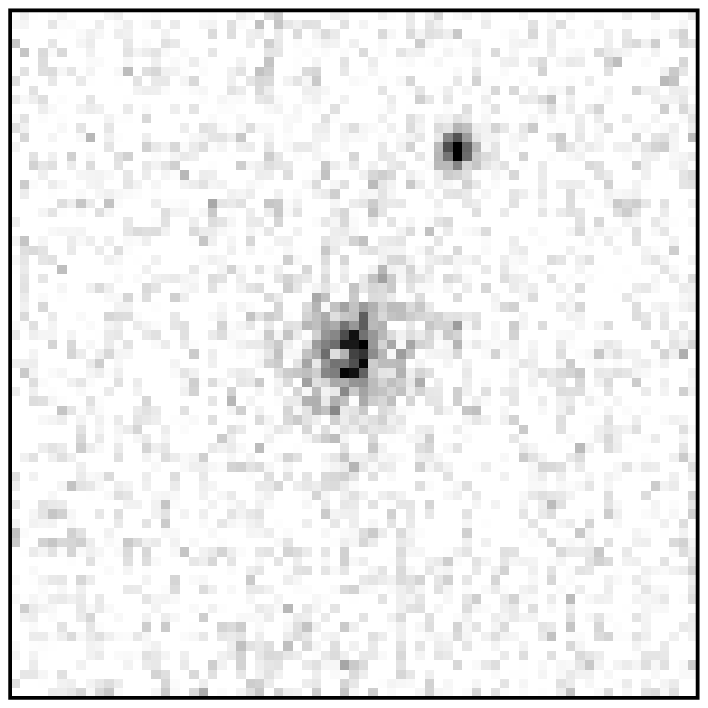}}
\put(3.0,3.0){HE\,1335--0847}
\put(5.7,2.85){%
\includegraphics[bb = 98 98 299 299,clip,angle=0,width=2.7cm]{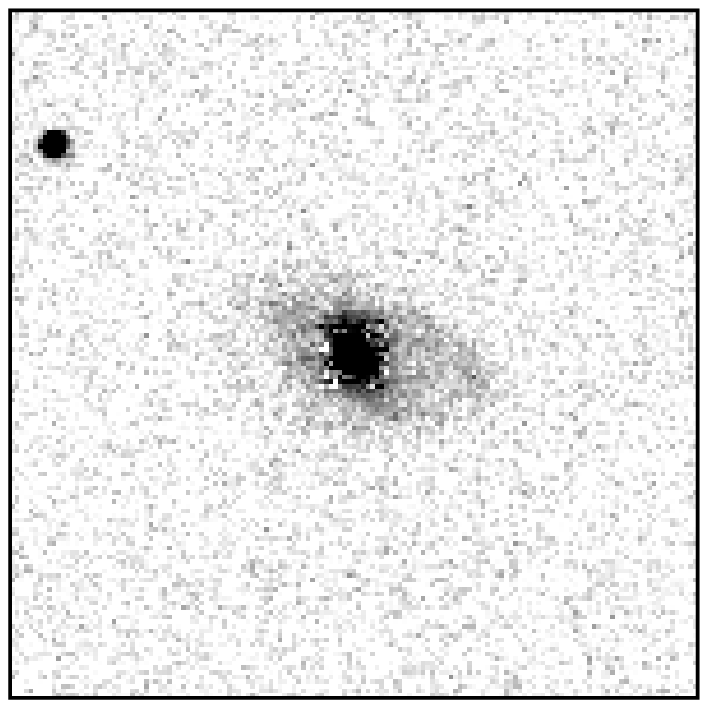}}
\put(5.85,3.0){HE\,1338--1423}
\put(8.55,2.85){%
\includegraphics[bb = 98 98 299 299,clip,angle=0,width=2.7cm]{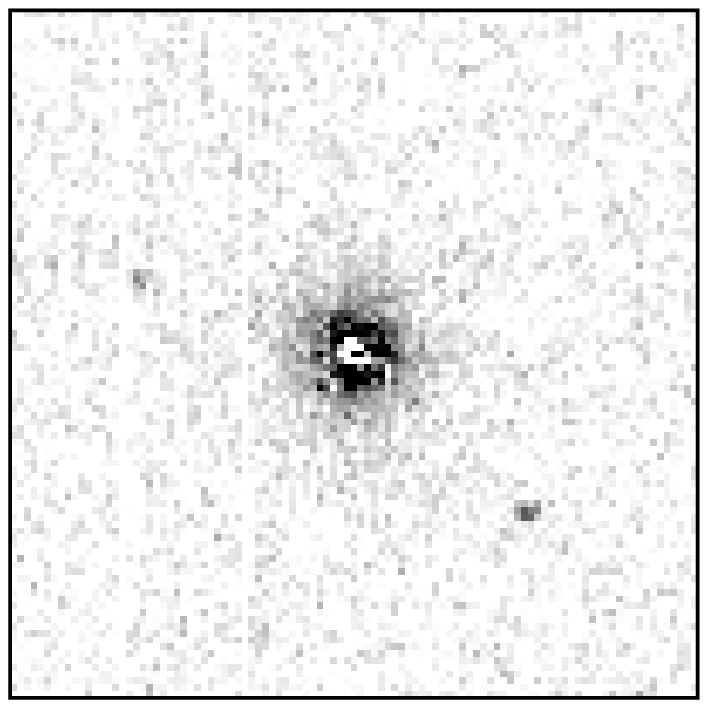}}
\put(8.7,3.0){HE\,1348--1758}
\put(11.4,2.85){%
\includegraphics[bb = 98 98 299 299,clip,angle=0,width=2.7cm]{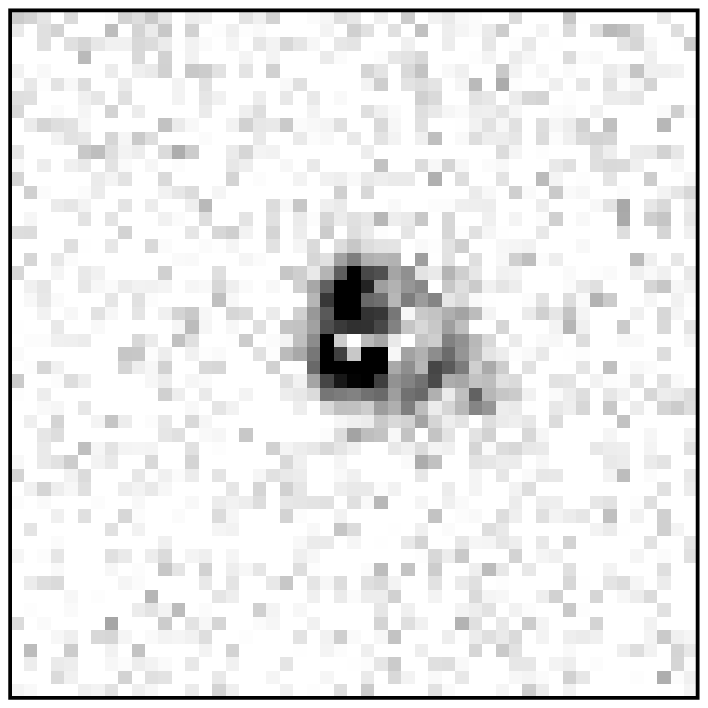}}
\put(11.55,3.0){HE\,1405--1545}
\put(14.25,2.85){%
\includegraphics[bb = 98 98 299 299,clip,angle=0,width=2.7cm]{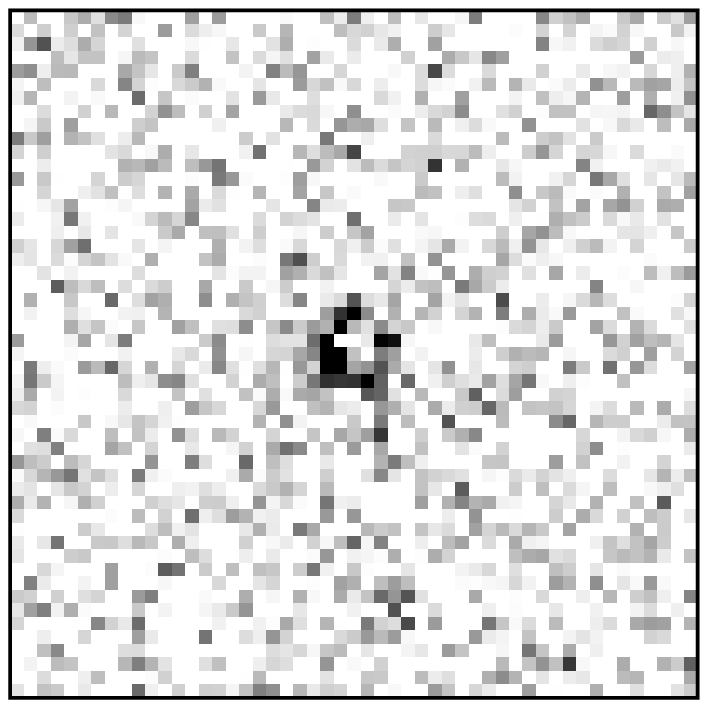}}
\put(14.4,3.0){PG\,1416--1256}
\put(0.0,0.0){%
\includegraphics[bb = 98 98 299 299,clip,angle=0,width=2.7cm]{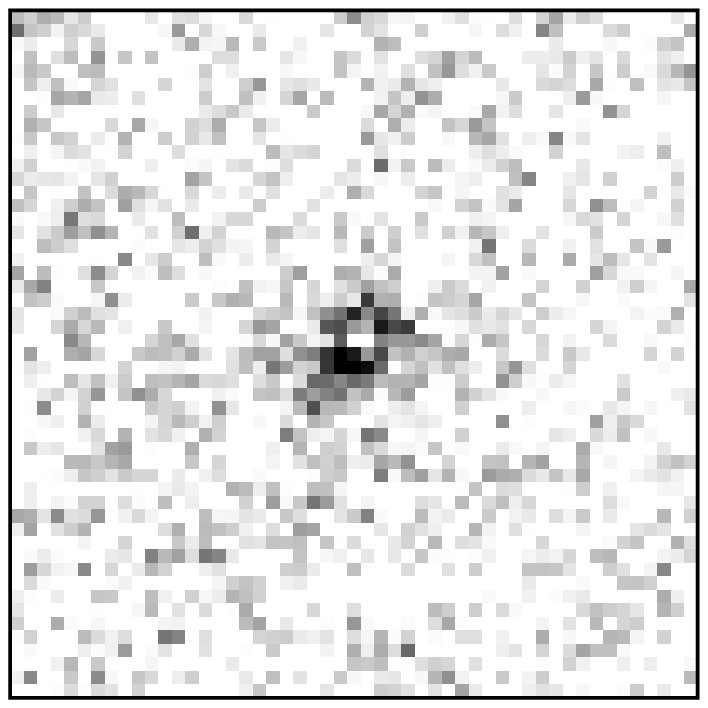}}
\put(0.15,0.15){HE\,1420--0903}
\put(2.85,0.0){%
\includegraphics[bb = 98 98 299 299,clip,angle=0,width=2.7cm]{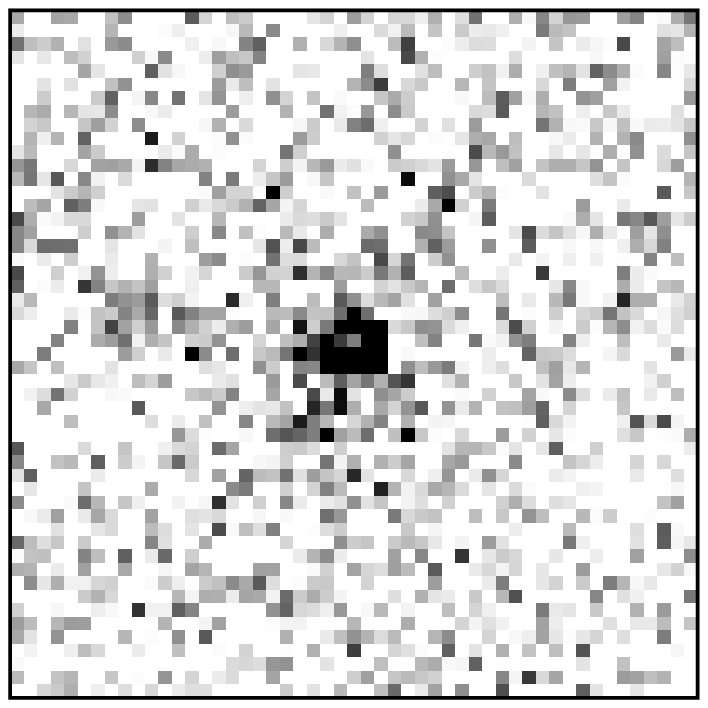}}
\put(3.0,0.15){HE\,1434--1600}
\put(5.7,0.0){%
\includegraphics[bb = 98 98 299 299,clip,angle=0,width=2.7cm]{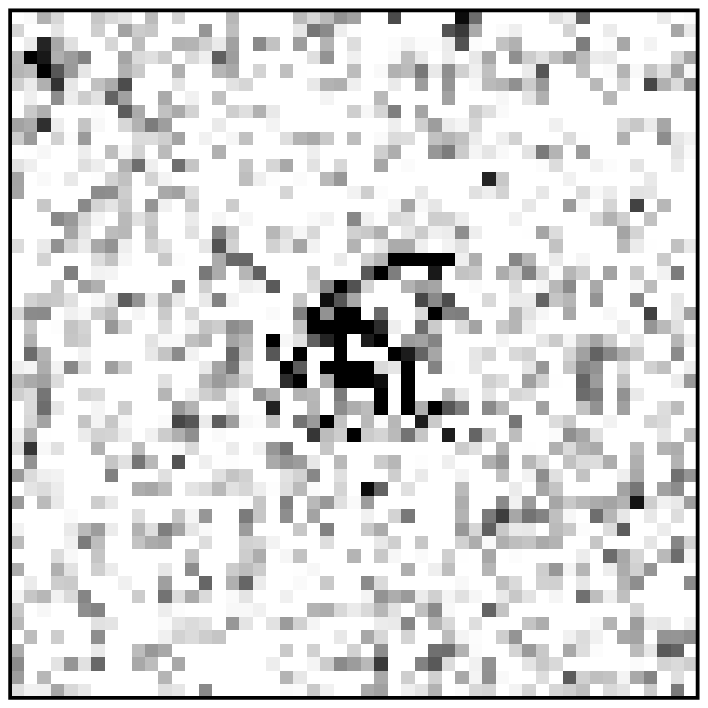}}
\put(5.85,0.15){HE\,1522--0955}
\put(8.55,0.0){%
\includegraphics[bb = 98 98 299 299,clip,angle=0,width=2.7cm]{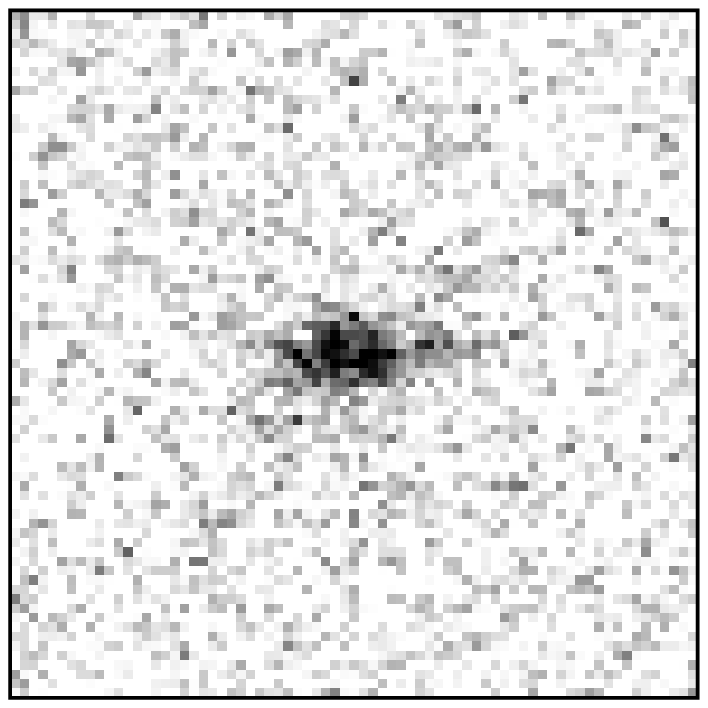}}
\put(8.7,0.15){RXJ\,04407--3442}
\put(11.4,0.0){%
\includegraphics[bb = 98 98 299 299,clip,angle=0,width=2.7cm]{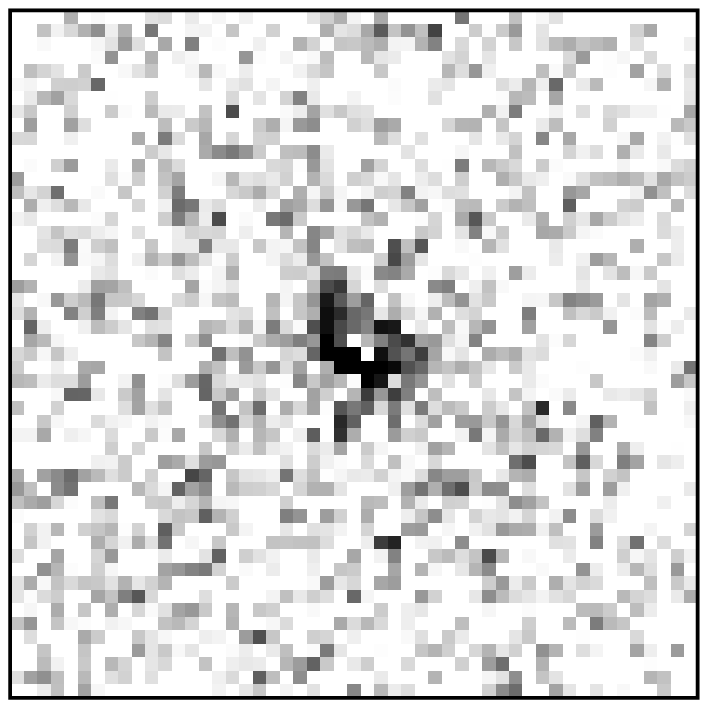}}
\put(11.55,0.15){RXJ\,04474--3309}
\put(14.25,0.0){%
\includegraphics[bb = 98 98 299 299,clip,angle=0,width=2.7cm]{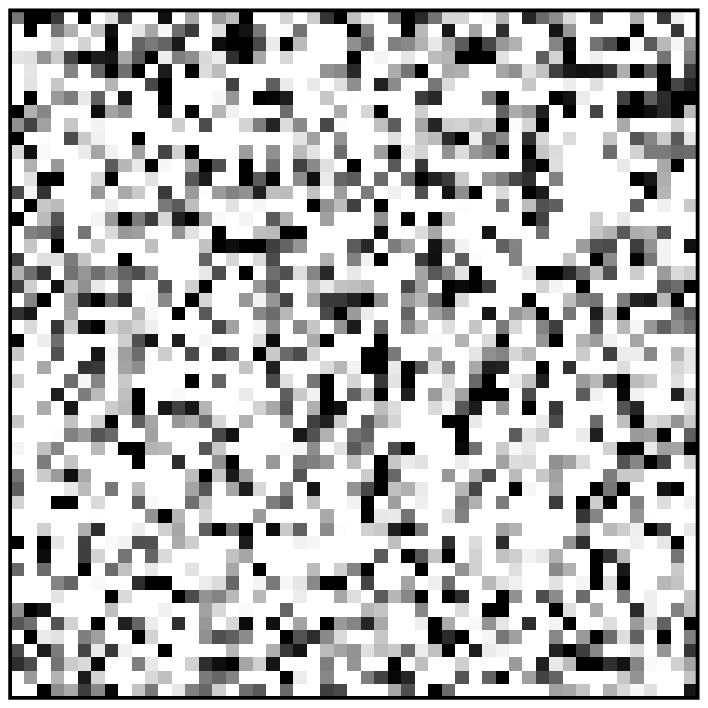}}
\put(14.4,0.15){RXJ\,04485--3346}
\end{picture}
\caption{
The extracted host galaxies: Shown are the residual host galaxy images
after subtraction of the scaled PSF star to remove the nuclear
contribution. The size of the individual images is given in the last
column of Table~\ref{tab:objekte1}. (Continued next page)
}\label{fig:allobj}  
\end{figure*}

\begin{figure*}
\setlength{\unitlength}{1cm}
\begin{picture}(16.95,5.55)

\put(0.0,2.85){%
\includegraphics[bb = 98 98 299 299,clip,angle=0,width=2.7cm]{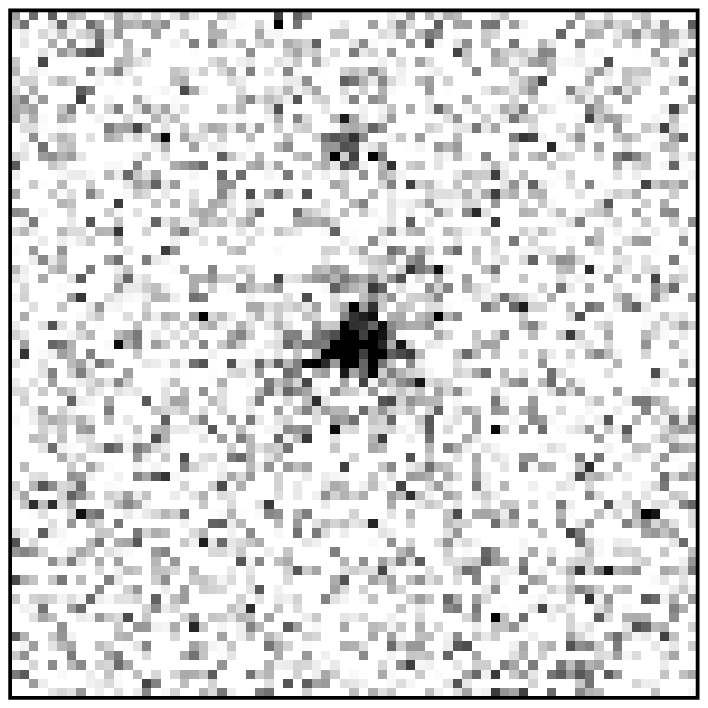}}
\put(0.15,3.0){RXJ\,05043--2554}
\put(2.85,2.85){%
\includegraphics[bb = 98 98 299 299,clip,angle=0,width=2.7cm]{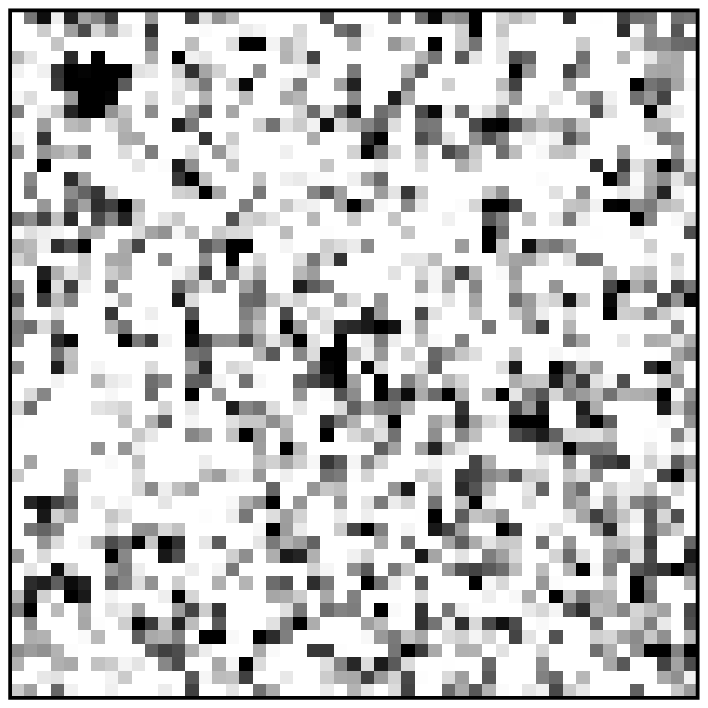}}
\put(3.0,3.0){RXJ\,05079--3411}
\put(5.7,2.85){%
\includegraphics[bb = 98 98 299 299,clip,angle=0,width=2.7cm]{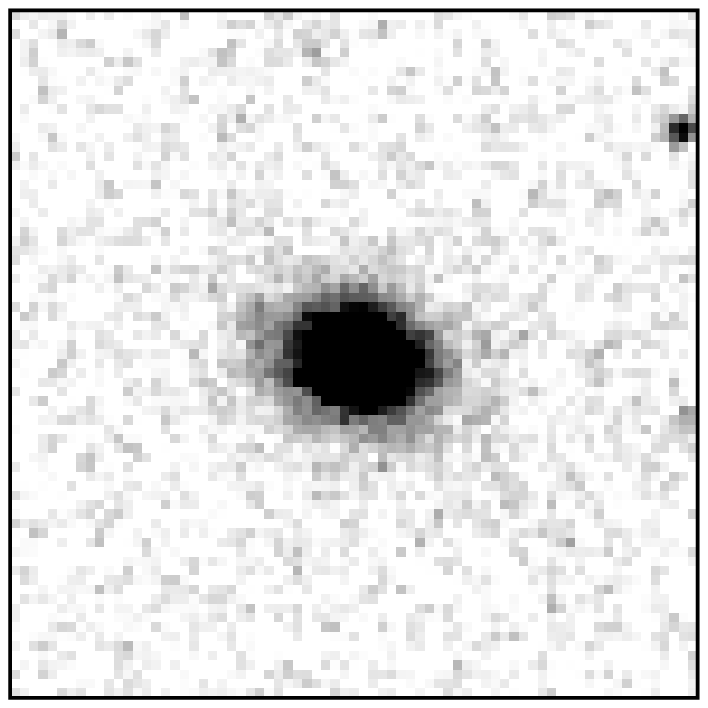}}
\put(5.85,3.0){RXJ\,10119--1635}
\put(8.55,2.85){%
\includegraphics[bb = 98 98 299 299,clip,angle=0,width=2.7cm]{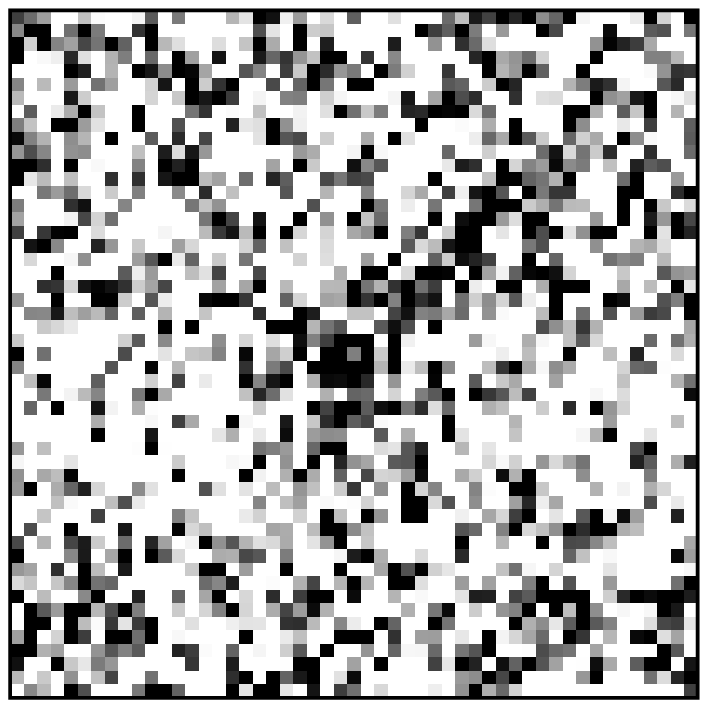}}
\put(8.7,3.0){RXJ\,10596--1506}
\put(11.4,2.85){%
\includegraphics[bb = 98 98 299 299,clip,angle=0,width=2.7cm]{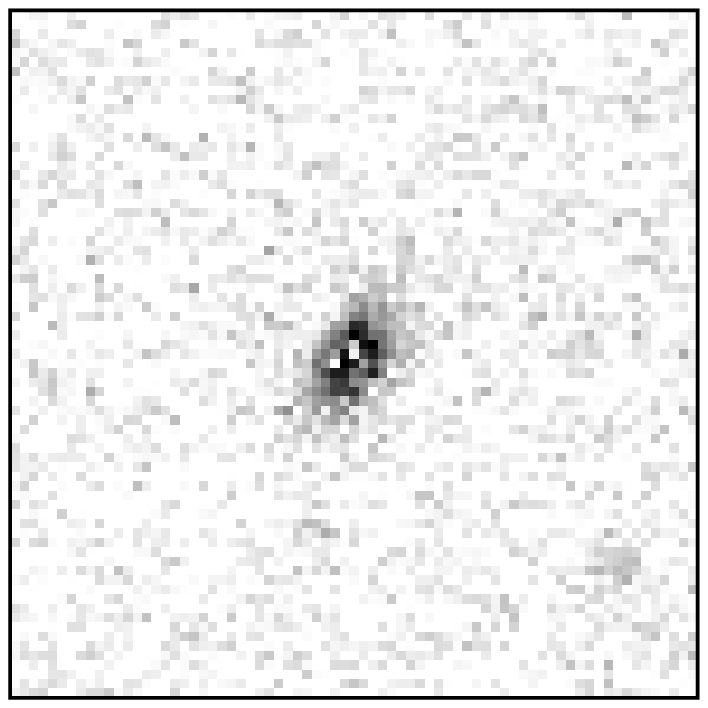}}
\put(11.55,3.0){RXJ\,11355--1328}
\put(14.25,2.85){%
\includegraphics[bb = 98 98 299 299,clip,angle=0,width=2.7cm]{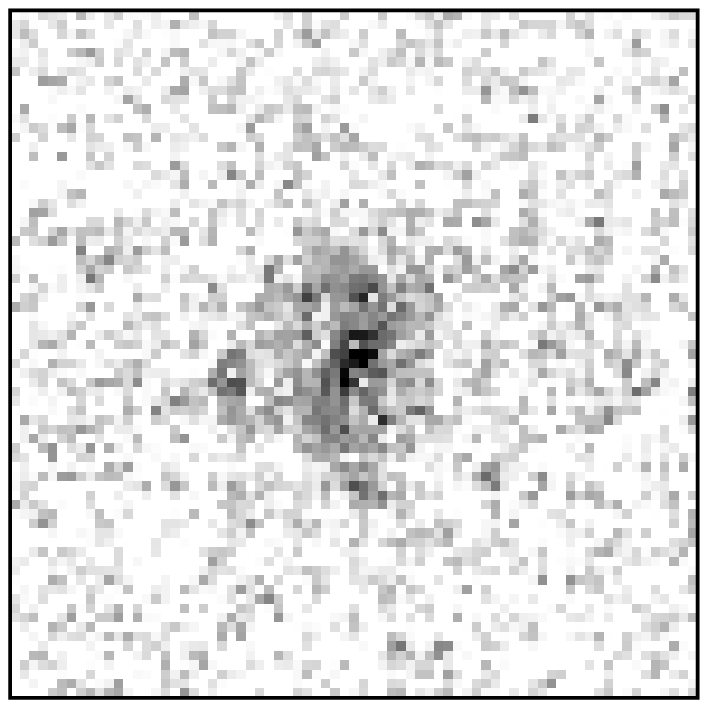}}
\put(14.4,3.0){RXJ\,12167--3035}
\put(0.0,0.0){%
\includegraphics[bb = 98 98 299 299,clip,angle=0,width=2.7cm]{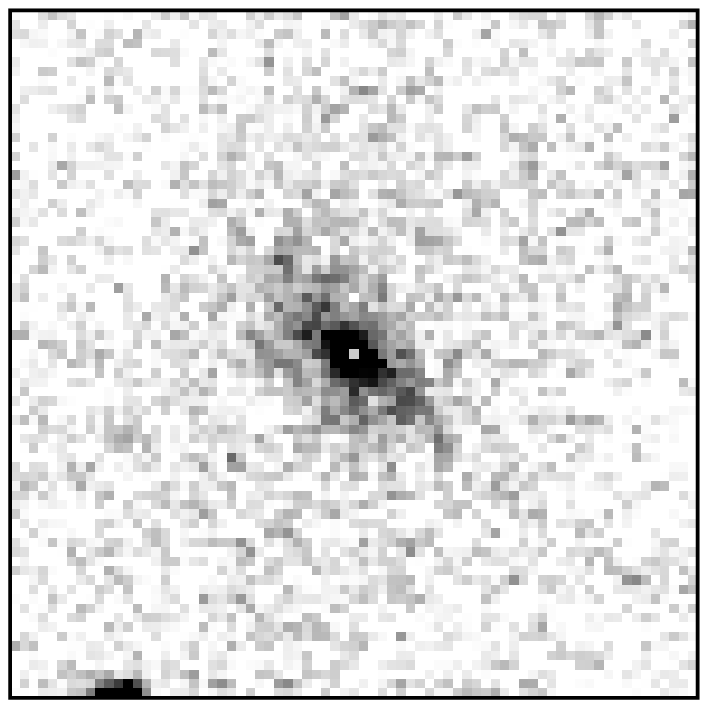}}
\put(0.15,0.15){RXJ\,12297--3153}
\put(2.85,0.0){%
\includegraphics[bb = 98 98 299 299,clip,angle=0,width=2.7cm]{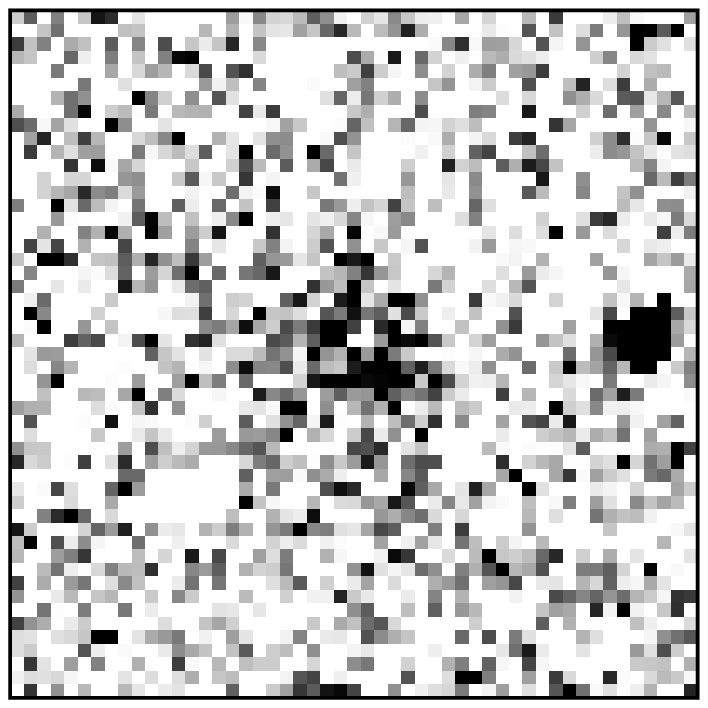}}
\put(3.0,0.15){RXJ\,12570--1409}
\put(5.7,0.0){%
\includegraphics[bb = 98 98 299 299,clip,angle=0,width=2.7cm]{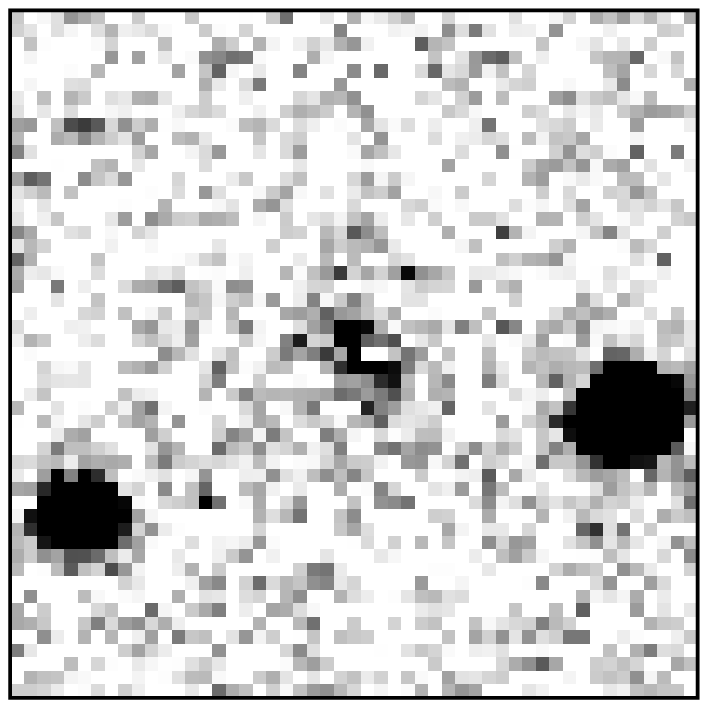}}
\put(5.85,0.15){RXJ\,13531--2802}
\end{picture}
\contcaption{}
\end{figure*}

Observations of photometric standard star sequences during the same
night were used to calibrate the instrumental magnitudes of point
sources in the CCD images \citep[cf.\ procedure described
by][]{reim96}. We did not perform corrections to the standard Bessell
or Johnson systems, but since any such corrections should be below
0.1~mag for the type of objects discussed here, this aspect may be
safely neglected. Large aperture measurements on the CCD frames gave
the total (AGN\,+\,host) magnitudes. Together with the redshifts
obtained from our low-resolution spectroscopy, we could place each
source in a Hubble diagram as shown in
Fig.~\ref{fig:B_z_distribution1}.

It can be seen that the majority of our objects is located in the
borderline region between high luminosity QSOs and lower luminosity
Seyfert~1s. Since there seems to be a continuity of properties in all
respects, we see no use in maintaining the historicaly grown arbitrary
luminosity division and refer to all these objects as `QSOs'
henceforth.

\section{Data analysis}\label{sec:analysis}

\subsection{Separation of stellar and nuclear light}\label{sec:psfsub}

To assess the galaxy apparent magnitudes, the nuclear and stellar
light components needed to be separated. This was done by subtracting
a scaled empirical two dimensional point-spread function (PSF) defined
from a bright star in the object frame \citep[cf.\ similar procedures
employed e.g.\
by][]{vero90,dunl93,mcle94a,mcle94b,roen96,scar00,sanc03}. Potential
PSF stars had to be bright yet unsaturated, undistorted by cosmics,
and spatially as close as possible to the QSO. The latter condition
was particular important for a focal-reducer type instruments such as
EFOSC where the shape of the PSF can vary significantly over the
frame.

Prior to PSF subtraction, QSO and PSF star were recentered with
respect to the pixel grid, so that their centroids coincided by better
than 1/10th of a pixel. To avoid artificial broadening of either QSO
or PSF star due to the necessary rebinning, both objects were shifted
by equal amounts towards each other.

Following the definition of a PSF star, the factor of nuclear light to
be subtracted had to be determined. An amount was subtracted as to
yield a smooth, flat-top radial profile for the residual image, with
no depression in the centre, a criterium sucessfully used by other
authors \citep{mcle94a,roen96}.  For this calculation we utilised
one-dimensional radial surface brightness profiles of the object and
PSF star images, azimuthally averaged over elliptical isophotes.

Since the exact amount the nuclear contribution is {\em a priori}
unknown, this procedure will usually lead to an underestimation of the
galaxy flux, again by an unknown amount. However, two properties can
be assumed valid for an arbitrary galaxy image: positive flux at all
radii, and a surface brightness decreasing monotonously with radius.
The former is always true; the latter is at least reasonable except
for the -- easily detectable -- cases of strong spiral arms or
circumnuclear starburst rings. The `monotony' criterion will yield a
higher flux for the residual galaxy than the `positive central flux'
criterion, and although it still leads to an oversubtraction of
nuclear light (see next section), the amount of this
oversubtraction will be minimised.

After subtraction of the AGN component, the residuals were always
inspected to check for possible PSF mismatches. Typical mismatch
patterns were `butterfly' residuals with paired regions of positive
and negative flux. In such cases it was first tried to apply slightly
different centroiding displacements; in case of no improvement, a
different star was used for the PSF. We finally succeeded to obtain
reasonable PSF models for all objects in the sample.

The residual galaxy images, after subtraction of the nuclei, are shown
in Fig.~\ref{fig:allobj}. The correction for oversubtraction discussed
in the next section was not applied to these images.

\subsection{Correction for oversubtraction} \label{sec:oversub}

As the overestimation of nuclear light in the PSF subtraction process
can result in a galaxy brightness too faint by several tenths of a
magnitude \citep{abra92,dunl93}, we estimated the amount of
oversubtraction from simulations. An array of artificial QSO+host
images was created and treated with the methods described above. The
image quality was chosen to match that of the actual data, and for
each model several realisations of random photon shot noise were
produced. Average correction terms and resulting uncertainties were
calculated to give the amount of oversubtraction as well as the
corresponding spread as a function of input model parameters.

The images were created as a function of redshift, galaxy
morphological type, half-light radius, seeing, and nuclear and galaxy
luminosity. For the morphological types two different models were
assumed: An exponential disc model with
\[
I(r) = I_0 \exp\left[{-1.68\frac{r}{r_0}}\right]
\]
for spiral and disc galaxies \citep{free70}, and a de Vaucouleurs'
model
\[
I(r) = I_0 \exp\left[{-7.67(\frac{r}{r_0})^\frac{1}{4}}\right]
\]
for elliptical or bulge dominated galaxies \citep{deva48}. Here $I(r)$
is the radial surface brightness distribution in mag~arcsec$^{-1}$,
$I_0$ the central surface brightness and $r_0$ the radius in seconds
of arc containing half of the total light. Because of the general
difficulties in the analysis of galaxies with bright nuclei, a further
breakdown in Hubble sub-types is usually not possible in ground-based
data and was not attempted in these simulations.

\begin{figure*}
\includegraphics[bb = 58 598 295 770,clip,angle=0,height=4.2cm]{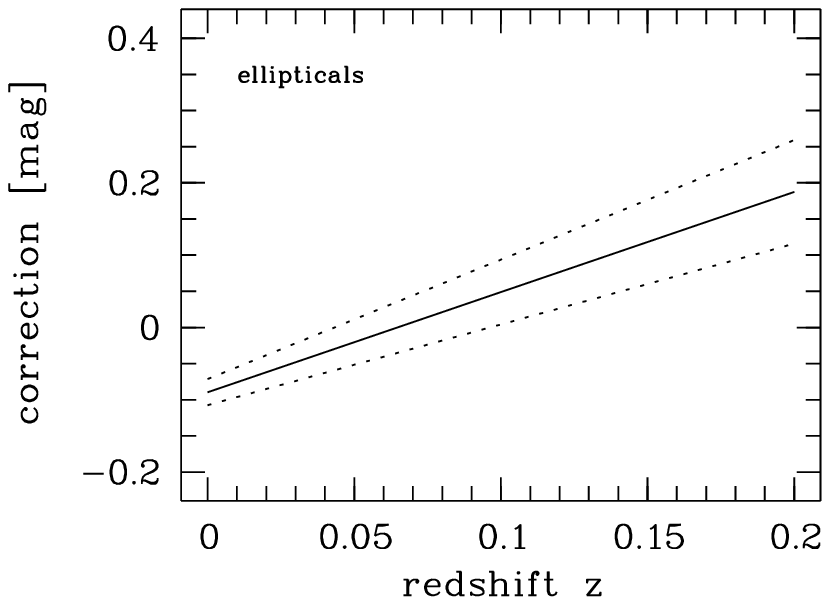}
\hfill
\includegraphics[bb = 80 598 295 770,clip,angle=0,height=4.2cm]{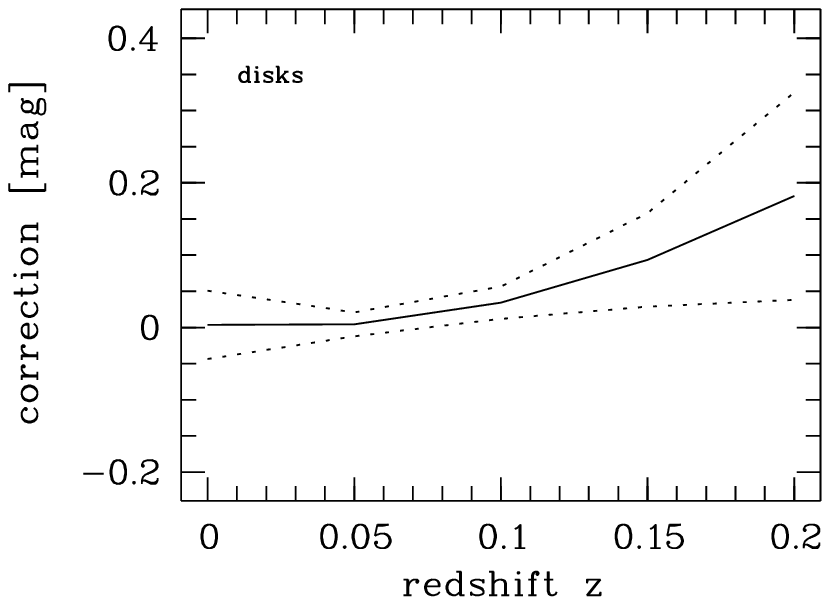}
\hfill
\includegraphics[bb = 80 598 295 770,clip,angle=0,height=4.2cm]{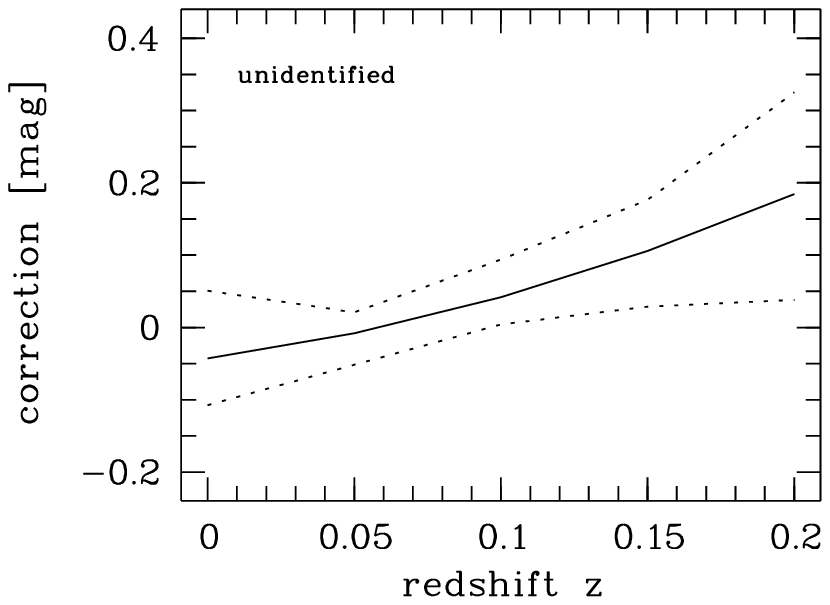}
\caption{
Correction terms for oversubtraction of the nuclear component as a
function of redshift as determined in numerical simulations.  The
solid lines represent approximate averages over the range of adopted
input parameters.  The left panel shows the values for elliptical, the
middle panel that for disc galaxies. The right panel is a combination
of the two, to be used for hosts with unidentified galaxy type.
Dotted lines bracket the range of values found in the simulations.
}
\label{fig:sim}
\end{figure*}                            

At each redshift, using two different half-light radii,
galaxy-to-nuclear luminosity ratios, seeing values, as well as various
photon shot noise realisations yielded a distribution of correction
values.  Figure~\ref{fig:sim} shows the resulting average correction
terms -- differences between `true' and reconstruced magnitudes --,
with suitable interpolation between the simulated redshifts. The range
covered by the distributions is indicated by the dotted lines. As the
distributions are considerably non-Gaussian, we have adopted the
min--max range rather than rms scatter as uncertainty envelope.

The simulations show that the correction term depends mainly on the
amount of angular extention, i.e.\ compactness, of the galaxy in
comparison to the shape of the point like nucleus. In the case of
relatively small seeing variations (full range $1\farcs4\pm0\farcs3$)
present in this data, the dominating parameters thus are the
morphological type of the galaxy, ellipticals being more compact than
discs, and the angular scale length, i.e.\ the combination of physical
scalelength and redshift. Intrinsically more compact and more distant
galaxies require larger corrections, thus also larger corrections for
ellipticals than for disc galaxies for the same range of half-light
radii. The effect from the range of scale lengths dominated over the
effect of different seeing and the different ratios of nuclear and
galaxy luminosities by a factor of two.

In the decomposition of quasar images, scale lengths are not a well
constrained parameter, except for the case of very high S/N data
\citep{abra92,tayl96} and it was also not possible to determine these
numbers reliably for our sample. However, the half-light radii used in
the simulations ($r_0 = 9$~kpc and 18~kpc for discs, 15~kpc and 30~kpc
for ellipticals) are rather in the upper range of what is expected in
the data. This leads to \emph{conservative} estimates for the
corrections: If an observed object has a host galaxy with a smaller
scale length, its luminosity tends to be \emph{under}estimated.

For many of our objects we could also not establish reliable
morphological types (see below). To be able to deal with these cases
we determined a correction term for hosts of unknown type by
combination of the two type-specific relations, and this is shown in
the right-hand panel of Fig.~\ref{fig:sim}. The range of possible
correction values is dominated by ellipticals for lower redshifts, and
by discs at the high end. Note that the systematic differences between
different models are negligible, considering the other error sources.
This is an important feature of our adopted procedure, as it indicates
that our PSF subtraction strategy allows us to estimate total host
galaxy luminosities nearly independently of morphological type.

\subsection{Host galaxy morphological types}\label{sec:galtypes}

The correction terms described above generally depend on galaxy
type. If the type is known, the intrinsic uncertainty associated with
the correction term is smaller than for the combined general
correction relation shown in the middle panel of Fig.~\ref{fig:sim}.
It is therefore useful to determine galaxy types for as many objects
as possible.

To determine individual galaxy types we fitted surface brightness
distribution laws to the one-dimensional luminosity profiles of the
nucleus-subtracted images, trying both the exponential disc and the
de~Vaucouleurs profiles as given above. The inner radius for the fit
was fixed at one FWHM of the seeing, the outer radius was set at a
S/N=1 in the galaxy profile.

To decide which model provided the better approximation, some authors
\citep[e.g.][]{roen96,tayl96} compared the $\chi^2$ values of both
model fits and adopted the model with the smaller $\chi^2$. This
strategy is questionable, as the statistical significance of a
decision is difficult to control, especially since real galaxies often
fail to perfectly trace the theoretical distributions, but display
spiral arms, knots, or other irregularities. More conservatively, we
considered a galaxy type only to be safely determined if it could not
be ruled out in a $\chi^2$ test at 95\% confidence level, while at the
same time the alternative model was rejected. Alltogether for 12 of
the 57 objects in the sample we could establish galaxy types in this
way -- four discs and eight ellipticals. Further three objects not
classified with the $\chi^2$ test were obviously spirals due to
visible spiral arms.

\subsection{Photometry}\label{sec:photometry}

Galaxy magnitudes were measured by simulated aperture photometry in
the PSF-subtracted residual images. Increasing the aperture radius in
steps of one pixel yielded a curve-of-growth for the integrated flux
with radius. Whenever the sky background was perfectly subtracted,
these curves converged; non-convergence of the curve-of-growth, within
the statistical uncertainties, indicated a nonperfect background
subtraction. This yielded an excellent tool to optimise the background
determination. The galaxy magnitude was adopted as the total
integrated flux, after local background removal, to which the
correction term for oversubtraction according to Fig.~\ref{fig:sim}
was applied.

Pixels in the vicinity of the galaxies showing obvious artefacts of
the CCD and cosmic ray events were eliminated from the analysis by
creating a mask for each image. Masked pixels were not used, but
effectively replaced by the average surface brightness value in the
corresponding annulus. Likewise, projected neighbours such as
foreground stars and background galaxies, but also physical companions
were excluded (see discussion in Sect.~\ref{sec:companioninfluence} and
Sect.~\ref{sec:peculiarities} for a few exceptions).

\subsection{Photometric uncertainties}\label{sec:uncert}

We considered three major sources of uncertainties that contributed to
the total error of the host galaxy magnitudes, apart from Poissonian
errors in the individual images:
\begin{enumerate}

\item Uncertainties in the determination of the total flux from the
curve-of-growth. The simulations show that in different shot noise
realisations the derived flux fluctuates by $\pm 0.05$~mag (rms) out
to $z=0.2$ for this image quality, independent of the galaxy type.

\item Errors in the correction for oversubtraction. According to
Fig.~\ref{fig:sim}, the range around the average is smaller than $\pm
0.1$~mag for $z < 0.1$ and $\pm 0.15$~mag for $0.1 <z< 0.2$.

\item The uncertainty estimates for the corrections are only valid if
the half-light radii of the objects in the sample are not much
different from the values of $r_0$ adopted in the simulations.  From
the curve-of-growth analyses we estimate half-light radii between 9
and 18~kpc for most of the objects. As our data do not really have the
angular resolution to probe into the central regions, it is likely
that these values are somewhat too high for several of the
objects. This unknown systematic error could push the luminosities of
the most compact hosts to yet higher values. We have chosen to be
conservative and to assume that the adopted corrections are, on
average, correct.
\end{enumerate}

The uncertainty estimates are summarised in Tab.~\ref{tab:errors},
broken up into two redshift regions. For all but two objects, where
photon shot noise dominates, these external errors determine the error
budget.

\begin{table}
  \caption{Photometric uncertainties for external error
    sources discussed in the text. The last column gives the combined
    uncertainty $\Delta B_{\rmn{err,gal}}$ for two redshift ranges.}
  \label{tab:errors}
\begin{center}
\begin{tabular}{c*{3}{r@{.}l}}
  \hline
  & \multicolumn{4}{c}{Error source} & 
  \multicolumn{2}{c}{$\Delta B_{\rmn{err,gal}}$}\\
  \cline{2-5}
  $z$-range & 
  \multicolumn{2}{c}{(i)} & 
  \multicolumn{2}{c}{(ii)} & 
  \multicolumn{2}{c}{[mag]}\\
  \hline\hline
\rule{0cm}{2.7ex}%
$0.0\ldots0.1$&         $\pm0$& 05 & $\pm0$&1  & $\pm0$&15 \\
$0.1\ldots0.2$& $\pm0$& 05 & $\pm0$&15 & $\pm0$&2
\rule[-1.5ex]{0cm}{1.5ex}\\
\hline
\end{tabular}
\end{center}
\end{table}

\subsection{Influence of companions, knots, tidal tails}\label{sec:companioninfluence}
Resolved or unresolved flux of companion objects, tidal tails or
`knots' of unknown nature in or around the host galaxies can
contribute a significant amount of flux. This poses the question as to
what constitutes `the host galaxy'? The answer is necessarily
subjective, and depends on spatial and spectral resolution as well as
S/N. An unspecific knot might be identified in high resolution,
multicolour or spectroscopic data as a small companion galaxy in the
process of being accreted, while this could not be inferred with lower
resolution, single band imaging. Even then there is a gradual
transition between a companion -- which is not part of the host galaxy
-- and a clump within the host. Similar arguments can be brought
forward in the case of tidal tails or bridges.

We took the approach to exclude the contribution of all visible
compact asymmetries from the photometry of the host
galaxies. Technically, as described in Sect.~\ref{sec:photometry}, we
used the underlying smooth host galaxy contribution instead. We
estimate that we exclude the contributions from all {\em compact}
sources that would contribute more than 0.1~mag. Contributions from
low surface brightness companions can not be ruled out, but even with
higher S/N data these would be difficult to judge. For unresolved
companions see also Sect.~\ref{sec:peculiarities}.

\subsection{Systematic effects in host galaxy extraction}\label{sec:syseffects}
We now investigate whether our brightness estimation method could be
biased by systematic effects that depend on the S/N of the data. Most
crucial is our adopted strategy of determining the appropriate PSF
scaling factor to subtract the nuclear component from the host
galaxy. Two aspects are important in this context: (1) We have
employed a criterion that, although it still should lead to an
oversubtraction, is decidedly less conservative than the more
frequently employed criterion of purely positive residuals.  (2)
Unlike several others, we have applied explicit corrections for
oversubtraction determined from simulations.

\begin{figure}
\includegraphics[bb = 47 85 302 272,clip,angle=0,width=\columnwidth]{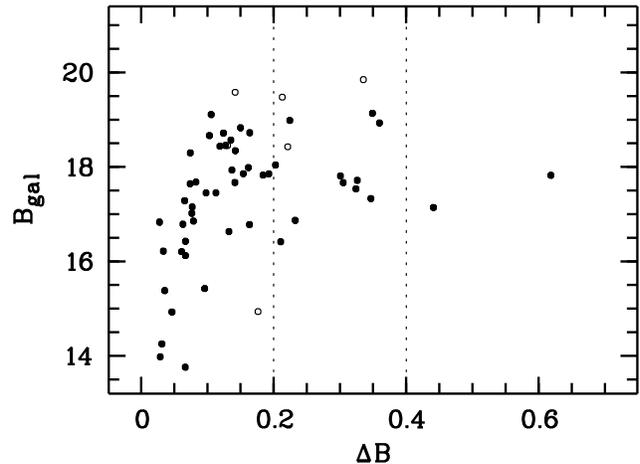}

\caption{Difference $\Delta m$ in derived brightness of the host
  galaxies between the `monotony' criterion for PSF subtraction used for
  analysis and the more conservative
  `positivity' criterion for the surface brightness of the residual
  profile. Only for two objects is $\Delta m > 0.4$. Open symbols
  mark the galaxies with higher uncertainties described above.
  \label{fig:diff_0m}} 
\end{figure}

To test the level by which our `monotony' criterion differs from the
more conservative `positivity' criteron, we have applied the latter to
our dataset and compared the corresponding (uncorrected) aperture host
magnitudes after PSF subtraction. The result of this exercise is shown
in Fig.~\ref{fig:diff_0m}, where the difference $\Delta B$ between
magnitudes determined with either method is plotted against apparent
host magnitude. $\Delta B$ is smaller than 0.4~mag for all but two
objects, and even smaller than 0.2~mag in over 70\,\% of the
cases. Concerning the faintest objects, being the most difficult for
separating host and nucleus, there is no systematic effect apparent
from this relation that would systematically over- or underestimate
these host galaxies in flux.

The correction for oversubtraction described in
Sect.~\ref{sec:oversub} and listed in Tab.~\ref{tab:objekte1} are $<
0.25$ mag for all objects. As discussed in Sections~\ref{sec:oversub}
and \ref{sec:uncert}, a possible systematic error in corrections could
occur if some of the observed host galaxies were more compact than
assumed in the simulations -- but in that case the adopted corrections
would be conservative again and the host luminosities rather under-
than overestimated.

\subsection{External test of PSF subtraction}\label{sec:externaltest}
The simulations described in secs.~\ref{sec:oversub} and
\ref{sec:uncert} demonstrate that our extracted host galaxy fluxes are
quite reliable, and provide realistic error margins. In the light of
the relatively low S/N of the data that this study is based on, we
attempted to support this statement by taking into account a second,
independent method and additional data.

In the course of a separate project with the aim to investigate the
stellar composition of quasar host galaxies \citep{jahn02a} we
obtained multiband photometry for 19 objects out of the sample
discussed here. New optical $V$, $R$ and $I$ band as well as
near-infrared data were obtained. To separate the nuclear and host
galaxy contributions for all of these images a new two-dimensional
modelling algorithm was used \citep{kuhl02}.

Modelling the two-dimensional surface brightness distribution is
usually superior to PSF scaling governed by radial profiles. However,
in the case of low S/N data such a method becomes increasingly
dominated by artefacts like PSF mismatch, asymmetries, and other
deviations from the model assumptions when fitting a full parameter
set (5--9 parameters). Thus without external constraints these methods
can not be applied to the data presented here.

To reduce the number of free parameters and to guarantee a
homogenenous treatment of all images of an object, the morphological
parameters of a given object were determined globally for all
bands. Using these parameters on the $B$ band images left only nuclear
and host fluxes as free parameters. With these additional constraints
modelling was successful for 12 of the 19 objects. In the remaining
cases the size of the PSF mismatch or low S/N prevented the
two-dimensional modelling to successfully converge.

\begin{figure}
\begin{center}
\includegraphics[bb = 67 450 396 772,clip,angle=0,width=\columnwidth]{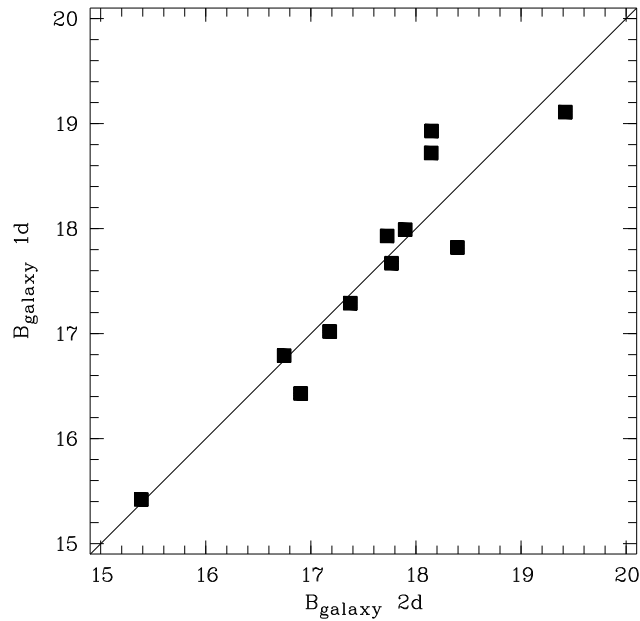}
\caption{
Apparent brightness of a 12 object `multicolour' subsample for which
two-dimensional modelling of the surface brightness distribution of
the images was possible. Plotted is the `1d' value of the host from
this study against the `2d' value. The solid line marks the 1:1
relation.
\label{fig:B_B_comparison}}
\end{center}
\end{figure}

Plotting the resulting apparent $B$ band host galaxy magnitudes
derived in this independent way against the values derived from the
PSF subtraction presented in this work yields the relation shown in
Figure~\ref{fig:B_B_comparison}. Individual deviations up to 0.75~mag
exist, but the values scatter symmetrically around the 1:1
correlation. The ensemble average difference is 0.00~mag
($\sigma=0.39$~mag).

The 12 objects of the `multicolour' subsample span both the full
redshift and brightness range of the present sample
(Fig.~\ref{fig:B_z_distribution2}). The absence of any trend with
apparent brightness suggests that the host galaxy fluxes are properly
recovered and -- for the upcoming discussion -- that there is neither
evidence that the luminosities of our objects are systematically
\emph{over}estimated, nor that any systematic effect is apparent to
boost the host galaxies of the lower S/N.

\section{Results} \label{sec:results}

\subsection{Apparent magnitudes} \label{sec:appmag}

For 55 of the 57 objects in the sample, significant flux was detected
in the PSF-subtracted image. In two cases the uncertainties are large
and the residuals are dominated by shot noise, manifest also in
significant fluctuations in the outer parts of the growth curves. Two
further objects were observed near full moon and also these images are
very noisy, although the host galaxies are clearly detected. In two
further cases the PSF was poorly defined, so that the PSF subtraction
must be considered quite uncertain. These six objects are marked by
the `L' flag in column 4 of Table~\ref{tab:objekte1} and by the open
symbols in the relevant figures below. They will be treated separately
in the subsequent discussion.

\begin{figure}
\begin{center}
\includegraphics[bb = 70 584 323 770,clip,angle=0,width=\columnwidth]{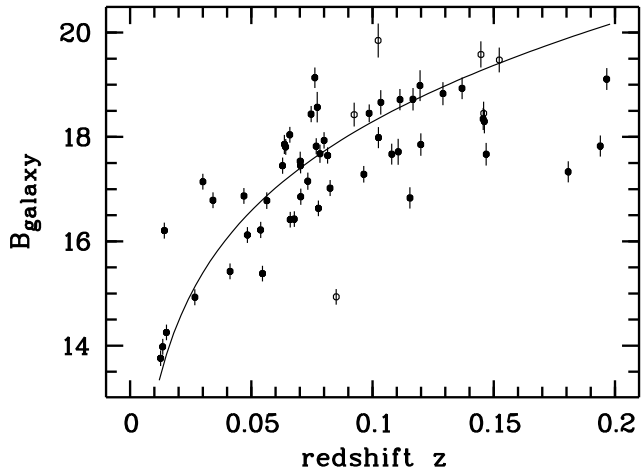}
\caption{
Distribution of the apparent galaxy magnitudes $B_{\rmn{gal}}$ with
redshift $z$ for the HES sample. Open symbols mark the objects with
higher uncertainty. Errors are the quadractic combination of internal
Poissonian and external errors. The solid line marks the apparent
brightness of a galaxy with Schechter luminosity $L^*$ including
galaxy $K$ correction for intermediate type (Sab) galaxy colours.
\label{fig:B_z_distribution2}}
\end{center}
\end{figure}

Redshift-dependent corrections for oversubtraction were applied using
the relation shown in Fig.~\ref{fig:sim}, type-specific in the case of
the 15 objects with determined galaxy types, type-averaged values for
the remaining ones.  The individual values of the corrections $\Delta
B_{\rmn{gal}}$ are given together with the corrected apparent
magnitudes $B_{\rmn{gal}}$ and $B_{\rmn{nuc}}$ in
Table~\ref{tab:objekte1}.  Figure~\ref{fig:B_z_distribution2} shows
the distribution of $B_{\rmn{gal}}$ with redshift $z$. Error bars were
derived from quadratic combination of internal Poissonian and external
errors. For comparison we plotted the brightness of a field galaxy
with Schechter luminosity \citep[$M_B = -21.0$;][]{zucc97} as a
function of $z$.

%

\begin{table*}
 \begin{small}
  \caption{
The observed sample: Properties and results. The first two columns
give object designations and redshifts, followed by a letter a--d
indicating the observing campaign as listed in Table. Column `DQ'
lists `L' if the data quality is considered to be low. Column `Type'
gives the morphological type when established (E: elliptical, D: disc,
S: spiral arms visible). The next columns list the estimated
correction for oversubtraction, the corrected apparent, and the
inferred absolute magnitudes (standard $K$ correction). The next to
last column lists the predicted $H$ band magnitude of the host galaxy,
based on `normal' galaxy colours. The last column gives the size of
the object's images in Fig.~\ref{fig:allobj}.
}
   \label{tab:objekte1}  
\begin{center}
\begin{tabular}{llccc*{7}{r@{.}l}c}
  \hline Object&$z$&C&DQ&Type& 
  \multicolumn{2}{c}{$\Delta\!B_{\mathrm{gal}}$} & 
  \multicolumn{2}{c}{$B_{\mathrm{nuc}}$} & 
  \multicolumn{2}{c}{$B_{\mathrm{gal}}$} &
  \multicolumn{2}{c}{$M_{B,\mathrm{nuc}}$} &
  \multicolumn{2}{c}{$M_{B,\mathrm{gal}}$} &
  \multicolumn{2}{c}{$M_{B,\mathrm{tot}}$} &
  \multicolumn{2}{c}{$M_{H,\mathrm{gal}}$} &
  Image size [\arcsec] \\
  \hline \hline
  HE\,0317--2638 & 0.078 & a &  & & $-$0&02 & 17&45 & 16&63 & $-$20&95 & $-$22&12 & $-$22&43& $-$25&72 &48\\
  IR\,03450+0055 & 0.030 & a &  & & $+$0&05 & 14&96 & 17&14 & $-$21&89 & $-$19&87 & $-$22&05& $-$23&47 &48\\
  HE\,0348--2226 & 0.111 & b &  & & $-$0&06 & 17&95 & 17&72 & $-$21&23 & $-$21&94 & $-$22&37& $-$25&54 &20\\
  HE\,0403--3719 & 0.056 & b &  &D& $-$0&02 & 16&56 & 16&78 & $-$21&11 & $-$21&16 & $-$21&88& $-$24&46 &38\\
  HE\,0414--2552 & 0.152 & c &L & & $-$0&11 & 17&34 & 19&48 & $-$22&55 & $-$21&16 & $-$22&81& $-$24&76 &38\\
  HE\,0427--2701 & 0.064 & a &  & &    0&00 & 17&60 & 17&81 & $-$20&48 & $-$20&60 & $-$21&30& $-$24&20 &38\\
  HE\,0444--3449 & 0.181 & a &  & & $-$0&15 & 16&24 & 17&33 & $-$23&93 & $-$23&68 & $-$24&53& $-$27&28 &38\\
  HE\,0444--3900 & 0.120 & c &  & & $-$0&07 & 17&45 & 18&98 & $-$21&89 & $-$20&93 & $-$22&26& $-$24&53 &48\\
  HE\,0507--2710 & 0.146 & c &  & & $-$0&11 & 17&39 & 18&30 & $-$22&38 & $-$22&15 & $-$23&00& $-$25&75 &48\\
  HE\,0517--3243 & 0.013 & a &  & & $+$0&07 & 15&69 & 13&76 & $-$18&33 & $-$20&89 & $-$21&05& $-$24&49 &80\\
  HE\,0526--4148 & 0.076 & c &  & & $-$0&01 & 17&95 & 19&14 & $-$20&50 & $-$19&70 & $-$20&92& $-$23&30 &38\\
  HE\,0529--3918 & 0.077 & b &  & & $-$0&01 & 17&35 & 18&57 & $-$21&13 & $-$20&30 & $-$21&54& $-$23&90 &20\\
  HE\,0952--1552 & 0.108 & d &  & & $-$0&06 & 16&82 & 17&67 & $-$22&55 & $-$22&21 & $-$23&13& $-$25&81 &48\\
  IR\,09595--0755& 0.055 & d &  & & $+$0&02 & 17&27 & 15&38 & $-$20&57 & $-$22&82 & $-$22&96& $-$26&42 &48\\
  HE\,1019--1414 & 0.077 & a &  & & $-$0&01 & 17&05 & 17&82 & $-$21&73 & $-$21&33 & $-$22&30& $-$24&93 &38\\
  PKS\,1020--103 & 0.197 & d &  & & $-$0&17 & 17&41 & 19&11 & $-$23&13 & $-$22&38 & $-$23&55& $-$25&98 &60\\
  HE\,1029--1401 & 0.085 & a &L & & $-$0&03 & 14&40 & 14&94 & $-$24&60 & $-$24&48 & $-$25&28& $-$28&08 &48\\
  HE\,1043--1346 & 0.068 & d &  &S& $-$0&01 & 17&99 & 16&43 & $-$20&39 & $-$22&25 & $-$22&42& $-$25&55 &48\\
  HE\,1106--2321 & 0.081 & d &  & & $-$0&02 & 15&28 & 17&64 & $-$23&49 & $-$21&54 & $-$23&66& $-$25&14 &48\\
  HE\,1110--1910 & 0.111 & d &  &E& $-$0&08 & 16&81 & 18&72 & $-$22&51 & $-$21&15 & $-$22&77& $-$25&15 &48\\
  Q\,1114--2846  & 0.070 & c &  & & $-$0&01 & 16&33 & 17&54 & $-$22&13 & $-$21&28 & $-$22&54& $-$24&88 &48\\
  PG\,1149--1105 & 0.048 & d &  & & $+$0&02 & 16&00 & 16&12 & $-$21&45 & $-$21&59 & $-$22&28& $-$25&19 &48\\
  HE\,1201--2409 & 0.137 & d &  & & $-$0&09 & 16&82 & 18&93 & $-$23&16 & $-$21&71 & $-$23&40& $-$25&31 &38\\
  HE\,1213--1633 & 0.064 & d &  & &    0&00 & 17&57 & 17&86 & $-$20&55 & $-$20&59 & $-$21&32& $-$24&19 &38\\
  HE\,1217--1340 & 0.070 & d &  &E& $-$0&20 & 16&73 & 17&45 & $-$21&69 & $-$21&23 & $-$22&19& $-$25&23 &48\\
  HE\,1228--1637 & 0.102 & d &  & & $-$0&05 & 16&91 & 17&99 & $-$22&23 & $-$21&64 & $-$22&71& $-$25&24 &48\\
  HE\,1235--0857 & 0.070 & d &  &D& $-$0&02 & 18&37 & 16&85 & $-$19&94 & $-$21&73 & $-$21&91& $-$25&03 &48\\
  HE\,1237--2252 & 0.096 & d &  & & $-$0&04 & 18&15 & 17&29 & $-$21&18 & $-$22&45 & $-$22&72& $-$26&05 &48\\
  HE\,1239--2426 & 0.082 & d &  & & $-$0&02 & 17&48 & 17&02 & $-$21&45 & $-$22&30 & $-$22&70& $-$25&90 &48\\
  HE\,1245--2709 & 0.066 & d &  &E& $-$0&20 & 17&73 & 18&04 & $-$20&83 & $-$20&72 & $-$21&46& $-$24&72 &48\\
  HE\,1248--1357 & 0.015 & d &  &S&    0&00 & 17&05 & 14&25 & $-$17&97 & $-$20&85 & $-$20&92& $-$24&15 &70\\
  IR\,12495--1308& 0.013 & d &  & & $+$0&07 & 15&65 & 13&98 & $-$18&73 & $-$20&88 & $-$21&08& $-$24&48 &130\\
  HE\,1300--1325 & 0.047 & a &  & & $+$0&03 & 16&52 & 16&87 & $-$20&95 & $-$20&85 & $-$21&66& $-$24&45 &38\\
  HE\,1309--2501 & 0.063 & c &  &E& $-$0&03 & 16&87 & 17&45 & $-$21&45 & $-$21&17 & $-$22&06& $-$25&17 &70\\
  PG\,1310--1051 & 0.034 & d &  & & $+$0&04 & 15&62 & 16&79 & $-$21&13 & $-$20&15 & $-$21&51& $-$23&75 &48\\
  HE\,1315--1028 & 0.099 & d &  &E& $-$0&10 & 16&95 & 18&45 & $-$22&02 & $-$20&98 & $-$22&35& $-$24&98 &48\\
  HE\,1328--2509 & 0.027 & d &  & & $+$0&05 & 16&35 & 14&93 & $-$19&80 & $-$21&58 & $-$21&81& $-$25&18 &48\\
  HE\,1335--0847 & 0.080 & d &  &D& $-$0&02 & 17&23 & 17&93 & $-$21&36 & $-$21&05 & $-$21&97& $-$24&35 &48\\
  HE\,1338--1423 & 0.041 & c &  & & $+$0&03 & 15&38 & 15&42 & $-$21&96 & $-$22&15 & $-$22&82& $-$25&75 &90\\
  HE\,1348--1758 & 0.014 & b &  & & $+$0&07 & 16&02 & 16&21 & $-$19&08 & $-$19&03 & $-$19&84& $-$22&63 &38\\
  HE\,1405--1545 & 0.194 & d &  &D& $-$0&24 & 16&97 & 17&83 & $-$23&78 & $-$23&79 & $-$24&49& $-$27&09 &38\\
  PG\,1416--1256 & 0.129 & c &  & & $-$0&08 & 16&70 & 18&83 & $-$23&16 & $-$21&66 & $-$23&39& $-$25&26 &38\\
  HE\,1420--0903 & 0.117 & d &  & & $-$0&07 & 18&20 & 18&72 & $-$21&37 & $-$21&38 & $-$22&11& $-$24&98 &38\\
  HE\,1434--1600  & 0.144 & c &  &E& $-$0&20 & 15&58 & 17&67 & $-$24&59 & $-$23&21 &$-$24&84 &$-$27&21 &38\\
  HE\,1522--0955  & 0.146 & d &  & & $-$0&11 & 15&92 & 18&34 & $-$24&33 & $-$22&61 &$-$24&52 &$-$26&21 &38\\
  RXJ\,04407--3442& 0.073 & c &  & & $-$0&01 & 18&39 & 17&15 & $-$19&89 & $-$21&47 &$-$21&69 &$-$25&07 &48\\
  RXJ\,04474--3309& 0.078 & c &  &E& $-$0&01 & 17&53 & 17&68 & $-$20&87 & $-$21&11 &$-$21&74 &$-$25&11 &38\\
  RXJ\,04485--3346& 0.145 & c &L & & $-$0&10 & 18&09 & 19&58 & $-$21&60 & $-$20&80 &$-$22&01 &$-$24&40 &38\\
  RXJ\,05043--2554& 0.120 & c &  & & $-$0&07 & 17&95 & 17&85 & $-$21&44 & $-$22&06 &$-$22&52 &$-$25&66 &48\\
  RXJ\,05079--3411& 0.102 & c &L & & $-$0&05 & 18&71 & 19&85 & $-$20&35 & $-$19&71 &$-$20&82 &$-$23&31 &38\\
  RXJ\,10119--1635& 0.066 & d &  &E& $-$0&20 & 18&41 & 16&41 & $-$20&05 & $-$22&37 &$-$22&49 &$-$26&37 &48\\
  RXJ\,10596--1506& 0.092 & c &L & & $-$0&03 & 18&75 & 18&43 & $-$20&28 & $-$21&01 &$-$21&44 &$-$24&61 &38\\
  RXJ\,11355--1328& 0.075 & c &  & & $-$0&01 & 18&26 & 18&44 & $-$20&18 & $-$20&36 &$-$21&02 &$-$23&96 &48\\
  RXJ\,12167--3035& 0.115 & c &  &S& $-$0&02 & 19&38 & 16&83 & $-$20&30 & $-$23&23 &$-$23&29 &$-$26&53 &48\\
  RXJ\,12297--3153& 0.054 & c &  & & $+$0&02 & 17&76 & 16&22 & $-$20&06 & $-$21&94 &$-$22&13 &$-$25&54 &48\\
  RXJ\,12570--1409& 0.146 & c &L & & $-$0&11 & 18&48 & 18&45 & $-$21&47 & $-$22&12 &$-$22&56 &$-$25&72 &38\\
  RXJ\,13531--2802& 0.103 & c &  & & $-$0&05 & 18&07 & 18&66 & $-$21&13 & $-$21&03 &$-$21&82 &$-$24&63 &38\\
  \hline
\end{tabular}
\end{center}
\end{small}
\end{table*}

\begin{figure*}
\includegraphics[bb = 60 486 451 772,clip,angle=0,width=12cm]{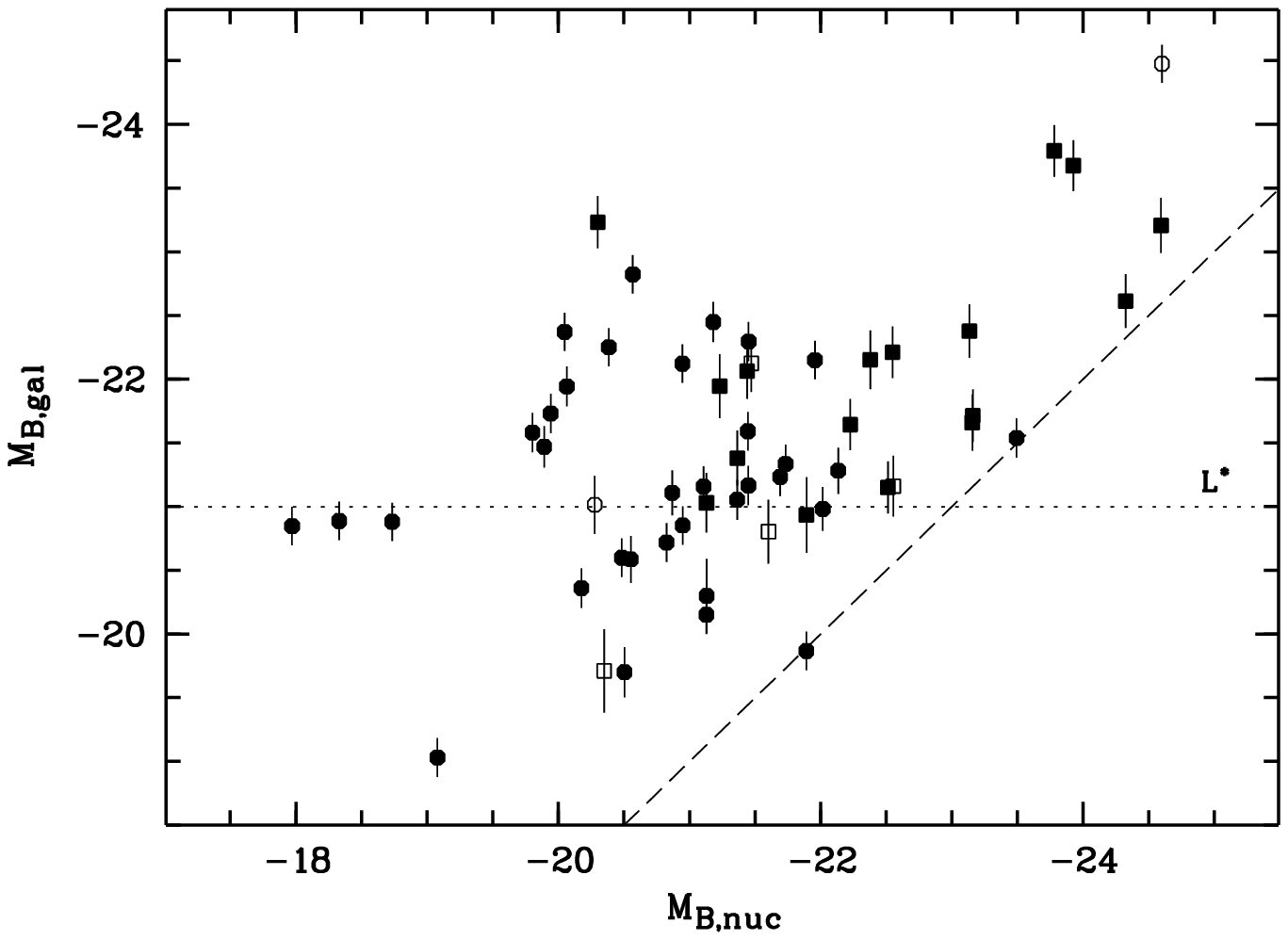}
\caption{
Host galaxy versus nuclear absolute magnitude in the $B$ band for all
57 objects in the sample. Redshift ranges are coded by different
symbols, circles: $z<0.1$, squares: $0.1<z<0.2$. Open symbols denote
the six objects with higher uncertainties marked in Table
\ref{tab:objekte1}. The dashed line is a lower limit with slope 1.0,
similar to the NIR relation of McLeod \& Rieke, assuming the AGN to
radiate with $L\sim0.1L_\mathrm{Edd}$ (see text in
Sect.~\ref{sec:discussion}). The horizontal dotted line marks the
brightness of galaxies with Schechter luminosity $L^*$ for field
galaxies.}
  \label{fig:MB_MB1}
\end{figure*}

\subsection{Absolute magnitudes}\label{sec:absmag}

To convert from apparent to absolute magnitudes we first corrected for
Galactic foreground extinction, using the distribution of neutral
hydrogen column density $N_H$ \citep{dick90} and adopting $A_{B} = 4.2
\times N_H / 58$ with $N_H$ given in units of
$10^{20}\,\mathrm{cm}^{-2}$ \citep[cf.][]{bohl78}. The AGN $K$
correction was based on the usual formula
\[
K_{\rmn{nuc}}(z) = -2.5\,(\alpha + 1)\,\log{(1+z)};
\] 
the value of the spectral index was taken as $\alpha=+0.45$.  This is
significantly different from the canonical value of $\alpha \approx
-0.5$, but appropriate for the low-redshift range under discussion
here \citep[cf.\ the detailed discussion by][]{wiso00b}. Note that
this $K(z)$ relation leads to slightly lower nuclear luminosities than
usually obtained.

For the galaxies we adopted an approximate $K$ correction valid 
for inactive intermediate type (Sab) galaxies
\[
K_{\rmn{gal}}(z) = 3.5\times z,
\]
\citep{fuku95}, valid for $z<0.3$. The $B$ band is particularly
sensitive to $K$ correction effects as the difference between the Sab
relation and the corresponding one for E galaxies is $\sim$0.35~mag at
$z=0.2$. As the spectral energy distributions (SEDs) of our host
galaxies is unknown, the $K$ correction adds a non-negligible
additional uncertainty to the estimated absolute magnitudes of
especially the objects near the upper redshift limit. Assuming a
type-dependent $K$ correction for the 15 galaxies with established
morphological types has little effects to our results, as these
objects are all located at very low redshifts and have small
corrections anyway.

Some previous studies \citep[e.g.,][]{smit86,mcle94a,roen96} have
claimed host galaxy colours bluer than normal for their
morphologically classified galaxy types. As will be discussed in
Sect.~\ref{sec:bluecolours}, this might also be the case for our
present sample. If that case the $K$ correction would be
overestimated. If the SEDs were similar to intermediate to late-type
spirals with ongoing star formation, the appropriate $K$ correction
would be $K_{\rmn{gal}}(z) \simeq 1.5\times z$, and a systematic error
up to 0.4~mag (at $z=0.2$) could occur.
\medskip

The estimated absolute magnitudes $M_{B,\rmn{nuc}}$ and
$M_{B,\rmn{gal}}$ are also listed in Table~\ref{tab:objekte1}.
Figure~\ref{fig:MB_MB1} shows the properties of the sample in terms of
nuclear and galaxy luminosities. The former range from $-18\ga
M_{\rmn{nuc}} \ga -24.5$ (mean $M_{\rmn{nuc}} = -21.3$), the galaxy
magnitudes range from $-19\ga M_{\rmn{gal}} \ga -24$ (mean
$M_{\rmn{gal}} = -21.5$). When the low-quality objects are included,
the mean value for the galaxy brightness remains unchanged, the mean
nuclear luminosity increases by a marginal $0.1$~mag.

When dividing the sample into two subsets along the conventional
though arbitrary division of Seyfert~1 and quasars at $M_{\rmn{tot}} =
-23$, a high-luminosity subsample of 11 objects and a low-luminosity
subsample of 43 objects are formed. The average galaxy magnitude for
the high-luminosity objects is $M_{\rmn{gal}} = -22.6$, or about one
magnitude brighter than for the whole sample (average nuclear
luminosity is $M_{\rmn{nuc}} = -23.1$). For the fainter subsample
these values are $M_{\rmn{gal}} = -21.2$ and $M_{\rmn{nuc}} = -20.8$,
respectively. Incorporating the low-quality objects leaves these
values practically unchanged.

Three features in the distribution are worth noting:

\begin{enumerate}
  \renewcommand{\theenumi}{(\arabic{enumi})}
\item The host galaxies are generally very bright. A large fraction of
the objects, all in the case of the high-luminosity subsample, is
brighter than an $L^*$ field galaxy, some by as much as a factor of
$\sim$10.
\item There is a clear correlation between $M_{\rmn{nuc}}$ and
$M_{\rmn{gal}}$, although this is dominated by the objects with the
very brightest nuclei. No significant correlation is found for the
low-luminosity subsample.
\item As expected from previous studies \citep[e.g.][]{mcle95b}, faint
galaxies with bright nuclei are missing in the distribution. This is
not a selection effect but solely due to a physical lower limit in
host galaxy luminosity at given nuclear luminosity, since all of the
host galaxies are detected.
\end{enumerate}

\subsection{Companions and morphological peculiarities}
\label{sec:peculiarities}

In nine of our images we identified possible close companion objects
($\le 40$~kpc from the nucleus), and ongoing mergers in three further
cases. Of course, we cannot dismiss the possibility that some of the
`companions' could be fore- or background objects. In general (unless
noted otherwise) these objects have been masked out for the analysis
of the host, excluding them from radial profile creation, model
fitting, and luminosity estimation. Here we give some descriptive
comments on the detected companions.
\begin{description}
\item[HE\,0317--2638:] Two extra knots or companions in the host,
$\sim$$6''$ ($\sim$14~kpc) from the center; included for brightness
estimate of host, contribution is $\sim$30\% of total luminosity.

\item[HE\,0517--3243] (= ESO 362-G18): Ongoing merging event, tidal
arm with bright knot or companion galaxy.

\item[IR\,09595--0755:] Spectacular merging event, several knots and arcs.
Object also known as MCG-01-26-011 or the `Sextans ring'.

\item[HE\,1110--1910:] Diffuse knot $\sim$$10''$ ($\sim$30~kpc) from
centre is a spectroscopically confirmed companion (Wisotzki et al., in
prep.). 25\% of total luminosity, included in host brightness.

\item[HE\,1217--1340:] Image contains also a large edge-on spiral
with no obvious interaction; probably a foreground galaxy. Two nearby
objects ($\sim$$12''$ and $\sim$$15''$ distance, or $\sim$24 and
$\sim$30~kpc, respectively) appear stellar.  All excluded from flux
integration.

\item[HE\,1228--1637:] Object at $\sim$$15''$ ($\sim$40~kpc), fainter
by 0.5~mag, no tidal connection visible. Not included. This object happened to
be observed during the spectroscopic follow-up; it is an emission-line
galaxy at the QSO redshift.

\item[HE\,1239--2426:] Confirmed companion at $\sim$$12''$
($\sim$27~kpc), included in luminosity estimate, 8\% of total flux.

\item[HE\,1245--2709:] Faint object $\sim$$7''$ ($\sim$13~kpc) from host
center, 6\% of total flux. Included in total host brightness.

\item[IR\,12495--1308:] Active merger with several
knots, tidal bridges and at least two pronounced nuclei. The second nucleus 
does not show any spectroscopic evidence for activity.

\item[HE\,1328--2509:] Knot well within the visible galaxy borders, 
$\sim$$12''$ ($\sim$9~kpc) from center, $<$5\% of total flux
(included). Could be a foreground star.

\item[HE\,1335--0847:] Object $\sim$$14''$ ($\sim$19~kpc) away from host,
no visible tidal connection. 1.7~mag fainter than host galaxy, not included.

\item[RXJ\,13531--2802:] Two point like objects ($\sim$$12''$ and
$\sim$$14''$ distance, or $\sim$$34$ and $\sim$$40$ kpc, respectively),
brighter by 1.7 and 3~mag than the host galaxy, are most likely
foreground stars.
\end{description}

Compared to high resolution imaging with HST
\citep{disn95,fish96,bahc97} we find that only a very small fraction
of objects has close companions. We believe this to be in part an
effect of data quality. Our data is less deep than the HST imaging and
our images do not resolve faint companions closer than $\sim 2-3''$ to
the nucleus. The close companion statistics by \citet{bahc97} shows
five out of their 20 luminous quasars to have close companions within
$3''$ (projected to $z=0.2$). Of these only three contribute more than
0.1~mag to the total luminosity, none more than 0.25~mag, and and all
but one of these quasars are radio-loud. An order-of-magnitude
estimate from the single radio-quiet quasar yields a total of a few
faint companions to be expected in the unresolved galaxy centers of
the objects in our sample.

\section{Comparison to existing studies} \label{sec:others}

Although numerous imaging surveys of QSO host galaxies have been
published, few of them provide sufficient sample size and cover a
similar region of the Hubble diagram to be directly comparable to our
sample. The list of comparison studies reduces to two when looking for
the $B$ band. At other wavelengths the studies by \citeauthor{mcle94a}
and \citeauthor{dunl03} are the most relevant for this sample.

\subsection{\textit{B} band studies}

A larger $B$ band CCD study of host galaxies at low redshifts was
published by \citet{hutc87} and \citet{hutc89}. A total of 50 objects
were analysed in both $B$ and $R$, each half of which were radio-loud
and radio-quiet QSOs, respectively. The RLQ and RQQ samples were
selected to `match' in their $z$ distributions. Only 11 RQQ and 8 RLQ
hosts could be detected in the $B$ band. 

There are no objects common to Hutchings' and our sample so that only
a statistical comparison is possible. The 11 detected RQQ host
galaxies are on average $\sim$1~mag more luminous in $B$ than an
inactive $L^*$ galaxy, similar to our results. However, the small
sample size and its spread in luminosity and redshift space precludes
a more detailed comparison.

More substantial is the study by \citet{scha00}. They observed a
sample of 76 low redshift ($z\le0.15$) AGN of low and intermediate
luminosities ($-24<M_{B,\mathrm{tot}}<-18$) with the HST and ground
based telescopes. The sample was X-ray selected from the Einstein
Extended Medium Sensitivity Survey, thus not prone to selection
effects from host galaxy morphology.

Ground based imaging was performed in the $B$ and $R$ bands. The 66
objects observed in the $B$ band show host galaxy properties similar
to our lower-luminosity subsample, with a host galaxy luminosity range
of $-23.0<M_{B,\mathrm{gal}}<-18.2$ and median of $-21.0$ (converted
to Vega zeropoint from their AB magnitude system).

The authors report morphologies and colours not different from
inactive galaxies. No strong merger events are observed, only central
bars in a number of objects. Individual $B-I$ colours of their host
galaxies show a large spread, which are, as the authors note, in part
due to the separation of nucleus and host galaxy components. The mean
colour $B-I\sim2.2$ of their galaxy components is very similar to that
of inactive galaxies without strong star formation.

\begin{table}
\caption{Optical and optical-NIR colours for ellipticals, Sbc, and Scd
spirals, and intermediate types for galaxies with unidentified type
\citep{fuku95,fioc99}.}
\label{tab:B-H}
\begin{center}
\begin{tabular}[t]{crrrr}
  \hline
  Typ& $B\!-\!V$& $B\!-\!R$ &$V\!-\!H$& $B\!-\!H$\\
  \hline \hline
\rule{0cm}{2.7ex}%
E&       0.96& 1.57& 3.05& 4.0\\
interm.& 0.77& 1.33& 2.85& 3.6\\
Sbc&     0.57& 1.09& 2.7 & 3.3\\
Scd&     0.50& 1.00& 2.4 & 2.9%
\rule[-1.5ex]{0cm}{1.5ex}\\
\hline
\end{tabular}
\end{center}
\end{table}

\begin{figure*}
\includegraphics[bb = 38 81 534 272,clip,angle=0,width=17.7cm]{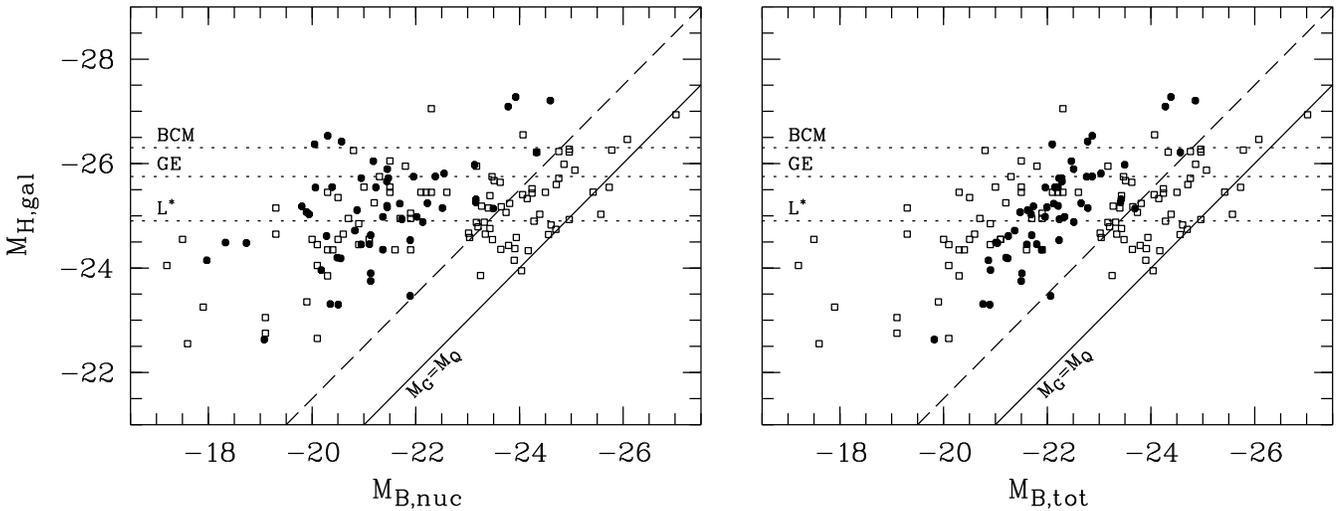}
\caption{Predicted $H$ band galaxy magnitudes vs.\ the nuclear (left)
and total (right) $B$ band magnitudes for the HES sample.  Open
symbols show the objects by \citet{mcle94a,mcle94b,mcle95b}.  The
horizontal lines are reference galaxy luminosities: Schechter $L^*$ of
field galaxies, giant ellipticals (GE), and brightest cluster members.
The solid diagonal line is the lower limit in galaxy $H$ band
magnitude as a function of nuclear magnitude
$M_{H,\mathrm{gal}}=M_{B,\mathrm{tot}}$ as first found by McLeod \&
Rieke, the dashed line is parallel to this relation but shifted by
1.5~mag to fit the HES sample.}
\label{fig:MH_MB_col}
\end{figure*}

\subsection{NIR data by McLeod \& Rieke}\label{sec:mcleod}

In three important papers, McLeod \& Rieke studied the hosts of a
total of $\sim$100 QSOs and Seyfert~1 galaxies in the near infrared
$H$ band \citep[hereafter MR1, MR2, MR3, or MR for the authors in
general]{mcle94a,mcle94b,mcle95b}. Together with \citet{scha00} their
studies are among the few where objects were selected in a well-defined
way. Attempting to build representative samples, MR used subsamples
from the Palomar-Green Bright Quasar Survey \citep{schm83} and the CfA
Seyfert sample \citep{huch92}.

The combined MR sample covers a somewhat larger range of nuclear
absolute magnitudes than the HES, but the Hubble diagram occupancy of
MR and HES samples is otherwise very similar. Unfortunately, there is
only one object common to both studies (see sect.~\ref{sec:common}
below), so again the only way to compare the two sets of observations
is by their statistical distribution of properties.

Instead of computing average statistical colours, we use our $B$ band
measurements to predict $H$ band magnitudes for our sample assuming
$B-H$ colours of normal inactive galaxies, depending on type if
applicable. Typical optical to near infrared colours were taken from
\citet{fuku95} and \citet{fioc99} and are listed in
Tab.~\ref{tab:B-H}.

For identified spirals and ellipticals we used $B-H = 3.3$ and 4.0,
respectively, for the remaining objects $B-H = 3.6$ was adopted. The
resulting predicted $H$ band luminosities are included in
Tab.~\ref{tab:objekte1}. Fig.~\ref{fig:MH_MB_col} shows the
distributions of observed MR and predicted HES $H$ band luminosities,
plotted against nuclear absolute $B$ magnitude. However, a fair
comparison between the HES and the MR samples is hampered by the fact
that the `nuclear' magnitudes quoted by MR are rather ill-defined:
While for the Palomar-Green BQS these are basically isophotal, i.e.\
in good approximation \emph{total} magnitudes, the Zwicky magnitudes
given for the CfA sample contain highly unequal mixtures of light from
stars in the central host galaxy regions and from the AGN. Being
unable to reconstruct similar compound magnitudes for our objects, we
have tried to bracket the correct distribution by plotting the HES
data in two versions, shown in the two panels of
Fig.~\ref{fig:MH_MB_col}: Total magnitudes (right-hand panel) are
reasonable for the comparison between HES and BQS objects from MR1 and
MR2; for the fainter AGN of the CfA sample galaxies (MR3), true
nuclear magnitudes are more appropriate.

The figure displays distinct differences between observed and
predicted $H$ band luminosities. In the MR data, several of the most
luminous hosts are among the CfA galaxies, whereas in the HES sample
the most luminous hosts also harbour very luminous nuclei. The
distributions appear to be consistent in the low-luminosity Seyfert
regime, but there is a strong mismatch at the high-luminosity end
which is independent of the choice of nuclear magnitude (in fact, the
mismatch increases when using `true nuclear' magnitudes for the HES
objects). Table~\ref{tab:disk_mcl1} quantifies this in terms of
average $M_{H,\rmn{gal}}$ values for the two luminosity bins (MR
restricted to $z\le0.2$, and using total magnitudes for the high-,
true nuclear magnitudes for the low luminosities). While for the
low-$L$ bin the samples are consistent within the expected
uncertainties, the discrepancy for the high-luminosity QSOs is
significant.

\begin{table*}
\begin{small}
\begin{center}
\caption{Comparison of predicted $M_{H,\rmn{gal}}$ as a function of
total luminosity, compared with values measured by McLeod \& Rieke.
The McLeod \& Rieke sample was restricted to $z\le0.2$ and the six
`low quality' objects have been removed from the HES sample. $N$ is
the number of objects in each subsample, $\overline{z}$ is the average
redshift, and $\Delta M$ is the difference in magnitudes between MR
and HES hosts.  Host galaxy colours of the HES objects were first
assumed to be `normal' (first set of columns), and then as similar to
star-forming Scd galaxies (last columns; see text for details).}
\begin{tabular}{ccr@{.}lr@{.}lr@{.}lccr@{.}lr@{.}lr@{.}lr@{.}lr@{.}lr@{.}l}
\hline
&
\multicolumn{7}{l}{MR} &
&
\multicolumn{5}{l}{HES} &
\multicolumn{4}{c}{normal} &
\multicolumn{4}{c}{Scd-like}
\\
\cline{2-8}\cline{10-22}\noalign{\smallskip}
$M_{B}$ & 
$N$ & 
\multicolumn{2}{c}{$\overline{z}$} & 
\multicolumn{2}{c}{$\overline{M}_B$} & 
\multicolumn{2}{c}{$\overline{M}_{H,\rmn{gal}}$} & 
&
$N$ &
\multicolumn{2}{c}{$\overline{z}$} & 
\multicolumn{2}{c}{$\overline{M}_B$} & 
\multicolumn{2}{c}{$\overline{M}_{H,\rmn{gal}}$} &
\multicolumn{2}{c}{$\Delta M$} &
\multicolumn{2}{c}{$\overline{M}_{H,\rmn{gal}}$} &
\multicolumn{2}{c}{$\Delta M_{\rmn{Scd}}$}\\
\hline
\hline
\rule{0cm}{2.7ex}%
$M_{B,\rmn{tot}}\le -23$     & 43 &0&13  &$-$24&0 & $-$25&1 && 11 & 0&14 & $-$23&8 &$-$26&1 &    1&0 & $-$25&3 &    0&2 \\
$-23<M_{B,\rmn{tot}}$        & 46 &0&019 &$-$20&5 & $-$24&7 && 43 & 0&07 & $-$20&8 &$-$24&8 & 0&1 & $-$24&0 & $-$0&7 \\
\hline
\end{tabular}
\label{tab:disk_mcl1}
\end{center}
\end{small}
\end{table*}

Before discussing for possible explanations, we identify one technical
point, often overlooked, that might reduce the discrepancy but cannot
account for the effect as such. The decomposition into nuclei and
hosts allowed us to apply appropriate $K$ corrections to both
components separately. Within the MR sample this was not possible, as
the contributions of galaxy light to the total $B$ magnitudes were not
known. Applying a QSO-type $K$ correction to such a composite $B$
magnitude, as done by MR, leads invariably to an overestimation of the
object's luminosity; the effect will be small when the nucleus
outshines the host, and largest for a low-luminosity nucleus in a
bright host at relatively high redshift. Using our own data we have
estimated that this effect could result in a bias of
$M_{B,\mathrm{tot}} \sim 0.5$~mag for extreme cases, however it will
generally be much smaller.

We note in passing that in preparing the comparison in
Fig.~\ref{fig:MH_MB_col} and Tab.~\ref{tab:disk_mcl1}, we adopted the
same QSO $K$ correction as MR, thus using $\alpha=-0.5$ rather than
the more previously used value of $\alpha=+0.45$.

\subsection{Studies by \protect{\citeauthor{dunl03}}}

\citet{dunl93}, \citet{tayl96}, \citet{mclu99}, \citet{hugh00},
\citet{nola01} and \citet{dunl03} compared the properties of RLQs and
RQQs at low redshifts. From the catalogue of \citet{vero91} they
selected samples of $\sim$10 objects for each class with $z\le 0.35$
and $M_{V,\rmn{tot}} \le -23$, to `match' in $z$ and $V$ distribution,
and obtained HST F675W ($\sim$$R$) band and ground based $K$ band
imaging as well as off-nuclear host galaxy spectra. Due to the
selection the sample has a smaller Hubble diagram coverage than our
sample.

From morphological analysis all except two RQQ host galaxies were
found to be large ellipticals. Both the derived $R-K$ colours and the
spectra showed SEDs consistent with the morphological classification,
i.e.\ a dominant evolved, red stellar population with no enhanced blue
component. They determined average values of $M_{K,\rmn{gal}}=-26.3$
and $M_{R,\rmn{gal}}=-23.5$ for their combined RQQ and RLQ samples,
the host galaxies of the RLQ being more luminous than their RQQ
counterparts by about 0.5~mag.

Their samples were primarily selected for the comparison of the
properties of the host galaxies of RQQs, RLQs and radio galaxies and
are not statistically complete or representative. We still want to use
their RQQ sample for comparison because of the high luminosity and the
reliable methods applied in extracting the host galaxy properties.

Our bright subsample has a very similar total luminosity,
$M_{V,\mathrm{tot}}=-23.8$ when assuming $M_V\simeq M_B$, compared to
$M_{V,\mathrm{tot}}=-24.0$ for the Dunlop RQQ sample. Their RQQ host
galaxies have an $R$ band luminosity of
$M_{R,\mathrm{gal}}=-23.3$. Converting again our average $B$ host
luminosity from Sect.~\ref{sec:absmag} above to the $R$ band, assuming
$B-R=1.3$, valid for intermediate type inactive galaxies \citep[see
Tab.~\ref{tab:B-H}]{fuku95}, we arive at a predicted
$M_{R,\mathrm{gal}}=-24.3$, much brighter than actually observed for
their sample.

\subsection{Comparison of individual objects}\label{sec:common}

The HES sample contains only two objects that were already studied by
others, in various photometric bands. We derive colours from the
measured apparent magnitudes and compare them to those of intermediate
type (Sab) inactive galaxies, including the $K$ corrections for both
colours taken from \citet{cole80}, \citet{fuku95}, and \citet{yosh88}.
\begin{description}
\item[PKS\,1020--103,] is an intermediately luminous RLQ at
$z=0.197$. It is one of the objects that we also analysed with
two-dimensional modelling, which in this case yielded a luminosity
less than 0.1~mag different. The object was observed by \citet{dunl03}
in the $R$ band. The resulting $B-R=2.0$ is identical to the
corresponding value expected for a normal galaxy at this
redshift. Thus this object shows no unusual colours, which is
confirmed by \citeauthor{dunl03} and \citeauthor{nola01}
\item[PG\,1416--1256,] $z=0.129$, has been observed by \citet{mcle94b}
in the $H$ band, yielding $B-H=3.45$ whereas a normal galaxy at this
redshift on average shows $B-H = 4.2$.
\end{description}

Thus one of these two objects appears much bluer than comparable
inactive early type spirals.

\section{Discussion} \label{sec:discussion}
In Sections~\ref{sec:uncert} through \ref{sec:externaltest} we
evaluated possible sources for systematic biases in extracted host
galaxy luminosities, and concluded that we accounted for all major
systematic effects. As previously found, there exists a lower limit in
host galaxy luminosity as a function of nuclear luminosity. 

As the main result of this analysis we find that high luminosity QSOs
are hosted by extraordinarily luminous (in the $B$ band)
galaxies. When comparing our data with other samples observed at
longer wavelengths, there is a significant discrepancy when assuming
host galaxy colours typical for inactive galaxies. We see two options
to resolve this discrepancy: Either our sample contains more luminous
galaxies than those of previous studies, possibly due to QSO sample
selection biases. Alternatively, the phenomenon could be explained by
unusually blue host galaxy colours. We now investigate each of these
options in more detail.

\subsection{Are previous samples biased?}\label{sec:bias}

If the HES selection procedure of low-redshift QSOs is largely
unbiased with respect to host galaxy properties, could then the
apparent excess of high-luminosity hosts be related to an
incompleteness of previous samples? Unfortunately there is very little
overlap between the sky coverage of the HES and other surveys, so the
direct route to test this on an object-to-object comparison is not
possible.

As already mentioned in the introduction, most optical QSO surveys
discriminate against objects with non-stellar morphology and are
thereby biased in a straightforward way. The amount of incompleteness
introduced into the samples depends on several factors: limiting
surface brightness of the survey imaging data (often photographic
plates); photometric band employed; definition of `extendedness'. For
the Palomar-Green survey \citep{schm83}, probably the most important
supplier of luminous low-$z$ QSOs and the principle source for MR,
\citet{koeh97} estimated an incompleteness of up to a factor of $\sim
3$--5 based on a comparison of measured number counts in the
HES. However, in a more recent study with the full HES, \citet{wiso00}
showed an incompleteness of only $\sim$1.5. It is not clear if this
incompleteness is brightness dependent but it is likely that high
luminosity QSOs suffer less than objects at the detection limit.
Overall, we can not completely rule out that the PG incompleteness
might have an effect, but probably the influence of any possible
selection bias would be small.

For the lower luminosity CfA Seyfert sample investigated by MR, which
is based on a dedicated galaxy catalogue, no selection bias against
extended objects is expected. Correspondingly, the $B-H$ host galaxy
colours in combination with the HES are very consistent with
`inactive' colours for the low-luminosity Seyferts.

\subsection{Blue colours due to younger population age?}
\label{sec:bluecolours}

After excluding selection biases as a dominant factor, we are left
with the implication that we see bluer than normal colours for some of
our host galaxies. In particular, there seems to be a division in host
galaxy colours that is correlated with luminosity. The low luminosity
QSO sample seems to show `normal' colours for the hosts and a wide
range of nuclear luminosities. These objects seem to be normal
galaxies in many aspects except for the active nuclei in their
centers.

The higher luminosity QSOs seem to be bluer than normal in
optical--NIR colours by $\sim$1~mag. Even though the number of objects
involved is small (eleven objects), the difference is not dominated by
e.g.\ the most luminous hosts in the sample. This unusual colour has
two effects: (1) The $(B-H)$ colour would numerically decrease, and
(2) the $K$ correction for the galaxy would become smaller. Both
effects operate in the same direction, making the predicted $H$ band
magnitudes fainter. The first effect causes a constant offset for all
objects, the second is redshift-dependent and largest for the
high-luminosity objects at $z\sim0.2$.

\begin{figure}
\begin{center}
\includegraphics[bb = 37 81 299 272,clip,angle=0,width=8.4cm]{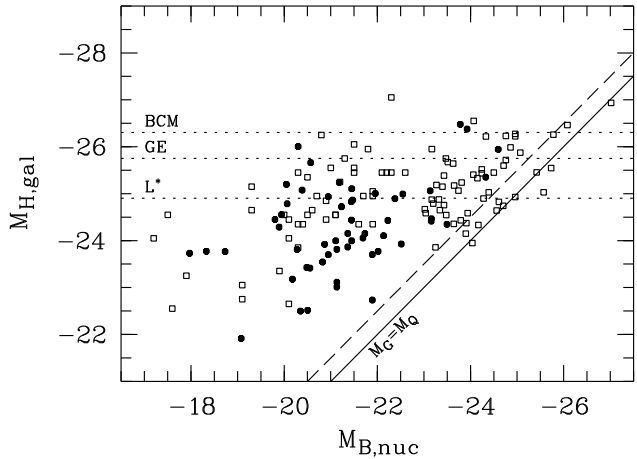}
\caption{Predicted HES and observed MR $H$ band luminosities, 
         computed with Scd-type colours and $K$ corrections. 
         Symbols and lines as in Fig.~\ref{fig:MH_MB_col}.
\label{fig:MB_gal_scd}}
\end{center}
\end{figure}

To illustrate the consequences, we used the Scd-type colours listed in
Tab.~\ref{tab:B-H}, together with appropriate $K$ corrections from
\citet{grie82} and \citet{fuku95}, and recalculated the predicted $H$
band magnitudes. The resulting colours of $(B-H)\sim 2.9$ and a $B$
band $K$ correction of $1.5\times z$ would result in an $H$ band
`dimming' of the brightest HES galaxies by $\sim$1~mag. Note that such
colours would be also bring the $B$ luminosity of PG\,1416--1256 from
Sect.~\ref{sec:common} into agreement with the results from MR.
Fig.~\ref{fig:MB_gal_scd} shows the changes in the comparison of MR
and HES samples when an Scd-like stellar population is assumed. The
strongest changes occur at relatively high $z$, affecting the most
luminous objects. The discrepancy between the bright end of the
samples is largely removed. Colours of $(B-H)\sim2.5$ are however more
typical of late type spirals. Such morphological types would be
contrary to the established fact that luminous AGN reside
predominantly in early type galaxies.

Conversely, if our luminous hosts are indeed mostly of early galaxy
type, such blue colour would be somewhat unexpected. Possible
explanations could be ongoing star formation at a rate substantially
higher than expected, an overall younger stellar population, or a
recent starburst.

The resulting composite spectral energy distributions would at short
wavelengths be dominated by the youngest stars, but the bulk of
radiation in the red would come from the old stellar population. Note
that virtually all of the bright HES objects are at the high redshift
end of our sample, $z\ga0.1$. For these objects we start to measure in
the rest-frame $U$ band rather than in $B$. Because of the strong
4000~\AA\ break displayed by old stellar populations, already a
relatively small fraction of very young stars could account for a
significant rest-frame $U$ band excess, boosting our measured $B$ band
magnitudes, but show relatively little effect in the other colours.

To give an order-of-magnitude estimate, we compute the effect of a
simple instantaneous burst population of young stars and the masses
nessessary to produce these blue colours. Adopting the simple
two-component model of an old and a young stellar population, we used
the colours and luminosities for starburst models given by
\citet{leit95}. The inferred mass of a burst population depends mainly
on the the assumed burst age, yielding values of $\sim$1\% for an age
of $10^{6.5}$~yrs and $\sim$5\% for an age of $10^7$~yrs. For a
$10^8$~yrs population already a significant fraction is required.

Are such blue colours compatible with previous results? Some earlier
studies have indicated bluer than normal colours for host galaxies,
however others find largely normal values:
\begin{itemize}
\item The abovementioned study by \citet{hutc89} finds $B-R = 0.94$
for their RQQ hosts and even $B-R = 0.29$ for the RLQ hosts, compared
to $B-R = 1.4$ for inactive Sab spirals and $B-R = 1.6$ for inactive
ellipticals.
\item \citet{roen96} and \citet{oern01} obtain rest frame $B-V$
colours for three elliptical RQQ host galaxies, bluer than normal by
0.6~mag. These object have similar luminosities than our luminous
objects.
\item The early spectrocopic studies by \citet{boro84} and
\citet{boro85} resulted in a range of colours, down to $B-V\simeq
0.2$, the colour of very late spirals. Five of six QSOs with $B-V<0.5$
resided in luminous early type galaxies
\citep{stoc91,hutc92,bahc97,marq01}, all of these RLQs.
\item A reanalysis of HST data of nine host galaxies by
\citet{mcle95a} shows a tendency for blue colours in $V-H$, an excess
by $\sim$0.5~mag compared to inactive Sab galaxies or $\sim$0.7~mag to
ellipticals.
\end{itemize}

So several studies find evidence for bluer colours. On the other hand,
as noted above, the more luminous sample investigated by
\citet{dunl03} and \citet{nola01}, all elliptical galaxies by
morphology, was found to show $R-K$ colours and spectra typical of old
evolved stellar populations, with ages $\ga 8$~Gyrs and no traces of
star formation. Similarly, normal colours have been reported for the
(lower-luminosity) X-ray selected objects by \citet{scha00} and for
the multicolour study of low-luminosity (mostly spiral) Seyfert
galaxies by \citet{koti94}. This agrees well with the results our low
luminosity subsample. Furthermore it should be expected that the most
luminous elliptical galaxies, with low remaining gas content, also
show old stellar populations and colours. If \citeauthor{dunl03} find
\textit{only} such objects, then this may or may not be an effect of
sample selection. On the other hand, at least in the luminosity regime
of our sample, there exists a range of host galaxies colours, from
normal to very blue.

Due to the current lack of detailed diagnostics, we prefer to refrain
from speculating on the cause for the enhanced blue colours at this
point. Obvious candidates are tidal interaction, or minor or major
mergers. If all or some nuclear activity is triggered by such events
then it is well possible that also star formation is induced in the
host galaxy by the same process. However our very limited data quality
does not allow a meaningful correlation of colour and either companion
statistics or a measurement of distortion as an indicator for
interaction. Deep higher resolution data and in particularly a well
defined comparison sample of inactive galaxies are required for this
task.

\section{Conclusions} \label{sec:conclusions}
We analysed a new sample of low-$z$ AGN, encompassing both luminous
QSOs and intermediate-luminosity Seyfert galaxies. The objects were
selected from the Hamburg/ESO survey, for which host galaxy dependent
selection biases are greatly reduced compared to other optical
surveys.

We found that the host galaxies in our sample are generally very
luminous, typically around $L^\star$ for the lower nuclear
luminosities, and several times brighter for high luminosity
QSOs. From a statistical comparison to studies in other bands we
conclude that the latter show signs of significantly bluer colours
($B-H\sim 2.5$) than inactive galaxies. We argue that the mismatch of
observed and predicted $H$ band magnitudes for the most luminous QSOs
in the Palomar-Green and HES samples is not likely to be explained by
selection effects.

These discrepancies can be resolved when invoking an additional
contribution from younger stars, adding blue light to the bulk
emission of an old(er) evolved stellar population. In case of a recent
starburst the inferred stellar masses would only need to comprise a
few percent of the total luminous mass. Such a starburst could be
induced e.g.\ by a minor or major merging event that at the same time
is triggering the nuclear activity in luminous QSOs.

More and different data is required to test the distribution and
origin of bluer colours. As one option multiband imaging would allow
homogenous photometry and direct measurement of colours. To identify
the source of an enhanced star formation rate, spectroscopy is
required. Only with spectra, timescales and masses of the star forming
processes become available. In fact, very little data involving either
direct multiband photometry or QSO host spectroscopy can be found in
the literature. Most of the few existing investigations reported host
galaxy colours that are bluer than those of normal galaxies,
consistent with our inferences.

\section*{Acknowledgements}
This work has profited from generous time allocation at ESO telescopes
for the Hamburg/ESO survey as ESO key programme 02-009-45K. KJ
gratefully acknowledges support by the \emph{Studien\-stiftung des
deutschen Volkes}.

\bibliographystyle{mn2e}
\bibliography{knuds}

\bsp
\label{lastpage}
\end{document}